\documentclass[fleqn,usenatbib]{mnras}
\usepackage{newtxtext,newtxmath}
\usepackage[T1]{fontenc}
\DeclareRobustCommand{\VAN}[3]{#2}
\let\VANthebibliography\thebibliography
\def\thebibliography{\DeclareRobustCommand{\VAN}[3]{##3}\VANthebibliography}
\usepackage{graphicx}	
\usepackage{amsmath}
\usepackage{lineno}
\usepackage{color}
\usepackage{graphicx}
\usepackage{longtable}
\usepackage{multirow}
\usepackage{url}
\usepackage{subfigure}
\usepackage{lipsum}
\usepackage{ulem}
\usepackage{bm}
\usepackage{threeparttable}
\usepackage{orcidlink}

\newcommand{\pccm}{pc\,cm$^{-2}$ }
\newcommand{\radm}{rad\,m$^{-2}$ }
\newcommand{\gpcyr}{Gpc$^{-3}$\,yr$^{-1}$ }
\defcitealias{SG03}{SG}
\defcitealias{Thompson2005}{TQM}

\title[FRBs in AGN disks]{Fast Radio Bursts in the Disks of Active Galactic Nuclei}

\author[Zhao et al.]{Z. Y. Zhao\orcidlink{0000-0002-2171-9861},$^{1}$
K. Chen\orcidlink{0000-0001-8955-0452},$^{1}$
F. Y. Wang\orcidlink{0000-0003-4157-7714},$^{1,2}$\thanks{E-mail:fayinwang@nju.edu.cn}
Zi-Gao Dai\orcidlink{0000-0002-7835-8585},$^{3}$
\\
$^{1}$School of Astronomy and Space Science, Nanjing University, Nanjing 210093, China\\
$^{2}$Key Laboratory of Modern Astronomy and Astrophysics (Nanjing University), Ministry of Education, Nanjing 210093, China\\
$^{3}$Department of Astronomy, School of Physical Sciences, University of Science and Technology of China, Hefei 230026, China
}
\date{Accepted XXX. Received YYY; in original form ZZZ}

\pubyear{2024}

\begin{document}
\label{firstpage}
\pagerange{\pageref{firstpage}--\pageref{lastpage}}
\maketitle

\begin{abstract}
Fast radio bursts (FRBs) are luminous millisecond-duration radio pulses with extragalactic origin, which were discovered more than a decade ago. Despite the numerous samples, the physical origin of FRBs remains poorly understood. FRBs have been thought to originate from young magnetars or accreting compact objects (COs). Massive stars or COs are predicted to be embedded in the accretion disks of active galactic nuclei (AGNs). The dense disk absorbs FRBs severely, making them difficult to observe. However, progenitors’ ejecta or outflow feedback from the accreting COs interact with the disk material to form a cavity. The existence of the cavity can reduce the absorption by the dense disk materials, making FRBs escape. Here we investigate the production and propagation of FRBs in AGN disks and find that the AGN environments lead to the following unique observational properties, which can be verified in future observation. First, the dense material in the disk can cause large dispersion measure (DM) and rotation measure (RM). Second, the toroidal magnetic field in the AGN disk can cause Faraday conversion. Third, during the shock breakout, DM and RM show non-power-law evolution patterns over time. Fourth, for accreting-powered models, higher accretion rates lead to more bright bursts in AGN disks, accounting for up to $1\%$ of total bright repeating FRBs.
\end{abstract}


%%%%%%%%%%%%%%%%%%%%%%%%%%%%%%%%%%%%%%%%%%%%%%%%%%
\section{Introduction}
Fast radio bursts (FRBs) are luminous millisecond-duration radio pulses with extragalactic origin, which was discovered by \cite{Lorimer2007}. Observations indicate that the properties of FRBs are complex and diverse, including repetition, energy distribution, host galaxy, polarization property and surrounding environment (see reviews of \citealt{Xiao2021,Petroff2022,ZB2023}). Despite the numerous samples \citep{CHIME/FRBCollaboration2021}, the physical origin of FRBs remains poorly understood. Among the many proposed models, the most promising ones are those related to active magnetars \citep{Lyubarsky2014,Katz2016,Murase2016,Beloborodov2017,Kashiyama2017,Metzger2017,Yang2018,Metzger2019,Kumar2020,Wang2020,Lu2020}, which is supported by the detection of FRB 200428 from the Galactic magnetar SGR J1935+2154 \citep{Bochenek2020,CHIME2020}. 

In magnetar models, based on the difference in emission regions \citep{ZB2020}, it can be roughly divided into `close-in' scenario ('pulsar-like' emission originates from the magnetosphere of magnetars, e.g., \citealt{Yang2018,Kumar2020,Lu2020}) and `faraway' scenario (gamma-ray bursts like, 'GRB-like' emission originates from relativistic outflows, e.g., \citealt{Lyubarsky2014,Beloborodov2017,Metzger2019}). Alternatively, FRBs have been argued to arise possibly from collisions of pulsars with asteroids or asteroid belts \citep{Geng2015,Dai2016}. \cite{Luo2020} found that the polarization position angle (PA) of FRB 180301 swings across the pulse 
profiles, which is consistent with a magnetospheric origin. Recently, the swing of PAs has also been reported in simulations of relativistic magnetized ion-electron shocks \citep{Iwamoto2024}. For the FRB which has been active for ten years \citep{Li2021}, the energy budget for the magnetar's magnetic energy becomes strained especially for the low-efficiency relativistic shock origin \citep{Wu2020}. One way to alleviate this problem is to seek other central engines.
The close-in models are only applicable to magnetar engines, while the far-away models are applicable to a wider range of scenarios as long as energy can be injected into the surrounding medium from the central engine. Accreting-powered models from compact objects (COs), e.g., black holes (BHs) and neutron stars (NSs), have been proposed \citep{Katz2020,Deng2021,LiQC2021,Sridhar2021,Sridhar2022}.

COs are predicted to be embedded in the accretion disk of active galactic nuclei (AGN). Magnetars could form from the core-collapse (CC) of massive stars. Massive stars in AGN disks are more likely to evolve rapid rotation \citep{Jermyn2021}, and eventually more magnetars are produced via the dynamo mechanism \citep{Raynaud2020}. In some cases, a magnetar can also be formed in the following process: binary neutron star (BNS) mergers \citep{Dai1998,Rosswog2003,Dai2006,Price2006,Giacomazzo2013}, binary
white dwarf (BWD) mergers \citep{King2001,Yoon2007,Schwab2016}, accretion-induced collapse (AIC) of WDs \citep{Nomoto1991,Tauris2013,Schwab2015} and neutron star-white dwarf (NSWD) mergers \citep{Zhong2020}. The AGN disk provides a favorable environment for these processes \citep[e.g.][]{Mckernan2020, Perna2021, Zhu2021a, Luo2023}. 
On the other hand, the accretion rate onto COs can be hyper-Eddington \citep[e.g.][]{Wang2021a, Pan2021b, Chen2023a}, which would release enough inflow energy for single FRB or successive ones \citep{Stone2017, Bartos2017}. Meanwhile, the accreting COs can potentially launch relativistic jets \citep[e.g.][]{Tagawa2022, Tagawa23, Chen2024}, providing an optimistic channel to drive FRBs. Overall, FRBs are expected to be produced by both young magnetars and accreting COs in AGN disks. Alternatively, FRBs have been argued to arise possibly from collisions of pulsars with asteroids or asteroid belts.

FRB propagation in a complex medium affects the observational properties, such as dispersion measures (DM), Faraday rotation measures (RM) and polarization property, scintillation and scattering, which have been studied in different environments, e.g., supernova remnant (SNR;\citealt{Yang2017,Piro2018}),  compact binary merger remnant \citep{Zhao2021a}, pulsar/magnetar wind nebula \citep{Margalit2018,YYH2019,Zhao2021b} and outflows from companion \citep{Wang2022,Zhao2023,Xia2023} or accretion disk \citep{Sridhar2021,Sridhar2022}. For some repeating FRBs, the large DMs or RMs \citep[e.g.][]{Michilli2018,Hilmarsson2021,Niu2022,Anna-Thomas2023} indicate that the sources are embedded in dense environments \citep{Katz2021,Katz2022}. A dense and magnetized plasma in AGN disks is naturally considered to contribute the large DMs or RMs.

% For FRB sources in AGN disk, the radio signals will also be dispersed or absorbed by the (circum-)disk medium. The magnetized disk causes Faraday rotation and conversion. 

\begin{figure*}
    \centering
    \includegraphics[width = 1\textwidth]{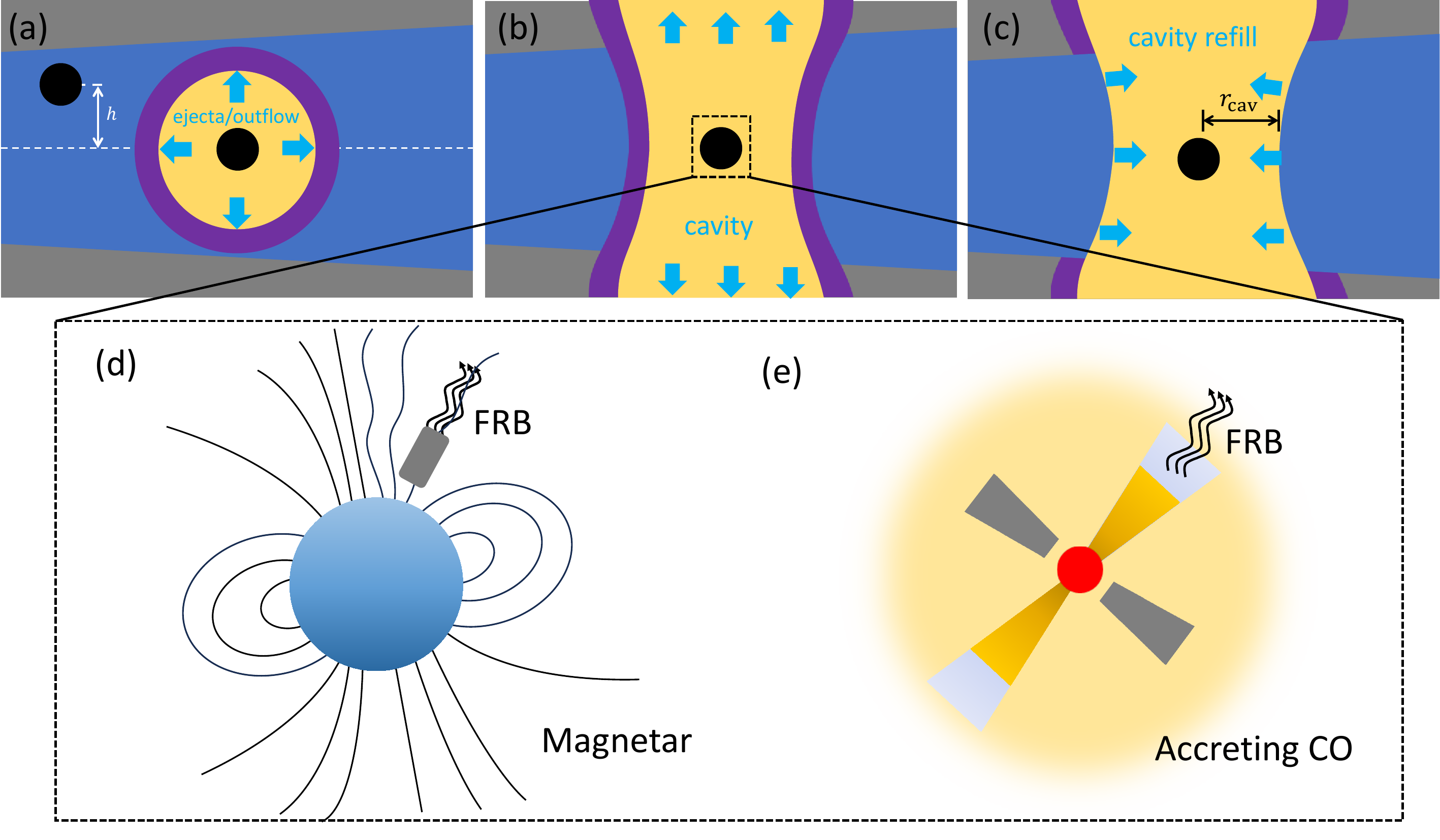}
    \caption{Schematic diagram of the generation and propagation of FRBs in AGN disk. (a) If the feedback of the source is weak, FRBs only can be detected from source sites at a few scale heights due to the absorption from the dense disk materials. If the feedback of the source is strong, a shock is formed during the interaction between the ejecta/outflows and the disk materials. The shock punches a cavity in the AGN disk. When $t\leq t_{\mathrm{bre}}$, the shock shell is almost spherical. (b) After the shock breaks out, the vertical propagation of the shock becomes easier because the density of gas in the AGN disk decreases rapidly. At this point, the shape of the cavity becomes ringlike. (c) When the shock decelerates to the speed of the local sound, the shock evolution ends. Finally, the cavity is refilled by the AGN disk material. In the case of progenitor ejecta, the cavity can only be opened once. But for the case of accretion outflows, the accretion rate can be restored to hyper-Eddington after the cavity is refilled. Then, the strong disk outflow forms the cavity again and the evolution process is periodic. (d) The `close-in' models. FRB originates from the magnetosphere of magnetars. The AGN disk environment only affects the propagation effect but not the radiation properties for magnetospheric origins. (e) The `far-away' models. FRB originates from relativistic outflows of accreting COs. In addition to the propagation effect, high accretion rates in the AGN disk lead to more bright bursts for accreting-powered origins. }
    \label{fig:model}
\end{figure*}

\begin{figure*}
    \centering
    \includegraphics[width = 1\textwidth]{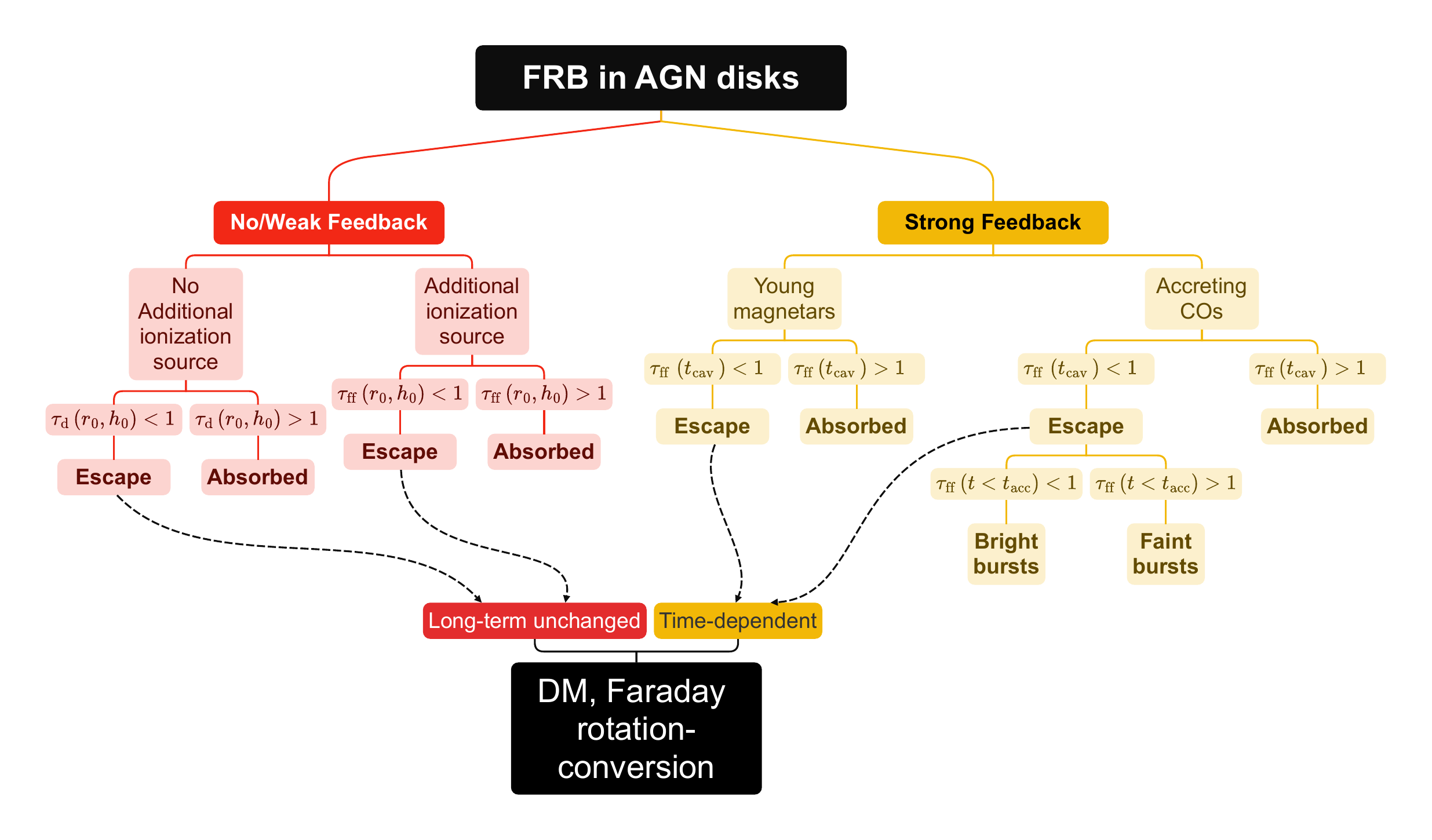}
    \caption{Related physical processes of FRBs in AGN disks. FRBs are thought to have originated from active magnetars or accreting compact COs. Progenitors’ ejecta or accretion disk outflows  interact with the disk material to form a cavity. If the feedback of the ejecta or outflows is weak, whether the FRB is absorbed or not depends only on the location of the source. In this case, the DM and RM from disk material are stable in the long term but fluctuate in the short term due to the turbulence. The toroidal magnetic field in the AGN disk can cause Faraday conversion. If the feedback is strong, as the cavity expands, at some point the optical depth drops below unity, allowing the FRB to be observed. The cavity expansion causes time-dependent DM and RM. For accreting-powered models, the burst luminosity depends on the accreting rate. The hyper-Eddington accreting of CO disks in AGN disks makes FRB brighter.}
    \label{fig:frb_agn}
\end{figure*}

In this paper, we investigate the generation and propagation of FRBs in AGN disks. Schematic diagrams of our model are shown in Figure \ref{fig:model}, and related physical processes are shown in Figure \ref{fig:frb_agn}. This paper is organized as follows. We assume that young magnetars or accreting COs in AGN disk can emit FRBs, and then progenitors’ ejecta or outflows of the accretion disk interact with the disk material to form a cavity. The existence of the cavity can reduce the absorption of materials in the dense disk, making the FRB easier to observe. If the feedback of the ejecta or outflows is weak, the DM, absorption and RM from the disk are presented in Section \ref{sec:dm_rm_disk}. For FRBs from the magnetosphere of young magnetars, the AGN disk environments do not affect the radiation properties. The propagating effects from the cavity opened by progenitors’ ejecta are presented in Section \ref{sec:mag}. For accreting-powered models, the burst luminosity depends on the accreting rate. The hyper-Eddington accreting of CO disks in AGN disks makes FRB brighter. In addition to burst luminosity, propagation effects from the cavity opened by outflows from CO disks are given in Section \ref{sec:co}. The inflections on turbulent disk materials and magnetic field governed by a dynamo-like mechanism in AGN disk are discussed briefly in Section \ref{sec:dis}. Finally, conclusions are given in Section \ref{sec:con}. In this work, we use the expression $Q_x = Q/10^x$ in cgs units unless otherwise noted.

\section{DM and RM contributed by AGN disks}\label{sec:dm_rm_disk}
In this work, we investigate the observational properties of FRBs in two different disk models, SG model \citep{SG03} and TQM model \citep{Thompson2005}. The disk structures of the SG/TQM model are presented in Appendix \ref{sec:disk_model}. Laterally propagating signals are easily absorbed by dense materials. Therefore, we can only receive signals from the vertical direction. The vertical structure of AGN disks has a Gaussian density profile
\begin{equation}\label{eq:rho_h}
    \rho_{\mathrm{d}}(r,h)= \rho_0(r) \exp \left(-h^2 / 2 H^2\right),
\end{equation}
where $\rho_0(r)$ is the mid-plane disk density and $H$ is the scale height, which is given by the solution of the disk model (see Appendix \ref{sec:disk_model}). 

For the inner disk, the free–free absorption optical depth is extremely high at the midplane of the disk due to dense ionized gas. We can only detect FRBs from source sites at a few scale heights, e.g. for the source at $\sim 10^{-3}$ pc, the disk becomes optically thin only for $h \gtrsim 3.9 H$ \citep{Perna2021}. However, the temperature is lower for the outer disk. The temperature for an SG disk is $\sim 10^3-10^4$ K when $r>10^4R_g$, where $R_g\equiv 2 G M/ c^2$ is the gravitational radius of the SMBH. While for a TQM disk, it is just $\sim 10^2-10^3$ K. If there is no extra ionization process (Ultraviolet/X-ray photons or shocks), gases are neutral at such temperatures. However, gases may also be ionized by radiation \citep{Dyson1980} or shocks associated with transients in AGN disk, such as supernova explosions \citep{Grishin2021,Moranchel-Basurto2021,Li2023},  accretion-induced collapse of an NS \citep{Perna2021} or a WD \citep{Zhu2021a} and accretion outflow feedbacks of compact stars \citep{Wang2021a,Chen2023a}, gamma-ray bursts/kilonovae \citep{Zhu2021b,Ren2022,Yuan2022}, and gravitational-wave bursts \citep{Wang2021b}. In this section, we assume that the radiation or shocks simply provide additional ionization and do not need to be associated with FRBs. In some models, FRBs are associated with young magnetars or accreting COs and we investigate the feedback (progenitors' ejecta or accreting outflows) in sections \ref{sec:mag} and \ref{sec:co} in detail.

Considering the ionization process in detail is complicated, we simply discuss two extreme cases \citep{Perna2021}: the material is fully ionized and there is no extra ionization source in the vertical direction. FRBs are dispersed or absorbed by the disk plasma (see Section \ref{sec:dm_disk}). The magnetic fields in the disk cause Faraday rotation-conversion (see Section \ref{sec:rm_disk}).

\subsection{DM and the optical depth from the disk}
\label{sec:dm_disk}
If a FRB source is located at height $h$, the DM from the disk for the ionized outer disk is
\begin{equation}
\begin{aligned}
    \mathrm{DM_{d}}(h)&=  \int_h^{\infty} \frac{\rho_0}{\mu m_\mathrm{p}} e^{-\frac{z^2}{2 H^2}} d z\\
        &= \mathrm{DM_0} \times \sqrt{\frac{\pi}{2}}  \mathrm{erfc}(\hat{h}/\sqrt{2}) ,
\end{aligned}
\end{equation}
where $\operatorname{erfc}(x)=\frac{2}{\sqrt{\pi}} \int_x^{\infty} e^{-y^2} d y$ is the complementary error function and $\hat{h}=h/H$. The mean molecular weight $\nu=0.62$ is taken in this work. 
If the FRB source is located in the midplane of the AGN disk, DM can be estimated as

\begin{equation}\label{eq:DM0}
\begin{aligned}
    \mathrm{DM_0} &=  \frac{\rho_0}{\mu m_p} H\\
&\approx  6\times10^6 ~\mathrm{pc~cm^{-2}}\rho_{0,-15}\left(\frac{H}{0.01 ~\mathrm{pc}}\right).
\end{aligned}
\end{equation}

The free–free absorption optical depth  is $\tau(\nu)=\int_{\mathrm{LOS}} \alpha_{\mathrm{ff}}(\nu) \mathrm{d} l$ \citep{Rybicki1986}, where $\alpha_{\mathrm{ff}}(\nu)=0.018 T^{-3 / 2} z_{\mathrm{i}}^2 n_{\mathrm{e}} n_{\mathrm{i}} \nu^{-2} \bar{g}_{\mathrm{ff}}$ is the free–free absorption coefficient with $\bar{g}_{\mathrm{ff}} \sim 1$ being the Gaunt factor. Here we assume $n_{\mathrm{e}}\sim n_{\mathrm{i}}$ and $z_{\mathrm{i}}\sim 1$. For the vertical direction, the free–free absorption optical depth is

\begin{equation}
\begin{aligned}
\tau_{\mathrm{ff,d}}(h) & =0.018 T^{-3 / 2} \nu^{-2} \times \int_h^{\infty}\left(\frac{\rho_0}{\mu m_p} e^{-\frac{z^2}{2 H^2}}\right)^2 d z \\
& =\tau_0 \times \frac{\sqrt{\pi}}{2} \operatorname{erfc}(\hat{h}),
\end{aligned}
\end{equation}
where $\tau_0$ is the free–free absorption optical depth from the midplane of the AGN disk

\begin{equation}\label{eq:tau0}
\begin{aligned}
\tau_0 & \approx  0.018 T^{-3 / 2} \nu^{-2}\left(\frac{\rho_0}{\mu m_p}\right)^2 H \\
&=  5 \times 10^8~T_4^{-3 / 2} \rho_{0,-15}^2\\
& \times \left(\frac{\nu}{1 \mathrm{~G Hz}}\right)^{-2}\left(\frac{H}{0.01~\mathrm{pc}}\right).
\end{aligned}
\end{equation}

\begin{figure*}
    \centering
    \includegraphics[width = 1\textwidth]{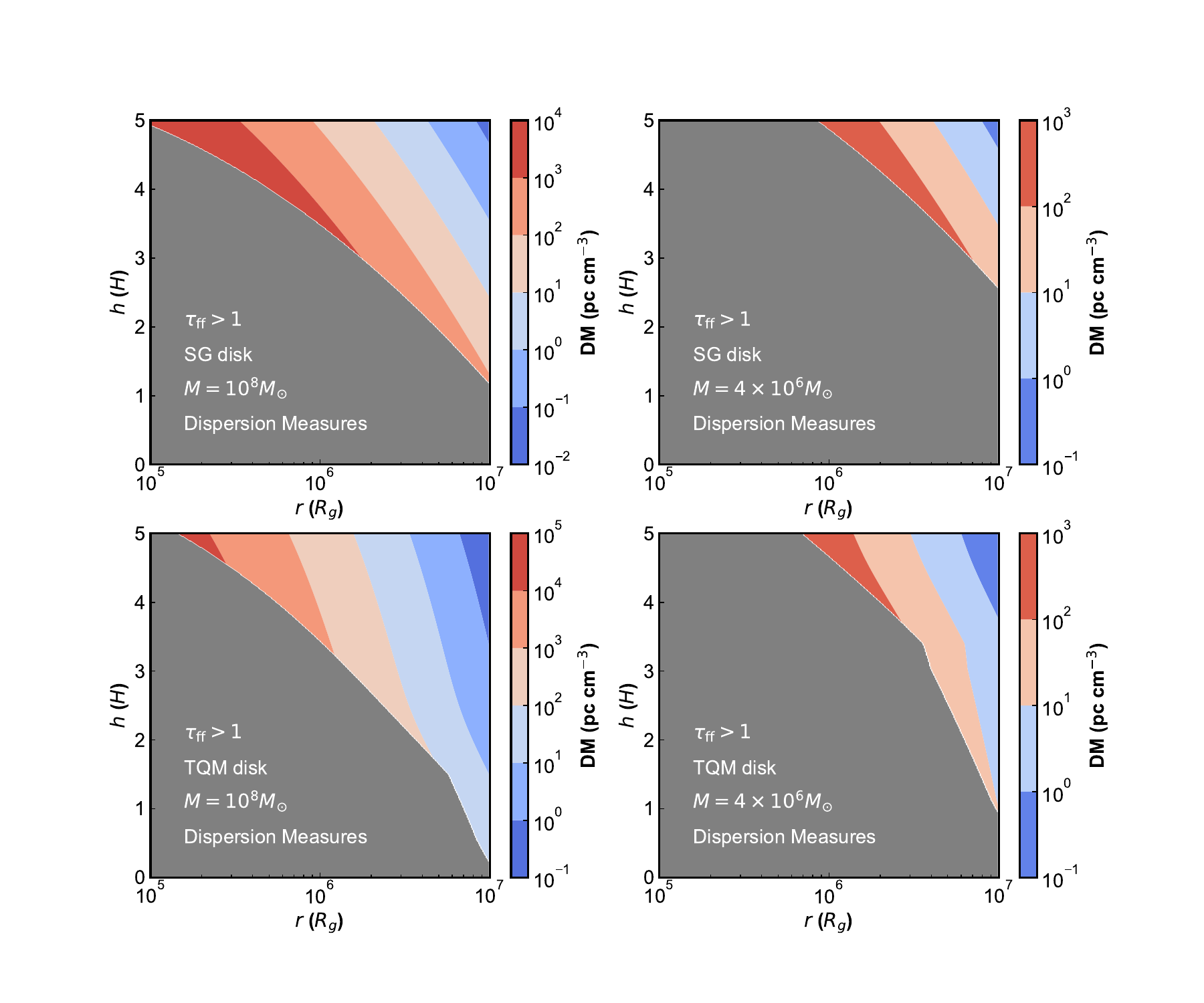}
    \caption{DMs form the AGN disk. The free–free absorption optically thick region for a signal with a frequency of 1 GHz is shown in gray. Due to absorption, FRBs can be observed only from source sited at a few scale heights. For the AGN disk of an SMBH with $M=10^8 M_{\odot}$, the largest DM contributed by the disk is about $10^4-10^5$ \pccm. For an SMBH with $M=4\times 10^6 M_{\odot}$, the largest DM contributed by the disk is about $10^3$ \pccm. It is almost the same as the DM of FRB 20190520B \citep{Niu2022}.}
    \label{fig:disk_dm}
\end{figure*}

\begin{figure*}
    \centering
    \includegraphics[width = 1\textwidth]{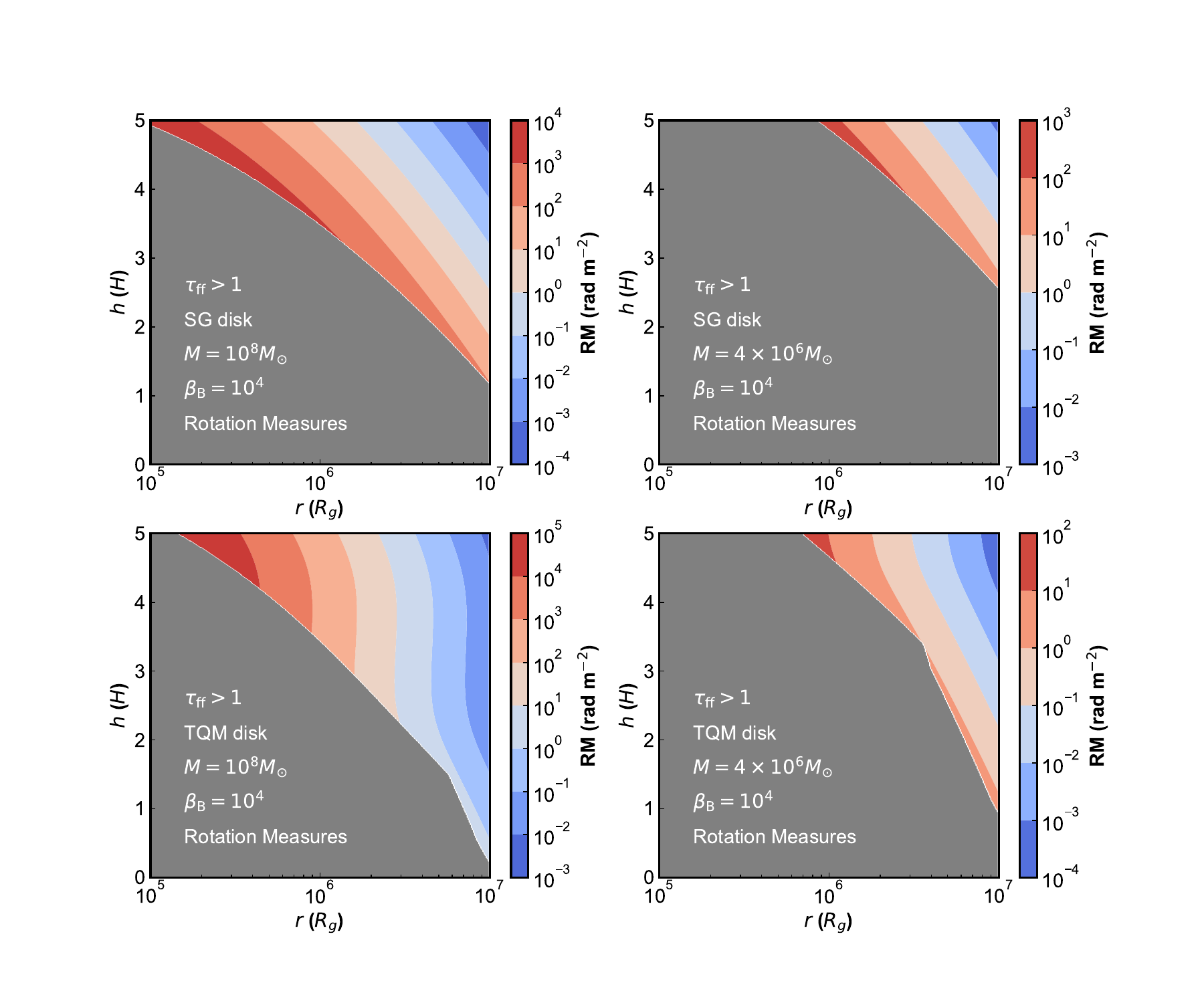}
    \caption{RMs contributed by the weak magnetization AGN disk ($\beta_{\mathrm{B}}\sim 10^4$, \citealt{Salvesen2016}). The free–free absorption optically thick region for a signal with a frequency of 1 GHz is shown in gray. For the AGN disk of an SMBH with $M=10^8 M_{\odot}$, the largest RM contributed by the disk is about $10^4-10^5$ \radm, which is in the same order of magnitude as the FRB with the extreme RM, e.g., FRB 20121102A (RM$\sim 10^{5}$ \radm) and FRB 20190520B (RM$\sim 10^{4}$ \radm). For an SMBH with $M=4\times 10^6 M_{\odot}$, the largest RM contributed by the disk is about $10^2-10^3$ \radm, which is consistent with the observed RMs of most FRB.}
    \label{fig:disk_rm}
\end{figure*}

DMs for different disk models are shown in Figure \ref{fig:disk_dm}. The free-free absorption optically thick region for a signal with a frequency of 1 GHz is shown in gray. Due to absorption, FRBs can be observed only from source sited at a few scale heights. For the AGN disk of an SMBH with $M=10^8 M_{\odot}$, the largest DM contributed by the disk is about $10^4-10^5$ \pccm. Although such value is larger than any DM of FRBs known to date, there are still possibilities that it can be detected by radio telescopes such as CHIME, which can detect DMs up to $\sim$13,000 \pccm \citep{CHIME2018}. For an SMBH with $M=4\times 10^6 M_{\odot}$, the largest DM contributed by the disk is about $10^3$ \pccm. It is similar to the DM of FRB 20190520B \citep{Niu2022}.

% \citetalias{SG03}

\subsection{Faraday rotation-conversion from the disk}\label{sec:rm_disk}
The radiation transfer equation of the Stokes parameters is \citep{Sazonov1969,Melrose1991}
\begin{equation}
\frac{d}{d s}\left(\begin{array}{c}
I \\
Q \\
U \\
V
\end{array}\right)=\left(\begin{array}{c}
\epsilon_I \\
\epsilon_Q \\
\epsilon_U \\
\epsilon_V
\end{array}\right)-\left(\begin{array}{cccc}
\eta_I & \eta_Q & \eta_U & \eta_V \\
\eta_Q & \eta_I & \rho_V & -\rho_U \\
\eta_U & -\rho_V & \eta_I & \rho_Q \\
\eta_V & \rho_U & -\rho_Q & \eta_I
\end{array}\right)\left(\begin{array}{c}
I \\
Q \\
U \\
V
\end{array}\right),
\end{equation}
where $\epsilon_A, A=I,Q,U,V$ are the emission coefficients, $\eta_A, A=I,Q,U,V$ are the absorption coefficients and $\rho_A, A=Q,U,V$ are the Faraday rotation-conversion coefficients. If there are no extra emission and absorption processes during the propagation of FRBs, the total intensity $I$ is conserved and can be assumed to be unity. Thus, the transfer equation can be simplified to
\begin{equation}\label{eq:transfer}
\frac{d}{d s}\left(\begin{array}{l}
Q \\
U \\
V
\end{array}\right)=\left(\begin{array}{rrr}
0 & -\rho_V & \rho_U \\
\rho_V & 0 & -\rho_Q \\
-\rho_U & \rho_Q & 0
\end{array}\right)\left(\begin{array}{l}
Q \\
U \\
V
\end{array}\right).
\end{equation}
The linear polarization degree is $L=\sqrt{Q^2+U^2}$, the circle polarization degree is $V$ and the total polarization degree is $P=\sqrt{L^2+V^2}$. The Faraday rotation rate of the electromagnetic wave with the angular frequency $\omega$ is \citep{Gruzinov2019}
\begin{equation}
     \rho_V =-\frac{1}{c} \frac{\omega_p^2 \omega_B}{\omega^2} \hat{B}_z,
\end{equation}
and the Faraday conversion rate is
\begin{equation}
    \rho_L=\rho_Q+i \rho_U =-\frac{1}{2 c} \frac{\omega_p^2 \omega_B^2}{\omega^3}\left(\hat{B}_x+i \hat{B}_y\right)^2,
\end{equation}
where $\hat{B}_x,\hat{B}_y,\hat{B}_z$ are the three components of the magnetic field unit direction vector $\hat{B}$. $\omega_p=\sqrt{4 \pi n_{e} e^2/m}$ and $\omega_B=eB/mc$ are plasma and Larmor frequencies, respectively.

Another description of the transfer Equation (\ref{eq:transfer}) is in the vector space: 
\begin{equation}\label{eq:transfer_vector}
    d \boldsymbol{S} / d z=\boldsymbol{\rho} \times \boldsymbol{S},
\end{equation}
where $\boldsymbol{S}=(Q, U, V)$ and $\boldsymbol{\rho}=\left(\rho_Q, \rho_U, \rho_V\right)$. The geometric interpretation of Equation (\ref{eq:transfer_vector}) is the Faraday rotation-conversion (or generalized Faraday rotation) on the Poincaré sphere. The Faraday rotation or Faraday conversion angle is $\theta_{\mathrm{FR}} \equiv-\int  \rho_{\mathrm{V}}dz$ and $\theta_{\mathrm{FC}} \equiv-\int  \rho_{\mathrm{L}}dz$, respectively. RM is defined as  
\begin{equation}
\begin{aligned}
\mathrm{RM} & \equiv \frac{\theta_{\mathrm{FR}}}{2 \lambda^2}=-\frac{1}{2 \lambda^2} \int \rho_Vd z \\
& =8.1 \times 10^5 \mathrm{rad~m^{-2}}\int  \left(\frac{n}{\mathrm{~cm}^{-3}}\right) \left(\frac{B_z}{\mathrm{G}}\right) \left(\frac{d z}{\mathrm{pc}}\right),
\end{aligned}
\end{equation}
which are only relevant to the properties of the Faraday screen. The importance of FR and FC can be evaluated by the radio of Faraday conversion rate and Faraday rotation rate
\begin{equation}
\begin{aligned}
\left|\frac{\rho_L}{\rho_V}\right| &\sim \frac{\omega_B}{\omega}\left|\frac{\left(\hat{B}_x+i \hat{B}_y\right)^2}{\hat{B}_z}\right|\\
& \sim 2.8\times 10^{-6} \left(\frac{B_{-3}}{\nu_{9}} \right) \left|\frac{\left(\hat{B}_x+i \hat{B}_y\right)^2}{\hat{B}_z}\right|.
\end{aligned}
\end{equation}
If $B_x \sim B_y \sim B_z$, we have $\rho_L/\rho_V\ll 1$, which means that the Faraday rotation dominates. In most cases, the Faraday conversion is negligible unless the magnetic field has a significant vertical component \citep{Melrose1994,Melrose2010,Gruzinov2019}, e.g., the magnetic field reversal region in a binary system \citep{Wang2022,LDZ2023,Xia2023} or a quasi-toroidal magnetic field in a supernova remnant (SNR; \citealt{Qu2023}).

\begin{figure*}
    \centering
    \includegraphics[width = 1\textwidth]{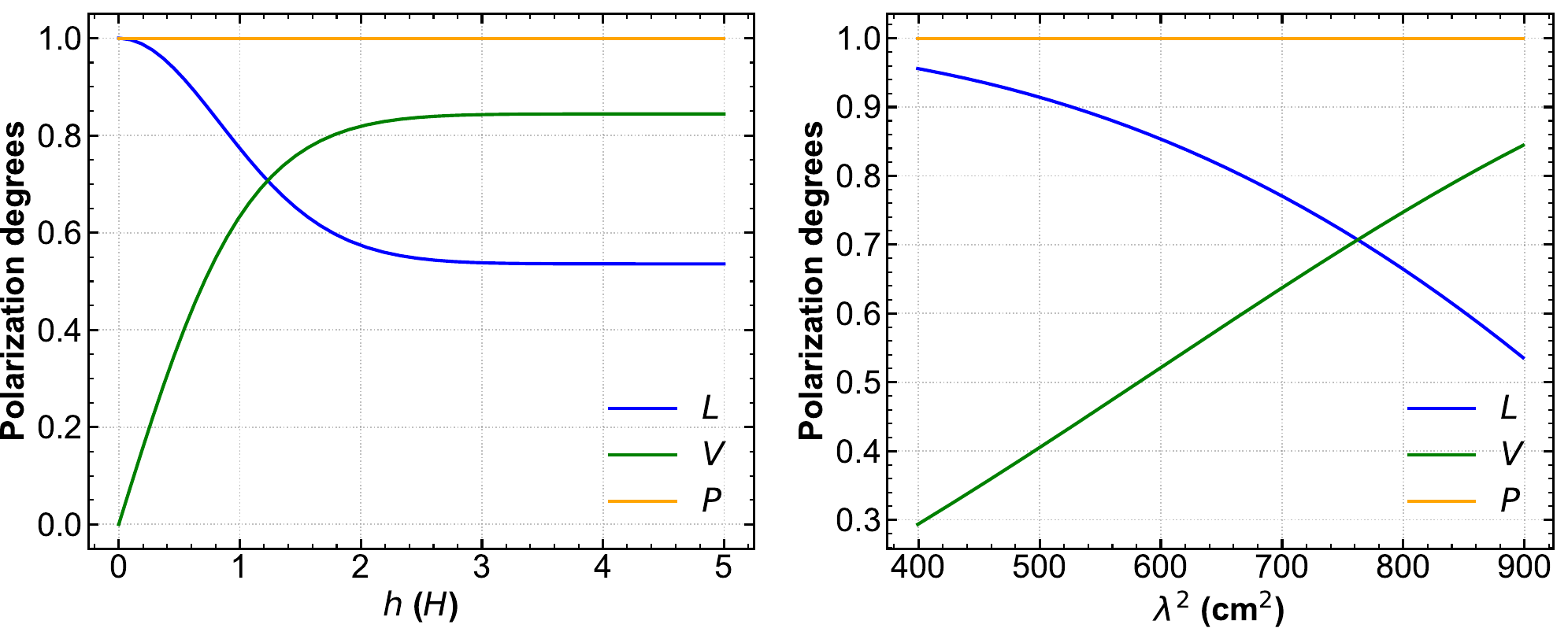}
    \caption{The polarization properties vary with height (left panel) and different bands (1-1.5 GHz, right panel) when radiation passes through a disk with a toroidal magnetic field. The intrinsic radiation of the source is 100\% linearly polarized. When radiation passes through a disk with a toroidal magnetic field, the total polarization degree ($P$, shown in orange lines) remains unchanged, and linear polarization ($L$, shown in blue lines) and circular polarization ($V$, shown in green lines) are converted into each other.}
    \label{fig:FC}
\end{figure*}

The magnetic field of the gas in the AGN disk can be estimated by the ratio of gas pressure to magnetic pressure
\begin{equation}\label{eq:B}
\beta_{\mathrm{B}}=\frac{p_{\mathrm{gas}}}{p_{\mathrm{B}}}=\frac{\frac{1}{2}\rho c_{\mathrm{s}}^2}{B^2/8\pi}.
\end{equation}
For the very strong magnetization disk, $\beta_{\mathrm{B}}\sim 10$, while for the very weak magnetization levels, $\beta_{\mathrm{B}}\sim 10^5$ \citep{Salvesen2016}. The magnetic field of the AGN disk has both toroidal and poloidal components. If we set the direction of the toroidal field is the $x$-axis and the direction perpendicular to the disk midplane (line of sight direction) is the $z$-axis. In this work, we only consider Faraday rotation and Faraday conversion in the case where poloidal and toroidal fields dominate and we assume that $\beta_{\mathrm{B}}$ does not change with height.

For the poloidal field, the parallel component is $\hat{B}_z(h)=h/\sqrt{R_0^2+h^2}$, where $R_0$ is the location of the source. If the outer disk is fully ionized, for an FRB emitted at a height $h$ from the midplane, its RM is

\begin{equation}
\begin{aligned}
\mathrm{RM_{d}}(h)& =8.1 \times 10^5 \mathrm{rad~m^{-2}}\int  \left(\frac{n}{\mathrm{~cm}^{-3}}\right) \left(\frac{B_z}{\mathrm{G}}\right) \left(\frac{d z}{\mathrm{pc}}\right)\\
&=\mathrm{RM}_0 \times \int_{\hat{h}}^{\infty}\left(e^{-\frac{z^2}{2}}\right)^{3/2} \frac{z}{\sqrt{(R_0/H)^2+z^2}} dz,
\end{aligned}
\end{equation}
where $\mathrm{RM}_0$ is the RM from the midplane
\begin{equation}
    \mathrm{RM}_0 \simeq  5.4 \times 10^8 \mathrm{rad~m^{-2}}~\beta_{\mathrm{B},4}^{-1/2}\rho_{0,-15}^{3/2} c_{\mathrm{s,5}} \left(\frac{H}{0.01 \mathrm{pc}}\right).
\end{equation}

RMs from the weak magnetization AGN disk ($\beta_{\mathrm{B}}\sim 10^4$, \citealt{Salvesen2016}) are shown in Figure \ref{fig:disk_rm}. The free–free absorption optically thick region for a signal with a frequency of 1 GHz is shown in gray. For the AGN disk of an SMBH with $M=10^8 M_{\odot}$, the largest RM contributed by the disk is about $10^4-10^5$ \radm, which is in the same order of magnitude as the FRB with the extreme RM, e.g., FRB 20121102A (RM$\sim 10^{5}$ \radm; \citealt{Michilli2018,Hilmarsson2021}) and FRB 20190520B (RM$\sim 10^{4}$ \radm; \citealt{Anna-Thomas2023}). For an SMBH with $M=4\times 10^6 M_{\odot}$, the largest RM contributed by the disk is about $10^2-10^3$ \radm, which is consistent with the RM of most FRBs.

If there is no extra ionization source, whether the FRB is absorbed or not depends only on the optical depth of the disk. For the TQM disk, the disk becomes optical thin at the distance $\sim 10^{3}-10^{4}R_g$, where the density is about $10^{-10}$ g cm$^{-3}$ and the temperature is about a few thousand Kelvin. At this temperature, the ionization degree of the gas is extremely low. For example, for a gas at 3000 K, the ionization degree given by the Saha equation is $\eta \sim 10^{-8}$. Although the DM contribution from the disk is negligible (DM$\sim 0.5$ \pccm), FC may still occur for toroidal fields. By solving Equation (\ref{eq:transfer}), the polarization properties as a function of height are shown in the left panel of Figure \ref{fig:FC}. We assume that the intrinsic radiation of the source is 100\% linearly polarized. When radiation passes through a disk with a toroidal magnetic field, the total polarization degree ($P$, shown in the orange line) remains unchanged, and linear polarization ($L$, shown in the blue line) and circular polarization ($V$, shown in the green line) are converted into each other. In the 1-1.5 GHz band, the changes in linear and circular polarization are shown in the right panel of Figure \ref{fig:FC}, which is similar to the polarization properties of FRB 20201124A \citep{Xu2022}.

\section{Young magnetars in AGN disks}\label{sec:mag}
The magnetar-powered models \citep{Lyubarsky2014,Murase2016,Beloborodov2017,Kashiyama2017,Metzger2017,Wang2017,Metzger2019,Kumar2020,Lu2020} is supported by the detection of FRB 200428 from the Galactic magnetar SGR J1935+2154 \citep{Bochenek2020,CHIME2020}. In this section, we only consider the  magnetospheric origin (e.g., \citealt{Yang2018,Kumar2020,Lu2020}) due to the energy budget constraints on the low-efficiency relativistic shock origin. 

Magnetars could form from the CC of massive stars or compact binary  mergers. Here we give the feedback of progenitors' ejecta in AGN disks. Considering that the shock propagation distance in the vertical direction may be much larger than the disk height, in addition to the disk material, the ejecta also interact with the surrounding material. Assuming the circum-disk material has a power-law density profile, the disk and circum-disk material can be modeled as \citep{Zhou2023}
\begin{equation}
\rho_{\mathrm{cd}}(r,h)= \begin{cases}\rho_0(r) \exp \left(-h^2 / 2 H^2\right), & h\leq h_{\mathrm{c}}, \\ \rho_0(r) \exp \left(-h_{\mathrm{c}}^2 / 2 H^2\right)\left(\frac{h}{h_{\mathrm{c}}}\right)^{-s}, & h>h_{\mathrm{c}}\end{cases}
\end{equation}
where $\rho_0(r)$ is the mid-plane disk density given in the solution of the disk model (see Appendix \ref{sec:disk_model}), and $s=1.5-3$ is the power-law index \citep{Zhou2023}. The critical height $h_{\mathrm{c}}$ represents the boundary between the disk and the circum-disk material. The critical density $\rho_{\mathrm{c}}=\rho_0(r) \exp \left(-h_{\mathrm{c}}^2 / 2 H^2\right)$ is the boundary density at $h_{\mathrm{c}}$. Observationally, the critical height is difficult to constrain. In this work, we assume that the transition between the disk and the circum-disk material occurs when $\rho(h)/\rho_0\sim10^{-3}$.

% If the transition between the disk and the circum-disk material occurs when $\rho(h)/\rho_0\sim10^{-2}$, then the corresponding critical height is $\hat{h}_{\mathrm{c}}=h_{\mathrm{c}}/H\sim3$. If this occurs when $\rho(h)/\rho_0\sim10^{-3}$, the critical height is $\hat{h}_{\mathrm{c}}\sim3.7$.

\begin{table}
%\vspace{-1.5cm}
\centering
\caption{Disk Parameters of Different Models at $R_0=1$ pc.}
\begin{tabular}{cccccc}
\hline
\#  & $M$ & Disk Model & $\rho$ & $H$ & $c_{\mathrm{s}}$ \\
  & ($M_{\odot}$) &  & (g cm$^{-3}$) & (pc) & (km s$^{-1}$) \\
\hline
a & $4\times10^{6}$ & SG &$4.3\times10^{-17}$ & 0.038 & 5 \\
b & $4\times10^{6}$ & TQM &$4.3\times10^{-17}$ & 0.01 & 1.2 \\
c & $10^{8}$ & SG &$10^{-15}$ & 0.02 & 14 \\
d & $10^{8}$ & TQM &$10^{-15}$ & 0.004 & 2.9 \\
\hline
\end{tabular}
\label{tab:disk}
\end{table}

\subsection{The cavity evolution}
After SN explosions or the merger of two COs, forward and reverse shock are generated during the interaction between the ejecta and the circumstellar medium (CSM). The DM and RM from the ejecta and the shocked CSM have been well studied before \citep{Yang2017,Piro2018,Zhao2021a,Zhao2021b}. 

\begin{figure}
    \centering
    \includegraphics[width = 0.5\textwidth]{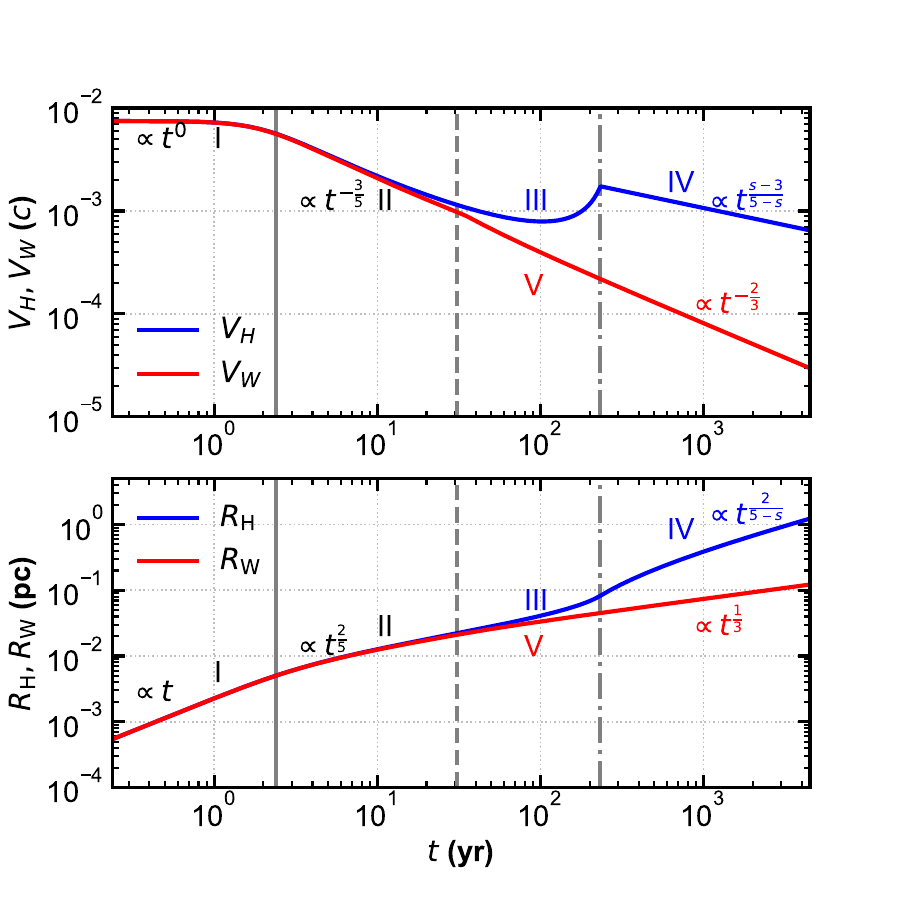}
    \caption{The shock velocity (top panel) and radius (bottom panel) evolution for $R_0$ = 1 pc, $E_0=10^{51}$ erg, $M_{\mathrm{ej}}=10M_{\odot}$ and $s=2$. The disk parameters are taken from Model c in Table \ref{tab:disk}. Timescales of FE duration, breakout and leaving the disk boundary are shown in gray vertical solid, dashed, and dash-dotted lines, respectively. The height expansion (blue lines) goes through the Free expansion phase (phase I, $t\leq t_{\mathrm{FE}}$), the ST phase (decelerated by disk medium, phase II, $t_{\mathrm{FE}}<t\leq t_{\mathrm{bre}}$), the Sakurai accelerating phase (phase III, $t_{\mathrm{bre}}<t\leq t_{\mathrm{c}}$) and the ST phase (decelerated by circum-disk medium, phase IV, $t_{\mathrm{c}}<t\leq t_{\mathrm{cav}}$) in sequence. The width expansion (red lines) goes through the Free expansion phase (phase I, $t\leq t_{\mathrm{FE}}$), the ST phase (phase II, $t_{\mathrm{FE}}<t\leq t_{\mathrm{bre}}$) and the SP phase (phase V, $t>t_{\mathrm{bre}}$) in sequence.}
    \label{fig:R}
\end{figure}

% For SN explosions in AGN disks, the shock is quenched by dense disk materials in the radial direction, but the shock can break out in the vertical direction. 

When the swept mass is much smaller than the ejected mass, the shock evolution is in the free expansion (FE) phase with the velocity $v_\mathrm{sh}=v_0 \equiv \sqrt{E_0 / M_{\mathrm{ej}}}$, where $E_0$ and $M_{\mathrm{ej}}$ are the explosion energy and the ejecta mass, respectively. When the swept mass is comparable to the ejected mass, the shock evolution enters the Sedov–Taylor (ST) phase $v_{\mathrm{sh}} \propto t^{-3 / 5}$ \citep{Taylor1946,Sedov1959}. The duration of the FE phase is
\begin{equation}
\begin{aligned}
    t_{\mathrm{FE}} & =\left(\frac{3 M_{\mathrm{ej}}}{4 \pi \rho_0 v_0^3}\right)^{\frac{1}{3}}\simeq 2.38 \mathrm{~yr}\left(\frac{\rho_0}{10^{-15} \mathrm{~g~cm}^{-3}}\right)^{-\frac{1}{3}} \\
    & \times \left(\frac{E_0}{10^{51} \mathrm{~erg}}\right)^{-\frac{1}{2}}\left(\frac{M_{\mathrm{ej}}}{10 ~M_{\odot}}\right)^{\frac{5}{6}}.
\end{aligned}
\end{equation}

In previous studies, the contribution to DM and RM from both shocked ejecta and shocked ISM has been considered because the age of the source is unknown \citep{Piro2018, Zhao2021a,Zhao2021b}. However, in the AGN disk, since the FE phase (ejecta-dominated phase) lasts only a few years, we only consider the evolution of the forward shock.

The conditions for shock breakout can be roughly estimated when the vertical propagation distance is comparable to the height of the disk $R_{\mathrm{H}}\simeq H$ \citep{Moranchel-Basurto2021,Chen2023a}. After the breakout, the shock propagates in a region where the density drops sharply, and the shock is accelerated (Sakurai accelerating phase $v_{\mathrm{sh}} \propto \rho^{-\mu}$, see \citealt{Sakurai1960}). All the above processes can be described by the following equation \citep{Matzner1999}
\begin{equation}\label{eq:vh1}
    v_\mathrm{H}(h)=\left(\frac{E_0}{M_{\mathrm{ej}}+M_{\mathrm{sw}}(h)}\right)^{1 / 2}\left(\frac{\rho_{\mathrm{d}}(h)}{\rho_0}\right)^{-\mu},
\end{equation}
where $\mu=0.19$ \citep{Matzner1999}, $M_{\mathrm{sw}}(h) \approx 4 \pi \int_0^h \rho\left(h^{\prime}\right) h^{\prime2} d h^{\prime}$ is the swept mass in the vertical direction. The FE phase, the ST phase and the Sakurai accelerating phase work when $M_{\mathrm{sw}}\ll M_{\mathrm{ej}}$, $M_{\mathrm{sw}}\gg M_{\mathrm{ej}}$ and $\rho(h)\gg \rho(0)$, respectively. The breakout timescale is
\begin{equation}
    t_{\mathrm{bre}}=\int_0^H \frac{d h^{\prime}}{v_\mathrm{sh}(h^{\prime})}
\end{equation}

When the shock reaches a critical height $h_{\mathrm{c}}$, the acceleration stops and is decelerated by the circum-disk material. The shock evolution re-enters the ST phase
\begin{equation}\label{eq:vh2}
    v_\mathrm{ST}(t)=v_{\mathrm{c}}\left(\frac{t}{t_{\mathrm{c}}}\right)^{\frac{s-3}{5-s}}, 
\end{equation}
for $t>t_{\mathrm{c}}$, where $v_{\mathrm{c}}$ and $t_{\mathrm{c}}$ is the shock velocity and time at $h=h_{\mathrm{c}}$. 

Although the size of the radial propagation of the shock is much smaller than that of the vertical propagation, the radial propagation determines how long the cavity exists. When $t\leq t_{\mathrm{bre}}$, the shock also experiences the FE and ST phase in the radial direction
\begin{equation}\label{eq:vw1}
    v_\mathrm{W}(w)=\left(\frac{E_0}{M_{\mathrm{ej}}+M_{\mathrm{sw}}(w)}\right)^{1 / 2}.
\end{equation}
The radial distance $w$ is usually much smaller than the location of the FRB source $R_0$, which makes radial density changes across the disk negligible. Therefore, the swept mass in the radial direction is $M_{\mathrm{sw}}(w) =4/3\pi w^3 \rho_{\mathrm{d}}(R_0)$. Also, since the density does not change much, the Sakurai accelerating phase is not considered in the radial direction. When the shock breaks out, the cavity depressurizes and the SNR becomes the shape of a ring-like shell. Then, the shock evolution enters a momentum-conserving snowplow (SP) phase
\begin{equation}\label{eq:vw2}
    R_{\mathrm{W}}^2 \dot{R}_{\mathrm{W}} \simeq H^2 \dot{R}_{\mathrm{W}}(H)
\end{equation}

When the shock decelerates to the speed of the local sound, the shock evolution ends. The radial width of the cavity in the AGN disk is
\begin{equation}
r_{\mathrm{cav}}=\left[\frac{H^2 \dot{R}_{\mathrm{W}}(H)}{c_{\mathrm{s}}}\right]^{1 / 2}
\end{equation}

The cavity formation timescale can be calculated from Equation (\ref{eq:vw2})
\begin{equation}
t_{\mathrm{cav}}=\frac{r_{\mathrm{cav}}^3-H^3}{3 H^2 v_{\mathrm{sh}}\left(t_{\mathrm{bre}}\right)}+t_{\mathrm{bre}}.
\end{equation}

In summary, shock velocity is given by 
\begin{equation}\label{eq:vh_sn}
\dot{R}_\mathrm{H}(t)=\left\{\begin{array}{lc}
v_\mathrm{H}(R_\mathrm{H}) &t\leq t_{\mathrm{c}} \\
v_{\mathrm{c}} \left(\frac{t}{t_c}\right)^{\frac{s-3}{5-s}} & t_\mathrm{c} <t\leq t_{\mathrm{cav}} \\
\end{array},\right.
\end{equation}
and 
\begin{equation}\label{eq:vw_sn}
\dot{R}_\mathrm{W}(t)=\left\{\begin{array}{lr}
v_\mathrm{W}(R_\mathrm{W}) & t\leq t_{\mathrm{bre}} \\
H^2 v_\mathrm{W}(t_{\mathrm{bre}}) / R_\mathrm{W}^2, & t_{\mathrm{bre}}<t\leq t_{\mathrm{cav}}.
\end{array}\right.
\end{equation}

The shock radius can be obtained by solving Equations (\ref{eq:vh_sn}) and (\ref{eq:vw_sn}). The radial and vertical shock evolution for the SNe explosion model with $R_0$ = 1 pc, $E_0=10^{51}$ erg, $M_{\mathrm{ej}}=10M_{\odot}$ and $s=2$ is shown in Figure \ref{fig:R}. The disk parameters are taken from Model c in Table \ref{tab:disk}. Timescales of FE duration, breakout and leaving the disk boundary are shown in gray vertical solid, dashed, and dash-dotted lines, respectively. The height expansion (blue lines)  goes through the Free expansion phase (phase I, $t\leq t_{\mathrm{FE}}$), the ST phase (decelerated by disk medium, phase II, $t_{\mathrm{FE}}<t\leq t_{\mathrm{bre}}$), the Sakurai accelerating phase (phase III, $t_{\mathrm{bre}}<t\leq t_{\mathrm{c}}$) and the ST phase (decelerated by circum-disk medium, phase IV, $t_{\mathrm{c}}<t\leq t_{\mathrm{cav}}$) in sequence. The width expansion (red lines) goes through the Free expansion phase (phase I, $t\leq t_{\mathrm{FE}}$), the ST phase (phase II, $t_{\mathrm{FE}}<t\leq t_{\mathrm{bre}}$) and the SP phase (phase V, $t>t_{\mathrm{bre}}$) in sequence.

Finally, the cavity is refilled by the AGN disk material with the speed of the local sound, and the refill timescale is estimated by
\begin{equation}
t_{\mathrm{ref}}=R_{\mathrm{W,cav}}/c_{\mathrm{s}}.
\end{equation}

\subsection{DM and RM Variations}
The long-term monitoring of DM and RM of repeating FRBs can reveal environments of the magnetar, so it is necessary to show the time-dependent DM and RM. The contributions from the disk are given before and are unimportant after the shock breakouts. Thus, we consider the contributions of DM mainly from two regions: the unshocked cavity and the shocked shell. In the shocked region, some shock energy converts into magnetic field energy. Therefore, RM is only contributed by the shocked shell.

\begin{figure*}
    \centering
    \includegraphics[width = 1\textwidth]{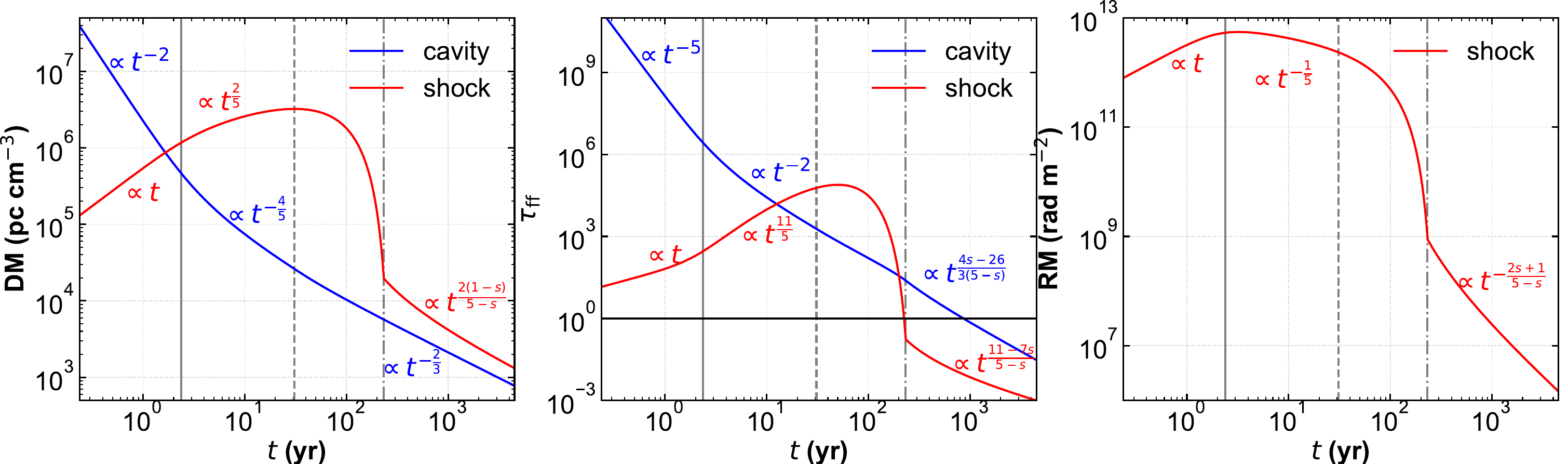}
    \caption{The DM, $\tau_{\mathrm{ff}}$ and RM evolution of the SNe model. The parameters are the same as Figure \ref{fig:R}. Blue and red lines represent the contribution from the cavity and shock shell, respectively. Timescales of FE duration, breakout and leaving the disk boundary are shown in gray vertical solid, dashed, and dash-dotted lines, respectively. Except for the Sakurai accelerating phase ($t_{\mathrm{bre}}<t\leq t_{\mathrm{c}}$), the approximate scaling laws for the remaining phases are given. In the Sakurai accelerating phase, DM $\tau_{\mathrm{ff}}$ and RM show non-power law evolution patterns over time, which is different from other environments, e.g., SNR \citep{Yang2017,Piro2018},  compact binary merger remnant \citep{Zhao2021a}, PWN/MWN \citep{Margalit2018,YYH2019,Zhao2021b}. Although the material in the AGN disk is very dense, as the cavity expands, it can become transparent around a thousand years after the birth of the magnetar. }
    \label{fig:dm_tau_rm}
\end{figure*}

\begin{table*}
\centering
\label{tab:timescale}
\caption{Evolution timescale of cavities in the AGN disk for different progenitor models}
\begin{threeparttable}
\begin{tabular}{cccccccccc}
\hline
Progenitors & $E_0$ & $M_{\mathrm{ej}}$ & $M$ &Disk Model & $t_{\mathrm{FE}}$ & $t_{\mathrm{bre}}$  & $t_{\mathrm{c}}$ & $t_{\mathrm{cav}}$ & $t_{\mathrm{ref}}$ \\
& (erg) & ($M_{\odot}$) & ($M_{\odot}$) &  & (yr) & (yr)  & (yr) & (yr) & (yr) \\
\hline
MS & $10^{51}$ & 10 & $10^{8}$ & SG & 2.38 & 30.9 & 232 & $4.49\times10^{3}$ & $1.12\times10^{4}$ \\
MS & $10^{51}$ & 10 & $10^{8}$ & TQM & 2.38 & 1.96 & 6.65 & $6.19\times10^{4}$ & $1.55\times10^{5}$ \\
MS & $10^{51}$ & 10 & $4\times10^{6}$ & SG & 6.78 & 25.9 & 164 & $7.13\times10^{4}$ & $1.78\times10^{5}$ \\
MS & $10^{51}$ & 10 & $4\times10^{6}$ & TQM & 6.78 & 4.21 & 12.8 & $7.69\times10^{5}$ & $1.92\times10^{6}$ \\
\hline
BWD/AIC\tnote{1}  & $\sim10^{50}$ & $\sim0.1$ & $10^{8}$ & SG & 0.16 & 88.9 & 724 & $2.08\times10^{3}$ & $5.21\times10^{3}$ \\ 
BWD/AIC & $\sim10^{50}$ & $\sim0.1$ & $10^{8}$ & TQM & 0.16 & 1.74 & 12.9 & $2.87\times10^{4}$ & $7.19\times10^{4}$ \\ 
BWD/AIC & $\sim10^{50}$ & $\sim0.1$ & $4\times10^{6}$ & SG & 0.46 & 60.5 & 489 & $3.31\times10^{4}$ & $8.27\times10^{4}$ \\ 
BWD/AIC & $\sim10^{50}$ & $\sim0.1$ & $4\times10^{6}$ & TQM & 0.46 & 2.83 & 19.6 & $3.57\times10^{5}$ & $8.93\times10^{5}$ \\ 
\hline
NSWD\tnote{2}  & $\sim10^{49}$ & $\sim0.01$ & $10^{8}$ & SG & 0.075 & 280 &  & 968 & $2.42\times10^{3}$ \\ 
NSWD & $\sim10^{49}$ & $\sim0.01$ & $10^{8}$ & TQM & 0.075 & 5.01 & 40.2 & $1.33\times10^{4}$ & $3.34\times10^{4}$ \\ 
NSWD & $\sim10^{49}$ & $\sim0.01$ & $4\times10^{6}$ & SG & 0.22 & 190 & $1.55\times10^{3}$ & $1.54\times10^{4}$ & $3.84\times10^{4}$ \\ 
NSWD & $\sim10^{49}$ & $\sim0.01$ & $4\times10^{6}$ & TQM & 0.22 & 7.63 & 60.3 & $1.55\times10^{5}$ & $4.14\times10^{5}$ \\ 
\hline
BNS\tnote{3}  & $\sim5\times10^{49}$ & $\sim0.001$ & $10^{8}$ & SG & $4.93\times10^{-3}$ & 125 & $1.02\times10^{3}$ & $1.65\times10^{3}$ & $4.14\times10^{3}$ \\ 
BNS & $\sim5\times10^{49}$ & $\sim0.001$ & $10^{8}$ & TQM & $4.93\times10^{-3}$ & 2.20 & 17.9 & $2.28\times10^{4}$ & $5.70\times10^{4}$ \\ 
BNS & $\sim5\times10^{49}$ & $\sim0.001$ & $4\times10^{6}$ & SG & 0.014 & 84.7 & 691 & $2.63\times10^{4}$ & $6.57\times10^{4}$ \\ 
BNS & $\sim5\times10^{49}$ & $\sim0.001$& $4\times10^{6}$ & TQM & 0.014 & 3.31 & 26.9 & $2.83\times10^{5}$ & $7.09\times10^{5}$ \\ 
\hline
\end{tabular}
\begin{tablenotes}
\item[1] References: \cite{Dessart2007}
\item[2] References: \cite{Zenati2019}
\item[3] References: \cite{Bauswein2013,Radice2018}
\end{tablenotes}
\end{threeparttable}
\end{table*}

\begin{figure*}
    \centering
    \includegraphics[width = 1\textwidth]{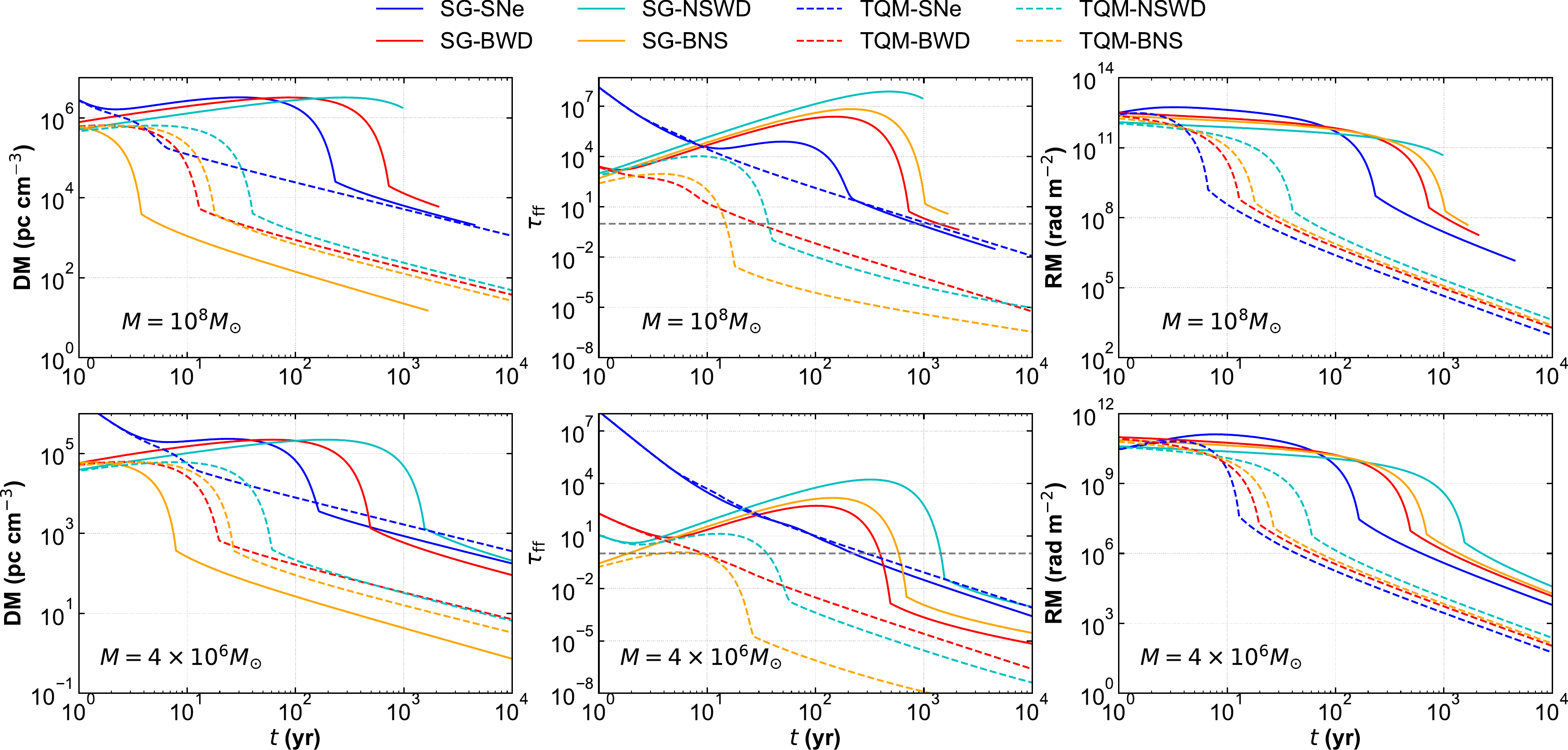}
    \caption{The DM, free-free absorption optical depth and RM evolution for different magnetar formation channels. The parameters and the evolution timescale of each model are listed in Table \ref{tab:timescale}. Solid and dashed lines represent the case of SG and TQM disk models, respectively. $\tau_{\mathrm {ff,sh}}=1$ are shown in gray dashed horizontal lines. For SG-NSWD and SG-BNS models with $M=10^8 M_{\odot}$, the optical depth is still greater than unity when the cavity stops expanding. In other cases, the cavity becomes transparent sometime after the magnetar is born. For binary mergers or AIC progenitors, this time is hundreds to thousands of years for the SG disk model, while it only takes a few decades for the TQM disk model. For massive star progenitors, the time become transport is about a thousand years for the SG disk model, while it is shortened to a few hundred years for the TQM disk model.}
    \label{fig:bin_dm_rm}
\end{figure*}

Before the shock breakouts, the difference in expansion between the radial and vertical directions can be ignored, so the cavity can be regarded as spherical (see Figure \ref{fig:R}). However, after the shock breakouts, because the density of gas in the AGN disk decreases rapidly in the vertical direction, the vertical propagation of the shock becomes easier. At this point, the shape of the cavity deviates from a spherical shape. For simplicity, we assume that the cavity is always a cylinder during expansion and the cavity volume is $V_{\mathrm{cav}}=\pi R_{\mathrm{H}}R_{\mathrm{W}}^2$, where $R_{\mathrm{H}}$ and $R_{\mathrm{W}}$ are the solutions on Equations (\ref{eq:vh_sn}) and (\ref{eq:vw_sn}).

The time-dependent DM from the cavity is 
\begin{equation}\label{eq:dm_cav}
\begin{aligned}
\mathrm{DM_{cav}}(t) & \simeq \eta \frac{M_{\mathrm{ej}}}{\mu m_p \pi R_{\mathrm{W}}(t)^2 R_{\mathrm{H}}(t)} \cdot(1-\xi) R_{\mathrm{H}}(t) \\
& \propto\left\{\begin{array}{cc}
t^{-2} & t\leq t_{\mathrm{FE}} \\
t^{-\frac{4}{5}} & t_{\mathrm{FE}}<t\leq t_{\mathrm{bre}} \\
t^{-\frac{2}{3}} & t_{\mathrm{bre}}<t\leq t_{\mathrm{cav}}
\end{array},\right.
\end{aligned}
\end{equation}
where $\xi$ is the thickness of the shock and $\eta$ is the ionization fraction. The shock thickness $1-\xi\sim 0.9$ is taken, which is consistent with results given by self-similar solutions of the SNR \citep{Chevalier1982}. From Equation (\ref{eq:dm_cav}), we find that $\mathrm{DM_{cav}}$ only depends on the radial expansion. For the FE phase, the cavity size is much smaller than the disk height ($R_{\mathrm{W}}\approx R_{\mathrm{H}} \ll H$).  At this time, the disk density can be regarded as a constant (see Equation (\ref{eq:rho_h})). Thus, DM from the cavity evolves as $\mathrm{DM_{cav}}\propto R_{\mathrm{W}}^{-2}\propto t^{-2}$. The approximate scaling laws for the remaining two phases in Equation (\ref{eq:dm_cav}) can be obtained in the same way.

The time-dependent free-free absorption optical depth from the cavity is

\begin{equation}\label{eq:tau_sn}
\begin{aligned}
\tau_{\mathrm{ff,cav}}(t)& \simeq 0.018 T_{\mathrm{ej}}^{-3 / 2} \nu^{-2} \eta^2 \\
& \times\left[\frac{M_{\mathrm{ej}}}{\mu m_p \pi R_{\mathrm{W}}(t)^2 R_{\mathrm{H}}(t)}\right]^2 \cdot \xi R_{\mathrm{H}}(t) \\
& \propto \begin{cases}t^{-5} & t\leq t_\mathrm{F E} \\
t^{-2} & t_\mathrm{FE}<t\leq t_\mathrm{bre} \\
t^{\frac{26-4 s}{3(5-s)}} & t_\mathrm{c}<t\leq t_{\mathrm{cav}}\end{cases}
\end{aligned}
\end{equation}
For the unshocked ejecta, a low ionization fraction $\eta=0.1$ and temperature $T_{\mathrm{ej}}=10^4$ K is taken \citep{Zhao2021a}. Except for the Sakurai accelerating phase ($t_{\mathrm{bre}}<t\leq t_{\mathrm{c}}$), where $R_\mathrm{H}$ depends on the numerical solution of Equation (\ref{eq:vh_sn}), the approximate scaling laws for the remaining phases is given in Equation (\ref{eq:tau_sn}).

For the shock region, the temperature can be estimated from
\begin{equation}\label{eq:Tsh}
    k_{\mathrm{B}} T_{\mathrm{sh}} \approx 9 m_{\mathrm{p}} v_{\mathrm{sh}}^2 / 16,
\end{equation}
 where $k_{\mathrm{B}}$ is the Boltzmann constant and the shock velocity $v_{\mathrm{sh}}$ is given in Equation (\ref{eq:vh_sn}). The high temperature makes gas fully ionized. The density of the shocked matter is $\rho_{\mathrm{sh}}=4\rho_{\mathrm{cd}}$ for strong shock waves. The time-dependent DM from the shocked shell is
 
\begin{equation}
\begin{aligned}
\mathrm{DM_{sh}}(t)& \simeq \frac{4 \rho_{\mathrm{cd}}\left[R_{\mathrm{H}}(t)\right]}{\mu m_\mathrm{p}} \cdot \xi R_{\mathrm{H}}(t) \\
& \propto \begin{cases}t & t\leq t_\mathrm{FE} \\
t^{\frac{2}{5}} & t_\mathrm{FE}<t\leq t_\mathrm{bre} \\
t^{\frac{2(1-s)}{5-s}} & t_\mathrm{c}<t\leq t_\mathrm{cav}\end{cases}
\end{aligned}
\end{equation}

The time-dependent free-free absorption optical depth from the shocked shell is

\begin{equation}
\begin{aligned}
\tau_{\mathrm {ff,sh}}(t)& \simeq 0.018 T^{-3 / 2} \nu^{-2} \left\{\frac{4 \rho_{\mathrm{cd}}\left[R_{\mathrm{H}}(t)\right]}{\mu m_\mathrm{p}}\right\}^2 \cdot \xi R_{\mathrm{H}}(t) \\
& \propto \begin{cases}t & t\leq t_{\mathrm{FE}} \\
t^{\frac{11}{5}} & t_{\mathrm{FE}}<t\leq t_{\mathrm{bre}}. \\
t^{\frac{11-7 s}{5-s}} & t_{\mathrm{c}}<t\leq t_{\text {cav }}\end{cases}
\end{aligned}
\end{equation}

The magnetic field in the shocked region is
\begin{equation}\label{eq:Bsh}
\frac{B^2}{8 \pi}=\epsilon_{\mathrm{B}} u_{\mathrm{th}},
\end{equation}
where $\epsilon_{\mathrm{B}}$ is the magnetic energy density fraction and $u_{\mathrm{th}}=9 \rho_{\mathrm{cd}} v_\mathrm{sh}^2 / 8$. 
The RM from the shocked shell is

\begin{equation}
\begin{aligned}
 \mathrm{RM_{sh}}(t) &\simeq8.1 \times 10^5 \mathrm{rad~m^{-2}}\\
    &\times \left(\frac{4n_{\mathrm{cd}}(t)}{\mathrm{~cm}^{-3}}\right) \left(\frac{B_\mathrm{sh}(t)}{\mathrm{G}}\right) \left(\frac{\xi R_{\mathrm{H}}(t)}{\mathrm{pc}}\right)\\
& \propto \begin{cases}t & t\leq t_{\mathrm{FE}} \\
t^{-\frac{1}{5}} & t_{\mathrm{FE}}<t \leq t_{\mathrm{bre}}. \\
t^{-\frac{2 s+1}{5-s}} & t_{\mathrm{bre}}<t\leq t_{\mathrm{cav}}\end{cases} \\
&
\end{aligned}
\end{equation}

\subsection{The influence of the progenitors}
The DM, $\tau_{\mathrm{ff}}$ and RM evolution of the SNe model are shown in Figure \ref{fig:dm_tau_rm}. The parameters are the same as Figure \ref{fig:R}. Blue and red lines represent the contribution from the cavity and shock shell, respectively. Timescales of FE duration, breakout and leaving the disk boundary are shown in gray vertical solid, dashed, and dash-dotted lines, respectively. Except for the Sakurai accelerating phase ($t_{\mathrm{bre}}<t\leq t_{\mathrm{c}}$), the approximate scaling laws for the remaining phases are given. In the Sakurai accelerating phase, DM $\tau_{\mathrm{ff}}$ and RM show non-power-law evolution patterns over time, which is different from other environments, e.g., SNR \citep{Yang2017,Piro2018},  compact binary merger remnant \citep{Zhao2021a}, PWN/MWN \citep{Margalit2018,YYH2019,Zhao2021b}. Although the material in the AGN disk is very dense, as the cavity expands, it can become transparent around a thousand years after the birth of the magnetar. 

Besides core-collapse (CC) explosions of massive stars, in some cases, a magnetar can also be formed in the following process: BNS mergers \citep{Dai1998,Rosswog2003,Dai2006,Giacomazzo2013}, BWD mergers \citep{King2001,Yoon2007,Schwab2016}, AIC of WDs \citep{Nomoto1991,Tauris2013,Schwab2015} and NSWD mergers \citep{Zhong2020}. The compact binary mergers or AIC of WDs can occur in the AGN disks \citep[e.g.][]{Mckernan2020, Perna2021, Zhu2021a, Luo2023}.

Although the merger dynamics are complicated, we can think of mergers as an explosion like a supernova in the long term, ejecting a certain amount of energy and material into the surroundings \citep{Zhao2021a}. Due to the lack of observations, the mass and energy of the ejecta after compact binary mergers are taken from the results of the numerical simulation (\citealt{Dessart2007,Bauswein2013,Radice2018,Zenati2019}; see Table \ref{tab:timescale}).

The DM, free-free absorption optical depth and RM evolution of different magnetar formation channels are shown in Figure \ref{fig:bin_dm_rm}. The parameters and the evolution timescale of each model are listed in Table \ref{tab:timescale}. Solid and dashed lines represent the case of SG and TQM disk models, respectively. $\tau_{\mathrm {ff,sh}}=1$ are shown in gray dashed horizontal lines. For SG-NSWD and SG-BNS models with $M=10^8 M_{\odot}$, the optical depth is still greater than unity when the cavity stops expanding. In other cases, the cavity becomes transparent sometime after the magnetar is born. For binary mergers or AIC progenitors, this time is hundreds to thousands of years for the SG disk model, while it only takes a few decades for the TQM disk model. For massive star progenitors, the time become transport is about a thousand years for the SG disk model, while it is shortened to a few hundred years for the TQM disk model.

\section{Accreting Compact Objects in AGN disks}
\label{sec:co}
The close-in models are only applicable to magnetar engines, while the far-away models apply to a wider range of scenarios as long as energy can be injected into the surrounding medium from the central engine. Motivated by the periodic repeating  FRBs \citep{Chime/FrbCollaboration2020,Rajwade2020,Cruces2021}, accreting-powered models from COs in ULX-like binaries have been proposed \citep{Sridhar2021}. In this model, FRBs are generated via synchrotron maser emission from the short-lived relativistic outflows (or “flares”) decelerated by the pre-existing (or “quiescent”) jet. For the BH engine, if the spin axis is misaligned with the angular momentum axis of the accretion disk, Lens-Thirring (LT) precession makes the FRB periodic. FRB from the precession jet encounters the disk wind which contributes variable DMs and RMs. The synchrotron radio emission from the ULX hypernebulae has been proposed to explain the persistent radio source (PRS) of FRBs \citep{Sridhar2022}, e.g., FRB 20121102A \citep{Chatterjee2017} and FRB 20190520B \citep{Niu2022}. 

To explain some most luminous FRBs, the COs should be undergoing hyper-Eddington mass transfer from a main-sequence star companion. We would like to point out that mass inflow rates of COs in the AGN disks are also possible to be extremely hyper-Eddington \citep{Chen2023a}. In this section, we investigate the accreting-powered models of FRBs in the AGN disks. In this work, we focus on the influence of the AGN disk environment, such as the burst properties and variable DMs and RMs from the disk wind (ignore the precession). Other properties can be found in \cite{Sridhar2021,Sridhar2022}.

\subsection{Burst properties}\label{sec:burst}
In this section, we briefly outline the intrinsic properties of FRBs from accreting COs (taking a BH as an example) in AGN disks. The extremely high accreting rate of COs in the AGN disk environment (see Section \ref{sec:outflow} in detail) has a great impact on the burst luminosity and peak frequency. 
When a BH is rotating, the spin energy can be extracted by forming ultra-relativistic jets (i.e., Blandford–Znajek (BZ) mechanism \citealt{Blandford1977}). The BZ luminosity is \citep{Tchekhovskoy2011}
\begin{equation}
L_{\mathrm{BZ}}=\eta_{\mathrm{BZ}} \dot{M} \cdot c^2 \approx \eta \dot{m} L_{\mathrm{Edd}},
\end{equation}
where $\eta_{\mathrm{BZ}}$ is the jet efficiency,  $\dot{m}=\dot{M} / \dot{M}_{\mathrm{Edd}}$ is the the dimensionless accretion rate, $\dot{M}_{\mathrm{Edd}}=L_{\mathrm{Edd}} / c^2$ is the Eddington limit accretion rate and $L_{\mathrm{Edd}}=4 \pi G M m_p c / \sigma_T=1.26 \times 10^{39} ~\mathrm{erg} \mathrm{~s}^{-1} M/10M_{\odot}$ is the Eddington limit accretion luminosity. The maximum jet efficiency is $\eta_{\mathrm{max}}\sim1.4$ for an extreme rotating BH with the dimensionless spin parameter $a = 0.99$ \citep{Tchekhovskoy2011}. For an accreting NS, the jet power can also be extracted from the rotational energy via a BZ-like mechanism \citep{Parfrey2016}. Taking $\eta \sim\eta_{\mathrm{max}}\simeq 1$, the maximum isotropic-equivalent FRB luminosity can  be estimated by 
\begin{equation}
\begin{aligned}
L_{\mathrm{FRB}}^{\max } & \approx f_\xi f_b^{-1} \eta_{\max } \dot{M} c^2 \sim f_{\xi,-3} f_{b,-2}^{-1}\left(\frac{M_{\mathrm{CO}}}{10 M_\odot}\right) \\
& \times \begin{cases}1.25 \times 10^{43} \mathrm{~erg} \cdot \mathrm{s}^{-1}\left(\frac{\dot{m}}{10^5}\right) & t\leqslant t_{\mathrm {acc}} \\
1.25 \times 10^{39} \mathrm{~erg} \cdot \mathrm{s}^{-1}\left(\frac{\dot{m}}{10}\right) & t> t_{\mathrm {acc }}\end{cases},
\end{aligned}
\end{equation}
where $M_{\mathrm{CO}}$ is the CO mass, $f_\xi$ is the radio emission efficiency (for synchrotron maser scenarios, $f_\xi\sim10^{-3}$, e.g., \citealt{Sridhar2021}) and $f_b$ is the beaming fraction. $t_{\mathrm {acc }}$ is the efficient accretion timescale of COs (see Section \ref{sec:wind_cav}). 

As mentioned before, FRBs are powered by the sudden accretion flare with the luminosity $L_{\mathrm{f}}=\eta  \dot{M} \cdot c^2$. The flare ejecta propagates into the cavity of the quiescent jet with the luminosity and bulk Lorentz factor being $L_{\mathrm{q}}=\eta_{\mathrm{q}} \dot{M} \cdot c^2$ and $\Gamma_{\mathrm{q}}$, respectively. For FRBs to escape, the quiescent jet should have a large bulk Lorentz factor $\Gamma_{\mathrm{q}} \gtrsim 100$ and a low jet efficiency $\eta_{\mathrm{q}} \ll 1$ \citep{Sridhar2021}. During the interaction between the flare ejecta and the gas in the quiescent jet, a forward shock generates with a radius \citep{Sari1995}
\begin{equation}
r_{\mathrm{dec}} \approx 2 \Gamma_{\mathrm{sh}}^2 c \cdot t_{\mathrm{f}} \approx 6 \times 10^{12} \eta_{\mathrm{q},-2}^{-1 / 2} t_{\mathrm{f},-3} \Gamma_{\mathrm{q}, 2}^2 \mathrm{~cm},
\end{equation}
where $t_{\mathrm{f}}$ is the duration of burst. The blast Lorentz factor  $\Gamma_{\mathrm{sh}}$ is given by pressure balance \citep{Beloborodov2017}

\begin{equation}
\Gamma_{\mathrm{sh}} \simeq \Gamma_{\mathrm{q}}\left(\frac{L_{\mathrm{f}}}{L_{\mathrm{q}}}\right)^{1 / 4} \simeq\Gamma_{\mathrm{q}}\eta_q^{-1 / 4} \approx 320\Gamma_{\mathrm{q}, 2}\eta_{q,-2}^{-1 / 4}.
\end{equation}
Electrons at the FRB emission radius $r_{\mathrm{FRB}}$ in the shocked gas gyrate with the Larmor radius $r_{\mathrm{L}}=\Gamma_{\mathrm{q}} m_{\mathrm{e}} c^2 / e B_{\mathrm{q}}$, where $B_{\mathrm{q}} \simeq\left(L_{\mathrm{q}} \sigma_{\mathrm{q}} / c r_{\mathrm{FRB}}^2\right)^{1 / 2}$ is the lab-frame measured magnetic field in high magnetization ($\sigma_{\mathrm{q}} \gtrsim 1$) upstream. The peak frequency from synchrotron maser emission is \citep{Gallant1992,Plotnikov2019}

\begin{equation}
\begin{aligned}
\nu_{\mathrm{pk}} & \sim \frac{\Gamma_{\mathrm{sh}} c}{2 \pi r_{\mathrm{L}}} \sim \frac{e \sigma_{\mathrm{q}}^{1 / 2}\left(L_{\mathrm{f}} L_{\mathrm{q}}\right)^{1 / 4}}{2 \pi m_{\mathrm{e}} c^{3 / 2} r_{\mathrm{FRB}}} \\
& \approx \frac{e \sigma_{\mathrm{q}}^{1 / 2} \eta^{1 / 4} \eta_{\mathrm{q}}^{3 / 4}L_{\mathrm{Edd}}^{1 / 2}\dot{m}^{1 / 2}}{4 \pi m_{\mathrm{e}} c^{5 / 2}t_{\mathrm{f}} \Gamma_{\mathrm{q}}^2} \left(\frac{r_{\mathrm{dec}}}{r_{\mathrm{FRB}}}\right)  \\
& \approx \left(\frac{r_{\mathrm{FRB}}}{r_{\mathrm{dec}}}\right)^{-1}\left(\frac{M_{\mathrm{CO}}}{10 M_\odot}\right)^{1 / 2} \frac{\eta^{1 / 4} \eta_{\mathrm{q},-2}^{3 / 4} \sigma_{\mathrm{q}}^{1 / 2}}{\Gamma_{\mathrm{q}, 2}^2 t_{-3}}\\
& \times \begin{cases}9.5~\mathrm{GHz} \left(\frac{\dot{m}}{10^5}\right)^{1 / 2}  & t\leqslant t_{\mathrm {acc}} \\
100~\mathrm{MHz}\left(\frac{\dot{m}}{10}\right)^{1 / 2}  & t> t_{\mathrm {acc }}\end{cases}.
\end{aligned}
\end{equation}
FRBs have been detected from 110 MHz \citep{Pleunis2021} to 8 GHz \citep{Gajjar2018}, which is about the same as the estimated frequency for $\Gamma_{\mathrm{q}}\sim 100$ and $\eta_{\mathrm{q}}\sim 0.01$.

When $t<t_{\mathrm {acc }}$, the CO accretion rate in AGN disks is extreme hyper-Eddington ($\dot{m}\sim 10^{5}-10^{10}$, see \citealt{Chen2023a}), which is enough to drive the most luminous FRBs till now. If the disk wind cavity can become optical thin before the efficient accretion stops (see Section \ref{sec:wind_cav}), high-frequency bright bursts are generated. However, when $t>t_{\mathrm {acc}}$, the accretion is weak in the low-density cavity ($\dot{m}\sim 10$, \citealt{Chen2023a}). At this time, we can only receive low-frequency faint bursts. The intrinsic burst properties for different accretion CO models are shown in Table \ref{tab:timescale2}, and the disk parameters are taken from the TQM disk model in Table \ref{tab:disk}.

\subsection{Accretion and outflow of COs}\label{sec:outflow}
We adopt descriptions of the accretion and outflow of COs in AGN disks from \citealt{Chen2023a}. Here, we list the relative equations briefly. The mass inflow rate of the CO accretion disk at the outer boundary ($r_{\mathrm{obd}}$) can be described based on the Bondi–Holye–Lyttleton (BHL) accretion rate (see \citealt{Edgar2004} for a review)
\begin{equation}\label{eq:M_dot_BHL}
\dot{M}_{\mathrm{BHL}}=\frac{4 \pi G^2 M_{\mathrm{CO}}^2 \rho_{\mathrm{CO}}}{\left(v_{\mathrm{rel}}^2+c_\mathrm{s}^2\right)^{3 / 2}},
\end{equation}
where $\rho_{\mathrm{CO}}$ is the gas density near the CO, and $v_{\mathrm{rel}}$ is the relative velocity between the CO and the gas in the AGN disk. However, in the AGN disk, taking into account the influence of SMBH gravity and the finite height of the AGN disk, the modified gas inflow rate is \citep{Kocsis2011}
\begin{equation}
\dot{M}_{\mathrm {inflow }}=\dot{M}_{\mathrm{BHL}} \times \min \left\{1, \frac{H}{r_{\mathrm{BHL}}}\right\} \times \min \left\{1, \frac{r_{\mathrm {Hill }}}{r_{\mathrm{BHL}}}\right\},
\end{equation}
where $r_{\mathrm {Hill }}=\left(M_{\mathrm{CO}} / 3 M\right)^{1 / 3} R_{\mathrm{CO}}$ is the Hill radius with $R_{\mathrm{CO}}$ being the radial location of the CO and $r_{\mathrm{BHL}}=G M_{\mathrm{CO}} /\left(v_{\mathrm{rel}}^2+c_\mathrm{s}^2\right)$ is BHL radius.

The outer boundary radius of the circum-CO disk can be approximated to the circularization radius ($r_{\mathrm{obd}}\sim r_{\mathrm{cir}}$). Under the assumption of the angular momentum conservation, the circularization radius of infalling gas is 
\begin{equation}
\sqrt{G M_{\mathrm{CO}} r_{\mathrm{cir}}}=v_{\mathrm{rel}}\left(r_{\mathrm{rel}}\right) r_{\mathrm{rel}},
\end{equation}
where $r_{\mathrm {rel }}=\min \left\{r_{\mathrm{BHL}}, r_{\mathrm {Hill }}\right\}$ is the radius of CO gravity sphere. Due to the differential rotation of the AGN disk, the captured gas exhibits varying velocity relative to the CO, which can be estimated as 
\begin{equation}
v_{\mathrm{rel}}\left(r_{\mathrm{rel}}\right)=V_{\mathrm{K}}\left(R_{\mathrm{CO}}\right)-V_{\mathrm{K}}\left(R_{\mathrm{CO}}+r_{\mathrm{rel}}\right) \approx \frac{1}{2} r_{\mathrm{rel}} {\Omega}_{\mathrm{K}},
\end{equation}
where $V_{\mathrm{K}}$ is the Keplerian velocity of the AGN disk. In the above Equation, we use the following approximation: $R_{\mathrm{CO}}\gg r_{\mathrm{rel}}$, which is obviously true for the CO location we are interested in. Though simplified, the resulting $r_{\mathrm {cir}}$ 
approximately matches the values obtained from numerical simulations \citep[e.g.][]{Tanigawa2012, Li2022}.

Besides capturing the nearby gas, an embedded CO can also exert a gravitational torque to repel the ambient gas, resulting in a reduction of gas density in the AGN disk annulus at $R_{\mathrm{CO}}$ \citep[e.g.][]{Ward1997}. Consequently, the gas density around the CO is set as
\begin{equation}
\rho_{\mathrm{CO}}=\left\{\begin{array}{ll}
\rho_{\mathrm{d}}, & R_{\mathrm{d}, \mathrm { gap }}\leq r_{\mathrm {rel }} \\
\rho_{\mathrm{d}, \mathrm { gap }}, & R_{\mathrm{d}, \mathrm { gap }}>r_{\mathrm {rel }}
\end{array} \right.,
\end{equation}
where the reduced density and half-width of the reduced region are \citep{Kanagawa2015,Kanagawa2016,Tanigawa2016}
\begin{equation}
\frac{\rho_{\mathrm{d}, \mathrm { gap }}}{\rho_{\mathrm{d}}}=1 /\left[1+0.04\left(M_{\mathrm{CO}} / M\right)^2\left(H / R_{\mathrm{CO}}\right)^{-5} \alpha^{-1}\right]
\end{equation}
and 
\begin{equation}
\frac{R_{\mathrm{d}, \mathrm {gap}}}{R_{\mathrm{CO}}}=0.21\alpha^{-1 / 4}\left(M_{\mathrm{CO}} / M\right)^{1 / 2}\left(H / R_{\mathrm{CO}}\right)^{-3 / 4}.
\end{equation}

The outer region of the circum-CO accretion disk becomes self-gravity unstable if the inflow mass rate is very high, leading to the reduction in the rate because of gas gravitational fragmentation \citep{Pan2021b,Tagawa2022}. The Toomre parameter of the circum-CO disk is $Q_{\mathrm{CO}}=\Omega_{\mathrm{K}} c_s / \pi G \Sigma \sim 2 \alpha_{\mathrm{CO}} h^3 v_\mathrm{K}^3 / G \dot{M}_{\mathrm {inflow }}$, where $\alpha_{\mathrm{CO}}$, $h=H_{\mathrm{CCOD}}/r$ and $v_{\mathrm{K}}=\sqrt{G M_{\mathrm{CO}}/r}$ is the viscosity parameter, the disk height ratio and the Keplerian velocity of the circum-CO disk, respectively. The modified mass inflow rate is 
\begin{equation}\label{eq:Mdot_obd}
\dot{M}_{\mathrm{obd}}=\left\{\begin{array}{ll}
\dot{M}_{\mathrm {inflow }}, & Q_{\mathrm{CO}}\left(r_{\mathrm{obd}}\right) \geqslant 1 \\
2 \alpha_{\mathrm{CO}} h^3 v_K^3 / G . & Q_{\mathrm{CO}}\left(r_{\mathrm{obd}}\right)<1
\end{array} .\right.
\end{equation}

The values of the inflow mass rate $\dot{M}_{\mathrm {inflow }}$ and the outer boundary radius of the circum-CO disk $r_{\mathrm{obd}}$ depend on Equations (\ref{eq:M_dot_BHL})-(\ref{eq:Mdot_obd}). It isn't easy to express with a simple analytical formula. The circum-CO accretion disk's self-gravity unstable conditions for different SMBH mass, CO locations and specific accreting models are studied in \cite{Chen2023a}. In this work, the value of $Q_{\mathrm{CO}}$ of different accreting models is shown in Table \ref{tab:timescale2}. For all the models we chose, the accretion of circum-CO disk is stable.

The initial mass inflow rates of COs in the AGN disk are given by Equation (\ref{eq:Mdot_obd}), which should be hyper-Eddington \citep{Chen2023a}. Photons are trapped for the extremely hyper-Eddington inflow and the trapping radius is $r_{\mathrm{tr}}=3 \dot{m} h r_g$ \citep{Kocsis2011}. The mass inflow rate is reduced inside the trapping radius due to the efficient outflow driven by the disk radiation pressure. In summary, the radius-dependent mass inflow rate is \citep{Blandford1999}
\begin{equation}
\dot{M}_{\mathrm{in}}(r)=\left\{\begin{array}{ll}
\dot{M}_{\mathrm{obd}}\left(r / r_{\mathrm{tr}}\right)^{p}, & r\leq r_{\mathrm{tr}} \\
\dot{M}_{\mathrm{obd}}, & r>r_{\mathrm{tr}}
\end{array},\right.
\end{equation}
where $p$ is the power-law index. The numerical simulations of \cite{Yang2014} show that $p\sim 0.4-1$, but the smaller value is also possible \citep{Kitaki2021}. 

The outflow of the circum-CO disk driven by radiation pressure can take away a fraction of the viscous heating. The luminosity of disk outflow is given by \citep{Chen2023a}
\begin{equation}
\begin{gathered}
L_{\mathrm{out}}=f_\mathrm{w} \dot{M}_{\mathrm{out}} c^2 \frac{r_g}{r_{\mathrm {in}}}\left(\frac{r_{\mathrm {out }}}{r_{\mathrm {in }}}\right)^{-p} \\
\times\left[\frac{1-\left(r_{\mathrm {out }} / r_{\mathrm {in }}\right)^{p-1}}{1-p}-\frac{1-\left(r_{\mathrm {out }} / r_{\mathrm {in }}\right)^{-3 / 2}}{3 / 2}\right],
\end{gathered}
\end{equation}
where $f_\mathrm{w}$ is the fraction of the heat taken away and $r_g\equiv 2 G M_{\mathrm{CO}}/ c^2$ is the gravitational radius of the CO. In this work, we take a constant $f_\mathrm{w}=0.5$ \citep{Chen2023a}. $\dot{M}_{\mathrm {out}}$ is the mass inflow rate at the outer boundary of the circum-CO disks $r_{\mathrm {out}}=\min\{r_{\mathrm {obd}}, r_{\mathrm {tr}}\}$. The inner boundary of the circum-CO disk is 
\begin{equation}
r_{\mathrm {in }}= \begin{cases}10 r_g, & \text { for BHs}  \\ \max \left\{r_{\mathrm{m}}, r_{\mathrm{NS}}, 10 r_g\right\}, & \text { for NSs} \end{cases}
\end{equation}
The inner boundary for BHs is $\sim 10r_g$. But for NSs, the strong magnetic field and the hard surface should also be considered \citep{Takahashi2018}. The circum-NS disk is truncated due to the magnetic stress of the NS magnetosphere at a radius
\begin{equation}
r_{\mathrm{m}}=k\left(\frac{\mu^4 r_{\mathrm{out}}^{2 p}}{G M_{\mathrm{NS}} \dot{M}_{\mathrm{out}}^2}\right)^{\frac{1}{7+2 p}},
\end{equation}
where the typical value of $k$ is $0.5-1$ \citep{Ghosh1979,Chashkina2019}. When the accretion flow hits the hard surface of the NS, additional energy with luminosity $L_{\mathrm{acc}} \simeq G \dot{M}_{\mathrm{in}}\left(r_{\mathrm{in}}\right) M_{\mathrm{NS}} / r_{\mathrm{NS}}$ is released, so the total energy injected is $L_{\mathrm {out }}+L_{\mathrm {acc}}$ \citep{Chashkina2019}. The velocity of the disk wind is
\begin{equation}
v_{\mathrm{w}} \simeq\left(\frac{L_{\mathrm{w}}}{\dot{M}_{\mathrm{w}}}\right)^{1 / 2}=\left(\frac{L_{\mathrm{w}}}{\dot{M}_{\mathrm{in}}}\right)^{1 / 2}.
\end{equation}
where the total luminosity the disk wind takes away is
\begin{equation}
L_{\mathrm{w}}= \begin{cases}L_{\mathrm {out }}, & \text { for BHs}  \\ L_{\mathrm {out }}+L_{\mathrm {acc}}, & \text { for NSs} \end{cases}
\end{equation}

In our calculations, the following typical values are used: $M_{\mathrm{NS}}=1.4~M_{\odot}$, $M_{\mathrm{BH}}=10~M_{\odot}$, $r_{\mathrm{NS}} =10$ km and $k=0.5$. 

\subsection{The cavity evolution}\label{sec:wind_cav}
The disk wind interacts with the disk material and forms a shocked shell, which is analogous to the evolution of the stellar-wind-driven interstellar bubbles \citep{Castor1975,Weaver1977}. At first, the wind expands freely until the mass released out $M_{\mathrm{out}}= \dot{M}_{\mathrm{w}} t$ in the wind is comparable to the swept mass in the disk $M_{\mathrm{sw}} =4/3\pi(v_{\mathrm{w}}t)^3 \rho_{\mathrm{d}}$ (in this phase, the cavity is almost spherical). Thus, the duration of the FE phase is
\begin{equation}\label{eq:t_EF_acc}
\begin{aligned}
t_{\mathrm{FE}} & =\left(\frac{3 L_\mathrm{w}}{4 \pi v_{\mathrm{w}}^5 \rho_{\mathrm{d}}}\right)^{\frac{1}{2}} \\
& \simeq 0.015 ~\mathrm{yr}~ L_{\mathrm{w}, 42}^{\frac{1}{2}} v_{\mathrm{w}, 9}^{-\frac{5}{2}} \rho_{\mathrm{d},-15}^{-\frac{1}{2}}
\end{aligned}
\end{equation}

For $t \gg t_{\mathrm{FE}}$, the shell evolution also goes through the adiabatic expansion phase and the radiative cooling phase in sequence. In both phases, the shock wave radius evolves as follows \citep{Weaver1977}:
\begin{equation}\label{eq:rt_acc}
r_{\mathrm {sh}}=\alpha \left(\frac{L_{\mathrm{w}} t^3}{\rho_{\mathrm{d}}}\right)^{1 / 5},
\end{equation}
where $\alpha \approx 0.88$ for the adiabatic expansion phase, but in the radiative cooling phase, swept gas collapses into a thin shell and makes $\alpha \approx 0.76$ \citep{Weaver1977}. In this work, we ignore the FE and radiative cooling phases because they have less impact on the shock evolution (see Equations (\ref{eq:t_EF_acc}) and (\ref{eq:rt_acc})). The shock velocity in the adiabatic expansion phase is \citep{Weaver1977}
\begin{equation}
v_{\mathrm{sh}}=0.53\left(\frac{L_{\mathrm{w}}}{\rho_{\mathrm{d}} t^2}\right)^{1 / 5}
\end{equation}
The adiabatic expansion phase lasts until efficient accretion of CO stops. The accretion timescale can be approximated by the viscous timescale of the accretion disk \citep{Chen2023a}
\begin{equation}
\begin{aligned}
& t_{\mathrm {acc}}=t_{\mathrm{vis}}\left(r_{\mathrm {obd}}\right)=\alpha_{\mathrm{CO}}^{-1} h_{\mathrm{CO}}^{-2} \Omega_{\mathrm{k,CO}}^{-1} \\
& \simeq 1100 \mathrm{~yr}~ \alpha_{\mathrm{CO,-1}}^{-1} r_{\mathrm{obd, 15}}^{\frac{3}{2}}\left(\frac{h_{\mathrm{CO}}}{0.5}\right)^{-2}\left(\frac{M_{\mathrm{CO}}}{10 ~M_\odot}\right)^{-\frac{1}{2}} \\
&
\end{aligned}
\end{equation}

Similar to Section \ref{sec:mag}, the propagation of shock in the AGN disk also needs to consider the vertical and radial directions respectively. The breakout timescale is

\begin{equation}
\begin{aligned}
t_{\mathrm{bre}} & =\left(\frac{H^5 \rho_\mathrm{d}}{\alpha^5 L_{\mathrm{w}}}\right)^{1 / 3} \\
& \simeq 12 ~\mathrm{yr}~ \left(\frac{\rho_{\mathrm{d},-15}}{L_{\mathrm{w}, 42}}\right)^{1 / 3}\left(\frac{H}{0.01~\mathrm{pc}}\right)^{\frac{5}{3}}
\end{aligned}
\end{equation}
After the shock breaks out, the shock is accelerated in the vertical direction \citep{Sakurai1960}. When the shock enters the circum-disk material, the shock evolution re-enters the adiabatic expansion phase. When the efficient accretion of CO stops, the shock transitions to the ST phase.

In the radial direction, the shock evolution enters a momentum-conserving snowplow phase (see Equation \ref{eq:vw2}) after the shock breaks out. The radial width of the cavity can be obtained when the shock velocity equates to the local sound speed
\begin{equation}
\begin{aligned}
& r_{\mathrm {cav}}=H \cdot \sqrt{\frac{v_\mathrm{bre}}{c_\mathrm{s}}} \simeq 0.70  H^{2 / 3}c_\mathrm{s}^{-1/2}\left(\frac{L_\mathrm{w}}{\rho_\mathrm{d}}\right)^{\frac{1}{6}} \\
& \simeq 0.1 ~\mathrm{pc}~\left(\frac{H}{0.01 ~\mathrm{pc}}\right)^{2 / 3}\left(\frac{c_\mathrm{s}}{5\times10^5 ~\mathrm{cm~s^{-1}}}\right)^{-\frac{1}{2}}\left(\frac{L_{\mathrm{w}, 42}}{\rho_{\mathrm{d},-15}}\right)^{\frac{1}{6}} \\
&
\end{aligned}
\end{equation}

Inspired by Equation (\ref{eq:vh1}), we describe the evolution of the cavity as
\begin{equation}\label{eq:vh_out}
\dot{R}_\mathrm{H}(t)=\left\{\begin{array}{lc}
0.53\left(\frac{L_{\mathrm{w}}}{\rho_{\mathrm{d}}(h) t^2}\right)^{1 / 5}\left(\frac{\rho_{\mathrm{d}}(h)}{\rho_0}\right)^{-\mu}, & t\leq t_{\mathrm{c}} \\
v_{\mathrm{c}} \left(\frac{t}{t_{\mathrm{c}}}\right)^{-\frac{2}{5}}, & t_\mathrm{c} <t\leq t_{\mathrm{acc}} \\
v_{\mathrm{acc}}\left(\frac{t}{t_{\mathrm{acc }}}\right)^{-\frac{3}{5}} & t_\mathrm{acc} <t\leq t_{\mathrm{cav}}
\end{array},\right.
\end{equation}
in the vertical direction, and
\begin{equation}\label{eq:vw_out}
\dot{R}_\mathrm{W}(t)=\left\{\begin{array}{lr}
0.53\left(\frac{L_\mathrm{w}}{\rho_{\mathrm{d}}t^2}\right)^{1 / 5}, & t\leq t_{\mathrm{bre}} \\
H^2 v_\mathrm{w}(t_{\mathrm{bre}}) / R_\mathrm{W}^2, & t_{\mathrm{bre}}<t\leq t_{\mathrm{cav}}
\end{array}\right.
\end{equation}
in the radial direction. Equation (\ref{eq:vh_out}) means that the shock leaves the disk boundary before the efficient accretion stops ($t_{\mathrm{c}}\leq t_{\mathrm{acc}}$). But if $t_{\mathrm{c}}>t_{\mathrm{acc}}$, the diabatic expansion phase (the second line in Equation (\ref{eq:vh_out})) will be skipped.

\begin{figure}
    \centering
    \includegraphics[width = 0.5\textwidth]{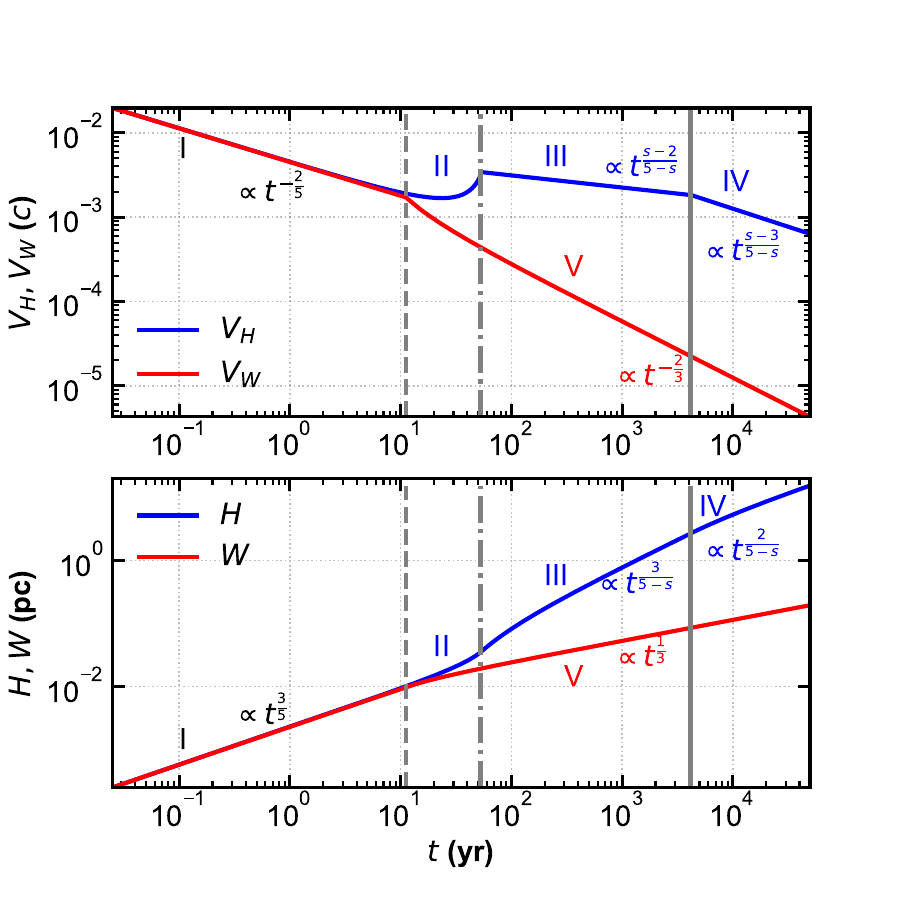}
    \caption{Solutions of Equations (\ref{eq:vh_out}) and (\ref{eq:vw_out}) for BH1. The model parameters are listed in the BH1 model in Table \ref{tab:timescale2}.  The shock velocity and radius evolution are shown in the top and bottom panels, respectively. Timescales of breakout, leaving the disk boundary and effective accretion are shown in gray vertical dashed, dash-dotted and solid lines, respectively. The height expansion (blue lines) goes through the adiabatic expansion phase in the disk (phase I, $t\leq t_{\mathrm{bre}}$),  the Sakurai accelerating phase (phase II, $t_{\mathrm{bre}}<t\leq t_{\mathrm{c}}$), the adiabatic expansion phase in the circum-disk medium (phase III, $t_{\mathrm{c}}<t\leq t_{\mathrm{acc}}$) and the ST phase (phase IV, $t_{\mathrm{acc}}<t\leq t_{\mathrm{cav}}$) in sequence. The width expansion (red lines) goes through the adiabatic expansion phase in the disk (phase I, $t\leq t_{\mathrm{bre}}$) and sp phase (phase V, $t>t_{\mathrm{bre}}$) in sequence.}
    \label{fig:R_co}
\end{figure}

\begin{figure}
    \centering
    \includegraphics[width = 0.5\textwidth]{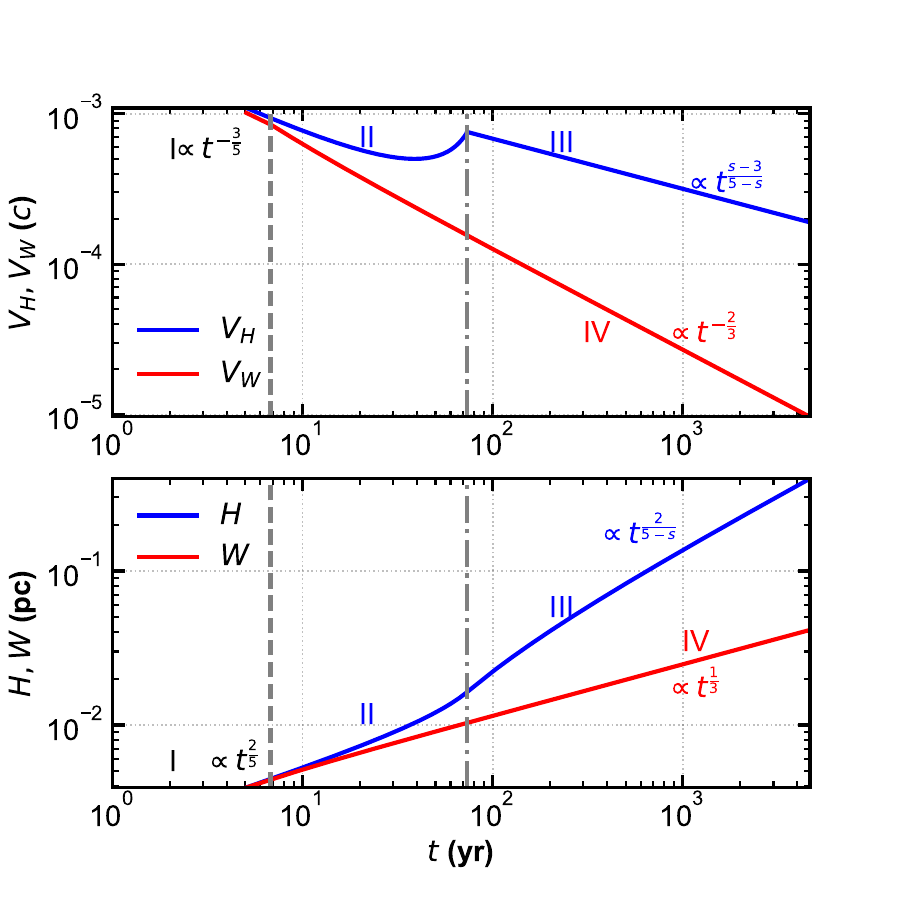}
    \caption{Solutions of Equations (\ref{eq:vh_Ew}) and (\ref{eq:vw_Ew}) for the NS2 model. The model parameters are listed in the NS2 model in Table \ref{tab:timescale2}. When $t_{\mathrm{bre}}>t_{\mathrm{acc}}$, the shock evolution is in analogy with that of the SN explosion (see Figure \ref{fig:R}). The shock velocity and radius evolution are shown in the top and bottom panels, respectively. Timescales of breakout and leaving the disk boundary are shown in gray vertical dashed and dash-dotted lines, respectively. The height expansion (blue lines) goes through the ST phase in the disk (phase I, $t\leq t_{\mathrm{bre}}$),  the Sakurai accelerating phase (phase II, $t_{\mathrm{bre}}<t\leq t_{\mathrm{c}}$) and the ST phase in the circum-disk medium (phase III, $t_{\mathrm{c}}<t\leq t_{\mathrm{cav}}$) in sequence. The width expansion (red lines) goes through the ST phase in the disk (phase I, $t\leq t_{\mathrm{bre}}$) and sp phase (phase IV, $t>t_{\mathrm{bre}}$) in sequence.}
    \label{fig:R_co2}
\end{figure}

Solutions of Equations (\ref{eq:vh_out}) and (\ref{eq:vw_out}) for BH1 are shown in Figure \ref{fig:R_co}. The model parameters are listed in the BH1 model in Table \ref{tab:timescale2}. The shock velocity and radius evolution are shown in the top and bottom panels, respectively. Timescales of breakout, leaving the disk boundary and effective accretion are shown in gray vertical dashed, dash-dotted and solid lines, respectively. The height expansion (blue lines)  goes through the adiabatic expansion phase in the disk (phase I, $t\leq t_{\mathrm{bre}}$),  the Sakurai accelerating phase (phase II, $t_{\mathrm{bre}}<t\leq t_{\mathrm{c}}$), the adiabatic expansion phase in the circum-disk medium (phase III, $t_{\mathrm{c}}<t\leq t_{\mathrm{acc}}$) and the ST phase (phase IV, $t_{\mathrm{acc}}<t\leq t_{\mathrm{cav}}$) in sequence. The width expansion (red lines) goes through the adiabatic expansion phase in the disk (phase I, $t\leq t_{\mathrm{bre}}$) and sp phase (phase V, $t>t_{\mathrm{bre}}$) in sequence.

When expansion ceases, the AGN disk material will refill the cavity. The refilled timescale is
\begin{equation}
\begin{aligned}
t_{\mathrm{ref}}&=\frac{r_{\mathrm {cav}}}{c_{\mathrm{s}}}\simeq 2\times 10^{4}~\mathrm{yr}\left(\frac{L_{\mathrm{w}, 42}}{\rho_{\mathrm{d},-15}}\right)^{\frac{1}{6}}\\
&\times \left(\frac{H}{0.01 ~\mathrm{pc}}\right)^{2 / 3}\left(\frac{c_\mathrm{s}}{5\times10^5 ~\mathrm{cm~s^{-1}}}\right)^{-\frac{3}{2}}.
\end{aligned}
\end{equation}
In the case of progenitor ejecta, the cavity can only be opened once. But for the case of accretion outflow, the accretion rate can be restored to hyper-Eddington after the cavity is refilled. Then, the strong disk outflow  forms the cavity again and the evolution process is circular. 

In Equation (\ref{eq:vw_out}), we assume that the accretion timescale is much longer than the breakout timescale ($t_{\mathrm{acc}} \gg t_{\mathrm{bre}}$). If $t_{\mathrm{bre}}>t_{\mathrm{acc}}$, we can treat the short-duration  accretion as an injection with the energy $E_\mathrm{w}\sim L_{\mathrm{w}}t_{\mathrm{acc}}$. The radius of the shock shell expands as \citep{Ostriker1988}
\begin{equation}\label{eq:rsh_Ew}
r_{\mathrm {sh,E}}=\left(\frac{E_{\mathrm{w}} t^2}{\rho_{\mathrm{d}}}\right)^{1 / 5},
\end{equation}
and the shock velocity is
\begin{equation}\label{eq:vshE}
v_{\mathrm {sh,E}}=0.4\left(\frac{E_{\mathrm{w}}}{\rho_{\mathrm{d}} t^3}\right)^{1 / 5}.
\end{equation}
In this case, the breakout timescale can be recalculated by Equation (\ref{eq:rsh_Ew})
\begin{equation}\label{eq:t_breE}
\begin{aligned}
t_{\mathrm {bre,E}} & =\left(\frac{H^5 \rho_\mathrm{d}}{E_\mathrm{w}}\right)^{1 / 2} \\
& \simeq 30 ~\mathrm{yr}\left(\frac{\rho_{\mathrm{d},-15}}{L_{\mathrm{w}, 42}}\right)^{1 / 2}\left(\frac{H}{0.01 ~\mathrm{pc}}\right)^{5 / 2}\left(\frac{t_{\mathrm{acc}}}{1 ~\mathrm{yr}}\right)^{-1 / 2}.
\end{aligned}
\end{equation}
The shock evolution equation for $t_{\mathrm{bre}}>t_{\mathrm{acc}}$ is
\begin{equation}\label{eq:vh_Ew}
\dot{R}_\mathrm{H}(t)=\left\{\begin{array}{lc}
0.4\left(\frac{E_{\mathrm{w}}}{\rho_{\mathrm{d}} t^3}\right)^{1 / 5}\left(\frac{\rho_{\mathrm{d}}(h)}{\rho_0}\right)^{-\mu}, & t_{\mathrm{acc}}\leq t\leq t_{\mathrm{c}} \\
v_{\mathrm{c}} \left(\frac{t}{t_{\mathrm{c}}}\right)^{-\frac{3}{5}}, & t_\mathrm{c} <t\leq t_{\mathrm{cav}} \\
\end{array}\right.
\end{equation}
in the vertical direction, and
\begin{equation}\label{eq:vw_Ew}
\dot{R}_\mathrm{W}(t)=\left\{\begin{array}{lr}
0.4\left(\frac{E_{\mathrm{w}}}{\rho_{\mathrm{d}} t^3}\right)^{1 / 5}, & t\leq t_{\mathrm{bre}} \\
H^2 v_\mathrm{w}(t_{\mathrm{bre}}) / R_\mathrm{W}^2, & t_{\mathrm{bre}}<t\leq t_{\mathrm{cav}}
\end{array}\right.
\end{equation}
in the radial direction. Solutions of Equations (\ref{eq:vh_Ew}) and (\ref{eq:vw_Ew}) for the NS2 model are shown in Figure \ref{fig:R_co2}. The model parameters are listed in the NS2 model in Table \ref{tab:timescale2}. When $t_{\mathrm{bre}}>t_{\mathrm{acc}}$, the shock evolution is in analogy with that of the SN explosion (see Figure \ref{fig:R}). The shock velocity and radius evolution are shown in the top and bottom panels, respectively. Timescales of breakout and leaving the disk boundary are shown in gray vertical dashed and dash-dotted lines, respectively. The height expansion (blue lines) goes through the ST phase in the disk (phase I, $t\leq t_{\mathrm{bre}}$),  the Sakurai accelerating phase (phase II, $t_{\mathrm{bre}}<t\leq t_{\mathrm{c}}$) and the ST phase in the circum-disk medium (phase III, $t_{\mathrm{c}}<t\leq t_{\mathrm{cav}}$) in sequence. The width expansion (red lines) goes through the ST phase in the disk (phase I, $t\leq t_{\mathrm{bre}}$) and sp phase (phase IV, $t>t_{\mathrm{bre}}$) in sequence.

\begin{table*}
\begin{center}
\caption{The intrinsic burst properties and evolution timescales of cavities in the AGN disk for different accretion CO models}
\begin{threeparttable}
\begin{tabular}{cccccccccccccccc}
\hline
\# & CO & $M_{\mathrm{CO}}$& $p$ & $M$ & $s$ & $Q_{\mathrm{CO}}$ & $L_{\mathrm{w}}$ & $\dot{m}$ & $L_{\mathrm{FRB}}^{\max}$ & $\nu_{\mathrm{pk}}$\tnote{1}&$t_{\mathrm{bre}}$  & $t_{\mathrm{c}}$ & $t_{\mathrm{acc}}$ &$t_{\mathrm{cav}}$ & $t_{\mathrm{ref}}$ \\
& &  ($M_{\odot}$) & &($M_{\odot}$) & & &($10^{41}$ erg $s^{-1}$) & ($10^4$)&($10^{42}$ erg $s^{-1}$) & (GHz)& (yr) & (yr)  & ($10^2$ yr) & (kyr) & ($10^4$ yr) \\
\hline
BH1 & BH & 10&0.2 & $4\times 10^{6}$ &1.5 & 2.31 & 2.93 & 4.7 & 5.87 &1.2 & 6.07 & 28.7  & 41.4 & 67.7 & 20.3 \\
BH2 & BH & 10&0.5 & $4\times 10^{6}$ &1.5 & 2.31 &0.46 & 4.7 & 5.87 &1.2 & 11.2 & 52.5  & 41.4 & 49.7 & 14.9 \\
NS1 & NS & 1.4&0.5 & $4\times 10^{6}$ &1.5 & 17.7 &0.48 & 4.28 & 0.74 &0.4 & 11.8 & 52.5  & 5.95 & 50.0 & 15.0 \\
BH3 & BH & 10&0.2 & $10^{8}$ &1.5 & 1.35 &16.3 & 40.1 &  50.1 &3.4 & 2.65 & 12.2  & 8.29 & 9.15 & 2.75 \\
BH4 & BH & 10&0.5 & $10^{8}$ &1.5 & 1.35 &1.37 & 40.1  &  50.1 &3.4 & 6.06 & 27.6  & 8.29 & 6.06 & 1.82 \\
NS2 & NS & 1.4&0.5 & $10^{8}$ &2 & 913 &0.69 & 9.73  &  1.7 &0.6 & 6.78 & 73.2  & 0.05 & 4.67 & 1.40 \\
\hline
\end{tabular}
\begin{tablenotes}
\item[1] The peak frequency of FRBs are calculated for $\Gamma_{\mathrm{q}}\sim 100$ and $\eta_{\mathrm{q}}\sim 10^{-3}$ \citep{Sridhar2021}.
\end{tablenotes}
\end{threeparttable}
\label{tab:timescale2}
\end{center}
\end{table*}

For the SG model, weaker outflows and a shorter cavity formation timescale are expected because the sound speed is higher. For a accreting BH in the SG disk with $M=4\times 10^{6}~M_{\odot}$ and $R_0=1$ pc (model a in Table \ref{tab:disk}), the wind luminosity is $L_\mathrm{w}=10^{40}$ erg s$^{-1}$ and the efficient accretion timescale is $t_{\mathrm{acc}}\approx 0.15$ yr. From Equation (\ref{eq:t_breE}), the  breakout timescale is $t_{\mathrm{bre}}\approx 3600$ yr. However, before the breakout happens, the shock is decelerated to the local speed of sound. Let $c_{\mathrm{s}}=v_{\mathrm{sh}}$ in Equation (\ref{eq:vshE}), the cavity formation timescale is $t_{\mathrm{cav}}\approx 2630$ yr. This means that for the SG model, the shock is less likely to break out. In the following section, we choose the TGM model to discuss the evolution of the cavity. The evolution timescales of cavities in the AGN disk for different accretion CO models are listed in Table \ref{tab:timescale2}. The disk parameters are taken from the TQM disk model in Table \ref{tab:disk}.

\subsection{DM and RM variations}

In this section, we investigate the DM and RM from the disk wind in the cavity and the shock shell. The mass released from the disk wind is
\begin{equation}\label{eq:Mw}
\begin{aligned}
& M_{\mathrm {w}}= \begin{cases}\dot{M}_\mathrm{w} t, & t\leq t_{\mathrm {acc}} \\
\dot{M}_\mathrm{w} t_{\mathrm {acc}} & t>t_{\mathrm{acc}}\end{cases} \\
& \simeq \begin{cases}0.02 ~M_\odot~ \dot{m}_5\left(\frac{M_{\mathrm{CO}}}{10~M_\odot}\right)\left(\frac{t}{10 ~\mathrm{yr}}\right), & t\leq t_{\mathrm {acc}} \\
2.2 ~M_\odot~ \dot{m}_5\left(\frac{M_{\mathrm{CO}}}{10~M_\odot}\right)\left(\frac{t_{\mathrm{acc}}}{1000 ~\mathrm{yr}}\right), & t>t_{\mathrm{acc}}\end{cases} \\
&
\end{aligned}
\end{equation}

Before the shock breakouts, the difference in expansion between the radial and vertical directions can be ignored, so the cavity can be regarded as spherical. However, after this, because the density of gas in the AGN disk decreases rapidly in the vertical direction, the vertical propagation of the shock becomes easier. At this point, the shape of the cavity deviates from a spherical shape. For simplicity, we assume that the cavity is always a cylinder during expansion. When $t_{\mathrm{bre}}<t_{\mathrm{acc}}$, the DM from disk wind in the cavity can be estimated by

\begin{equation}\label{eq:DMw}
\begin{aligned}
\mathrm{DM_w}(t) &= \eta \frac{M_\mathrm{w}(t)}{\mu m_{\mathrm{p}}\pi R_\mathrm{W}(t)^2 R_\mathrm{H}(t)} \cdot(1-\xi) R_{\mathrm{H}}(t) \\
& \propto \begin{cases}t^{-\frac{1}{5}} & t\leq t_{\mathrm {bre}} \\
t^{\frac{1}{3}} & t_{\mathrm {bre}}<t\leq t_{\mathrm {acc}} \\
t^{-\frac{2}{3}} & t_{\mathrm {acc}}<t\leq t_{\mathrm {cav}}\end{cases}.
\end{aligned}
\end{equation}
The shock thickness $1-\xi=0.86$ is taken from \citealt{Weaver1977}. The free–free absorption optical depth from disk wind is
\begin{equation}
\begin{aligned}\label{eq:tau_w}
\tau_{\mathrm{ff,w}}(t) &\simeq 0.018 T^{-3 / 2} \nu^{-2} \eta^2\\ &  \times  \frac{M_{\mathrm{w}}(t)^2}{\mu^2 m_\mathrm{p}^2\pi^2 R_\mathrm{W}(t)^4 R_\mathrm{H}(t)^2} \cdot(1-\xi) R_\mathrm{H}(t) \\
& \propto\left\{\begin{array}{ll}
t^{-1} & t\leq t_{\mathrm {bre}} \\
t^{\frac{1-2 s}{3(5-s)}} & t_\mathrm{c}<t\leq t_{\mathrm {acc}} \\
t^{-\frac{26-4 s}{3(5-s)}} & t_{\mathrm {acc }}<t\leq t_{\mathrm {cav }}
\end{array}\right.
\end{aligned}
\end{equation}

\begin{figure*}
    \centering
    \includegraphics[width = 1\textwidth]{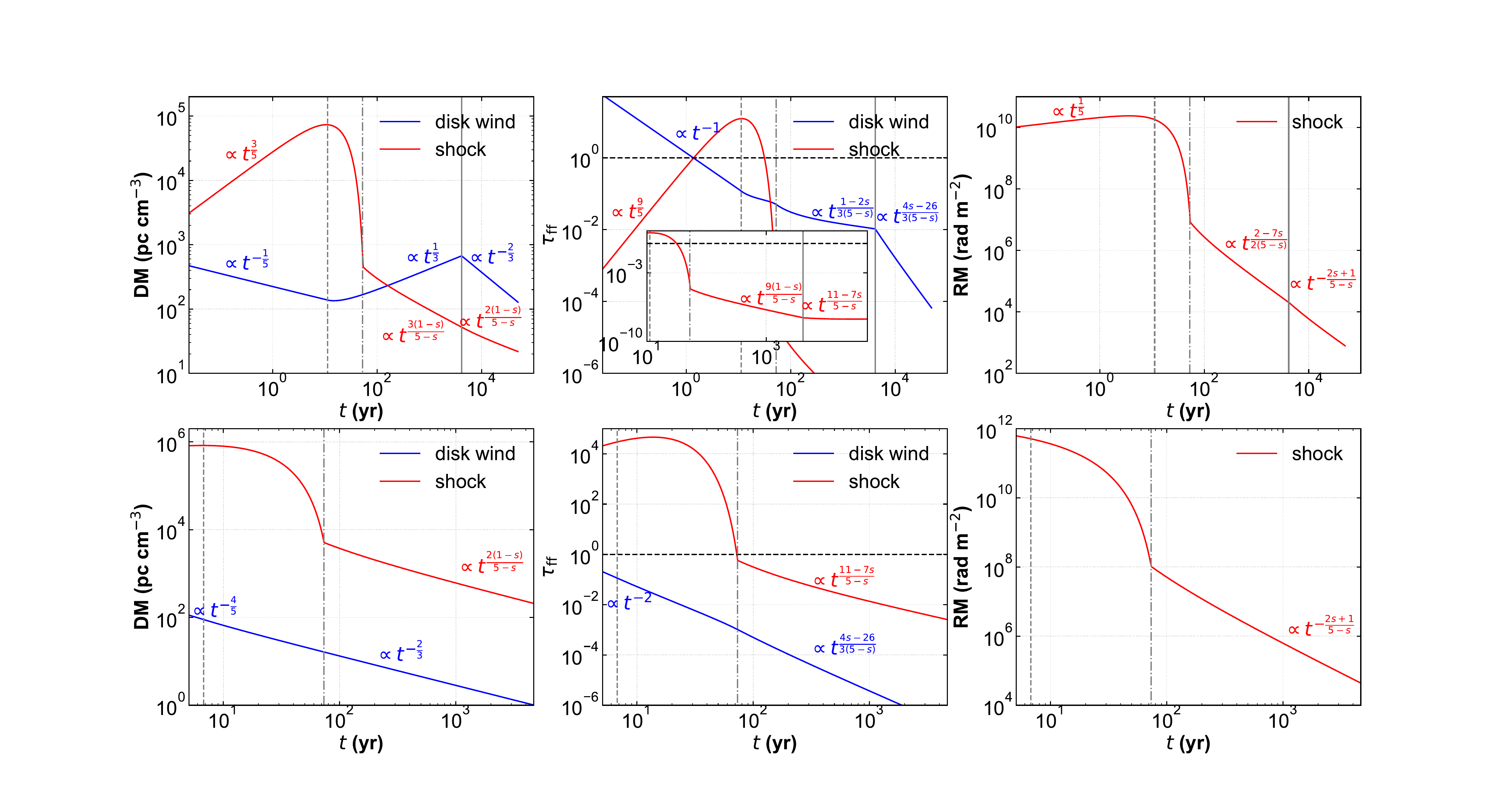}
    \caption{The DM, free-free absorption optical depth and RM evolution of accretion COs model for BH1 (top panel) and NS2 (bottom panel). The model parameters are listed in Tabel \ref{tab:timescale2}. The contributions from the unshocked cavity and the shocked shell are shown in blue and red lines, respectively. Same as the young magnetar models, DM $\tau_{\mathrm{ff}}$ and RM show non-power-law evolution patterns over time in the Sakurai accelerating phase, which is different from other environments, e.g., SNR \citep{Yang2017,Piro2018},  compact binary merger remnant \citep{Zhao2021a}, PWN/MWN \citep{Margalit2018,YYH2019,Zhao2021b}. \textbf{Top panels}: The time-dependent results are based on numerical solutions of Equations (\ref{eq:vh_out}) and (\ref{eq:vw_out}). Timescales of the breakout, leaving the disk boundary and effective accretion are shown in gray vertical dashed, dash-dotted and solid lines, respectively. For BH1, the cavity and the shocked shell are opaque for FRBs with $\nu=1$ GHz for a few decades (see the black dashed line for $\tau_{\mathrm{ff}}=1$). After the shock breakouts, the DM from the shocked shell and the free-free absorption become neglected compared to the disk wind. \textbf{Bottom panels}: The time-dependent results are based on numerical solutions of Equations (\ref{eq:vh_Ew}) and (\ref{eq:vw_Ew}). Timescales of the breakout and leaving the disk boundary are shown in gray vertical dashed and dash-dotted lines, respectively. For NS2, the cavity and the shocked shell are opaque for FRBs with $\nu=1$ GHz for a few decades (see the black dashed line for $\tau_{\mathrm{ff}}=1$). The results when $t_{\mathrm{acc}}<t \leq t_{\mathrm{bre}}$ do not satisfy the approximation expression given by Equations (\ref{eq:DMshE}) - (\ref{eq:RMshE}). For NS2, the cavity radius is close to $H$ when $t\sim t_{\mathrm{acc}}$, then it is no longer possible to assume that the density is constant as in the previous discussion (see Equation (\ref{eq:rho_h})). Thus, we only show the approximate expression when $t>t_{\mathrm{c}}$ for the shocked shell.}
    \label{fig:dm_co}
\end{figure*}

For the shocked region, gas is fully ionized because of the high temperature (see Equation (\ref{eq:Tsh})). The DM from the shock shell is
\begin{equation}
\begin{aligned}
\mathrm{DM_{sh}}(t) & =\frac{4\rho_{\mathrm{cd}}[R_\mathrm{H} (t)]}{\mu m_\mathrm{p}} \cdot \xi R_\mathrm{H}(t) \\
& \propto \begin{cases}t^{\frac{3}{5}} & t\leq t_{\mathrm {bre}} \\
t^{\frac{3(1-s)}{5-s}} & t_c<t\leq t_{\mathrm {acc }}. \\
t^{\frac{2(1-s)}{5-s}} & t_{\mathrm {acc }}<t\leq t_{\mathrm {cav }}\end{cases}
\end{aligned}
\end{equation}

The free–free absorption optical depth from shock shell is
\begin{equation}
\begin{aligned}
\tau_\mathrm{s h}(t) & =  0.018 T(t)^{-3/ 2} \nu^{-2} \frac{16\rho_{\mathrm{cd}}[R_\mathrm{H} (t)]^2}{\mu^2 m_p^2} \cdot \xi R_\mathrm{H} (t) \\
& \propto\left\{\begin{array}{ll}
t^{\frac{9}{5}} & t\leq t_{\mathrm {bre }} \\
t^{\frac{9(1- s)}{5-s}} & t_\mathrm{c}<t\leq t_\mathrm{acc}. \\
t^{\frac{11-7s}{5-s}} & t_{\mathrm {acc }}<t\leq t_{\mathrm {cav }}
\end{array}\right.
\end{aligned}
\end{equation}

In this work, we assume that only the shocked region is magnetized, and the magnetic field in the shocked shell is given in Equation (\ref{eq:Bsh}). The RM from the shocked shell is
\begin{equation}\label{eq:RMw}
\begin{aligned}
    \mathrm{RM_{sh}}(t) &\simeq8.1 \times 10^5 \mathrm{rad~m^{-2}}\\
    &\times \left(\frac{4n_{\mathrm{cd}}(t)}{\mathrm{~cm}^{-3}}\right) \left(\frac{B_\mathrm{sh}(t)}{\mathrm{G}}\right) \left(\frac{\xi R_{\mathrm{H}}(t)}{\mathrm{pc}}\right)\\
    & \propto \begin{cases}t^{\frac{1}{5}} 
    & t\leq t_{\mathrm {bre}} \\
    t^{\frac{2-7s}{2(5-s)}} & t_{\mathrm{c}}<t\leq t_{\mathrm {acc}}. \\
    t^{-\frac{2s+1}{5-s}} & t_{\mathrm{acc}}<t\leq t_{\mathrm {cav}} \\
\end{cases}.
\end{aligned}
\end{equation}

For the case $t_{\mathrm{bre}}>t_{\mathrm{acc}}$
, the DM from the disk wind in the cavity is 
\begin{equation}\label{eq:DMwE}
\begin{aligned}
\mathrm{DM_w}(t) &= \eta \frac{M_\mathrm{w}(t)}{\mu m_{\mathrm{p}}\pi R_\mathrm{W}(t)^2 R_\mathrm{H}(t)} \cdot(1-\xi) R_{\mathrm{H}}(t) \\
& \propto \begin{cases}t^{-\frac{4}{5}} & t\leq t_{\mathrm {bre}} \\
t^{-\frac{2}{3}} & t_{\mathrm {bre}}<t\leq t_{\mathrm {cav}} \end{cases},
\end{aligned}
\end{equation}
and the free–free absorption optical depth from disk wind is
\begin{equation}
\begin{aligned}\label{eq:tauwE}
\tau_{\mathrm{ff,w}}(t) &\simeq 0.018 T^{-3 / 2}\nu^{-2}  \eta^2  \\
& \times \frac{M_{\mathrm{w}}(t)^2}{\mu^2 m_\mathrm{p}^2\pi^2 R_\mathrm{W}(t)^4 R_\mathrm{H}(t)^2} \cdot(1-\xi) R_\mathrm{H}(t) \\
& \propto\left\{\begin{array}{ll}
t^{-2} & t\leq t_{\mathrm {bre}} \\
t^{-\frac{26-4s}{3(5-s)}} & t_{\mathrm {c }}<t\leq t_{\mathrm {cav }}
\end{array}\right.
\end{aligned}
\end{equation}
The same as the case $t_{\mathrm{bre}}<t_{\mathrm{acc}}$, we can only give the time evolution scaling law for the phases I and III, which is given in Equations (\ref{eq:DMwE}) and (\ref{eq:tauwE}). 

The DM, free-free absorption optical depth and RM from the shocked shell evolve as 

\begin{equation}\label{eq:DMshE}
\begin{aligned}
\mathrm{DM_{sh,E}}(t) & =\frac{4\rho_{\mathrm{cd}}[R_\mathrm{H} (t)]}{\mu m_\mathrm{p}} \cdot \xi R_\mathrm{H}(t) \\
& \propto \begin{cases}t^{\frac{2}{5}} & t\leq t_{\mathrm {bre}} \\
t^{\frac{2(1-s)}{5-s}} & t_{\mathrm {c}}<t\leq t_{\mathrm{cav}}\end{cases},
\end{aligned}
\end{equation}

\begin{equation}
\begin{aligned}
\tau_\mathrm{sh,E}(t) & =  0.018 T(t)^{-3/ 2} \nu^{-2} \frac{16\rho_{\mathrm{cd}}[R_\mathrm{H} (t)]^2}{\mu^2 m_p^2} \cdot \xi R_\mathrm{H} (t) \\
& \propto\left\{\begin{array}{ll}
t^{\frac{11}{5}} & t\leq t_{\mathrm {bre }} \\
t^{\frac{11-7s}{5-s}} & t_{\mathrm{c}}<t\leq t_{\mathrm {cav }}
\end{array}\right.
\end{aligned},
\end{equation}
and
\begin{equation}\label{eq:RMshE}
\begin{aligned}
    \mathrm{RM_{sh,E}}(t) &\simeq8.1 \times 10^5 \mathrm{rad~m^{-2}}\\
    &\times \left(\frac{4n_{\mathrm{cd}}(t)}{\mathrm{~cm}^{-3}}\right) \left(\frac{B_\mathrm{sh}(t)}{\mathrm{G}}\right) \left(\frac{\xi R_{\mathrm{H}}(t)}{\mathrm{pc}}\right)\\
    & \propto \begin{cases}t^{-\frac{1}{5}} 
    & t\leq t_{\mathrm {bre}} \\
    t^{-\frac{2s+1}{5-s}} & t_{\mathrm{c}}<t\leq t_{\mathrm {cav}} \\
\end{cases}.
\end{aligned}
\end{equation}

\begin{figure*}
    \centering
    \includegraphics[width = 1\textwidth]{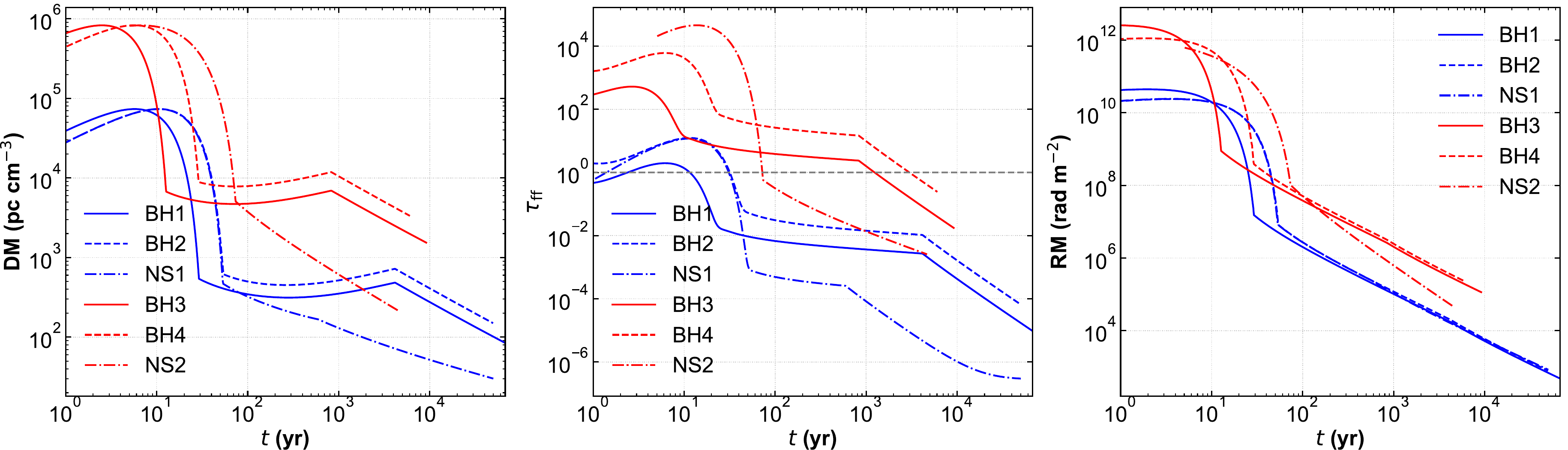}
    \caption{The total DM, $\tau_{\mathrm{ff}}$ and RM (including the contributions of the cavity and the shocked shell) from different accretion CO models. The model parameters are listed in Tabel \ref{tab:timescale2}. For the model BH1, BH2, BH3, BH4 and NS1, the effective accretion timescale is much longer than the shock breakout timescale, whose cavity evolution is governed by Equations (\ref{eq:vh_out}) and (\ref{eq:vw_out}). However, for NS2, the duration of effective accretion is very short compared to the shock breakout timescale, whose cavity evolution is governed by Equations (\ref{eq:vh_Ew}) and (\ref{eq:vw_Ew}). The condition under which FRBs with $\nu=1$ GHz can be observed ($\tau_{\mathrm{ff}}=1$) is represented by the gray horizontal dashed line. For $M=4\times 10^{6}~M_{\odot}$ (shown in blue lines), the cavity formation timescale is about tens of thousands of years. For the model of accreting BHs (BH1 and BH2), the effective accretion lasts about a few thousand years. When $t_{\mathrm{c}}<t\leq t_{\mathrm{acc}}$, the DM$_{\mathrm{sh}}$ decrease as $\propto t^{\frac{2(1-s)}{5-s}}$ but DM$_{\mathrm{w}}$ increase as $\propto t^{1/3}$. Thus, the total DM ($\mathrm{DM}=\mathrm{DM_{sh}}+\mathrm{DM_{w}}\sim10^{2}-10^{3}$ \pccm) shows a relatively stable evolutionary trend at this phase. When $t>t_{\mathrm{acc}}$, the DM is dominated by the disk wind in the cavity and decreases as $\propto t^{-2/3}$. For NS1, the effective accretion only lasts about a few hundred years. Owing to less disk outflow materials accumulating in the cavity, the DM and $\tau_{\mathrm{ff}}$ mainly come from the shocked shell. After the shock breakouts, the shocked shell becomes transparent. The DM ($\mathrm{DM}\simeq\mathrm{DM_{sh}}$) evolves as $\propto t^{\frac{3(1-s)}{5-s}}$ for $t_{\mathrm{c}}<t\leq t_{\mathrm{acc}}$ and $\propto t^{\frac{2(1-s)}{5-s}}$ for $t_{\mathrm{acc}}<t\leq t_{\mathrm{cav}}$. For $M=10^{8}~M_{\odot}$ (shown in red lines), the cavity formation timescale is about several thousands of years. For the model of accreting BHs (BH3 and BH4), the DM, $\tau_{\mathrm{ff}}$ and RM profiles are similar to the case of $M=4\times 10^{6}~M_{\odot}$, but the value is greater. For NS2, the DM and $\tau_{\mathrm{ff}}$ are dominated by the shocked shell (see Figure \ref{fig:dm_co}). For the accreting CO models, the RM is large and decreases with time, which is similar to the RM of FRB 20121102A \citep{Michilli2018,Hilmarsson2021}.}
    \label{fig:dm_co_tot}
\end{figure*}

The DM, free-free absorption optical depth and RM evolution of accretion COs model for BH1 (top panel, based on numerical solutions of Equations (\ref{eq:vh_out}) and (\ref{eq:vw_out})) and NS2 (bottom panel, based on numerical solutions of Equations (\ref{eq:vh_Ew}) and (\ref{eq:vw_Ew})) are shown in Figure \ref{fig:dm_co}. The model parameters are listed in Tabel \ref{tab:timescale2}. The contributions from the unshocked cavity and the shocked shell are shown in blue and red lines, respectively. Same as the young magnetar models, DM $\tau_{\mathrm{ff}}$ and RM  show non-power-law evolution patterns over time in the Sakurai accelerating phase, which is different from other environments, e.g., SNR \citep{Yang2017,Piro2018},  compact binary merger remnant \citep{Zhao2021a}, PWN/MWN \citep{Margalit2018,YYH2019,Zhao2021b}. For BH1, timescales of the breakout, leaving the disk boundary and effective accretion are shown in gray vertical dashed, dash-dotted and solid lines, respectively. The cavity and the shocked shell are opaque for FRBs with $\nu=1$ GHz for a few decades (see the black dashed line for $\tau_{\mathrm{ff}}=1$). After the shock breakouts, the DM from the shocked shell and the free-free absorption become neglected compared to the disk wind. For NS2, the time-dependent results are based on numerical solutions of Equations (\ref{eq:vh_Ew}) and (\ref{eq:vw_Ew}). The cavity and the shocked shell are opaque for FRBs with $\nu=1$ GHz for a few decades (see the black dashed line for $\tau_{\mathrm{ff}}=1$). The results when $t_{\mathrm{acc}}<t \leq t_{\mathrm{bre}}$ do not satisfy the approximation expression given by Equations (\ref{eq:DMshE}) - (\ref{eq:RMshE}). For NS2, the cavity radius is close to $H$ when $t\sim t_{\mathrm{acc}}$, then it is no longer possible to assume that the density is constant as in the previous discussion (see Equation (\ref{eq:rho_h})). Thus, we only show the approximate expression when $t>t_{\mathrm{c}}$ for the shocked shell.

The total DM, $\tau_{\mathrm{ff}}$ and RM (including the contributions of the cavity and the shocked shell) from different accretion CO models are shown in Figure \ref{fig:dm_co_tot}. The model parameters are listed in Tabel \ref{tab:timescale2}. For the model BH1, BH2, BH3, BH4 and NS1, the effective accretion timescale is much longer than the shock breakout timescale, whose cavity evolution is governed by Equations (\ref{eq:vh_out}) and (\ref{eq:vw_out}). However, for NS2, the duration of effective accretion is very short compared to the shock breakout timescale, whose cavity evolution is governed by Equations (\ref{eq:vh_Ew}) and (\ref{eq:vw_Ew}). The condition under which FRBs with $\nu=1$ GHz can be observed ($\tau_{\mathrm{ff}}=1$) is represented by the gray horizontal dashed line. For $M=4\times 10^{6}~M_{\odot}$(shown in blue lines), the cavity formation timescale is about tens of thousands of years. For the model of accreting BHs (BH1 and BH2), the effective accretion lasts about a few thousand years. When $t_{\mathrm{c}}<t\leq t_{\mathrm{acc}}$, the DM$_{\mathrm{sh}}$ decrease as $\propto t^{\frac{2(1-s)}{5-s}}$ but DM$_{\mathrm{w}}$ increase as $\propto t^{1/3}$. Thus, the total DM ($\mathrm{DM}=\mathrm{DM_{sh}}+\mathrm{DM_{w}}\sim10^{2}-10^{3}$ \pccm) shows a relatively stable evolutionary trend at this phase. When $t>t_{\mathrm{acc}}$, the DM is dominated by the disk wind in the cavity and decreases as $\propto t^{-2/3}$. For NS1, the effective accretion only lasts about a few hundred years. Owing to less disk outflow materials accumulating in the cavity, the DM and $\tau_{\mathrm{ff}}$ mainly come from the shocked shell. After the shock breakouts, the shocked shell becomes transparent. The DM ($\mathrm{DM}\simeq\mathrm{DM_{sh}}$) evolves as  
$\propto t^{\frac{3(1-s)}{5-s}}$ for $t_{\mathrm{c}}<t\leq t_{\mathrm{acc}}$ and $\propto t^{\frac{2(1-s)}{5-s}}$ for $t_{\mathrm{acc}}<t\leq t_{\mathrm{cav}}$. For $M=10^{8}~M_{\odot}$ ( shown in red lines), the cavity formation timescale is about several thousands of years. For the model of accreting BHs (BH3 and BH4), the DM, $\tau_{\mathrm{ff}}$ and RM profiles are similar to the case of $M=4\times 10^{6}~M_{\odot}$, but the value is greater. For NS2, the DM and $\tau_{\mathrm{ff}}$ are dominated by the shocked shell (see Figure \ref{fig:dm_co}). For the accreting CO models, the RM is large and decreases with time, which is similar to the RM of FRB 20121102A \citep{Michilli2018,Hilmarsson2021}.

\section{Discussion}\label{sec:dis}
In addition to the DM, absorption, Faraday rotation-conversion and burst luminosity discussed above, other possible observation properties should also be studied and tested in the future in the frame of the AGN disk.

\subsection{The distribution of COs in AGN disk}
Both magnetars and accreting COs are expected to be embedded in AGN disks. The distribution of FRB sources in AGN disk depends on the formations and migration of COs. COs can be formed via the core collapse of massive stars or AIC/compact binary merger. The outer disk is self-gravity unstable and is ongoing effective star formation. Thus, the SNe formation channel mainly occurs in the outer disk. The AIC of NSs preferentially occurs in the outer disk, while the binary NS merger preferentially occurs at $r \lesssim 10^{-3}$ pc from the SMBH \citep{Perna2021}. The migration of COs in the disk is subject to large uncertainty \citep[e.g.][]{Bellovary2016, Pan2021, Grishin2023}. Migration traps caused by gas torque are thought to occur at tens to hundreds of times the gravitational radius \citep{Bellovary2016}. However, migration traps can also exist at larger distances ($\sim 10^{3-5}R_{\mathrm{g}}$) if thermal torques are taken into consideration \citep{Grishin2023}.

A detailed study of CO distribution is beyond the scope of this article. Although the radial distribution of COs is uncertain, FRBs can be generated across the entire disk. If the FRB sources are in the inner disk, we can only detect signal sites at a few scale heights because of the absorption of the dense gas. However, most COs locate close to the midplane of the disk \citep{Perna2021}. The feedback (progenitors' ejecta or accreting outflows) of magnetars or accreting COs should be considered for emission from the midplane (see Sections \ref{sec:mag} and \ref{sec:co}). The local sound speed for the inner disk is too high for the shock to break out. However, the feedback cavity in the outer disk can reduce the absorption by the dense disk materials, making FRBs detectable.

\subsection{Event rate}
The event rate in AGN disks can be estimated as \citep{Tagawa2020}
\begin{equation}
    \mathcal{R}=\int \frac{d n_{\mathrm{AGN}}}{d M} \frac{f_{\mathrm{FRB}} N_{\mathrm{CO}}}{t_{\mathrm{AGN}}} d M, 
\end{equation}
where $N_{\mathrm{CO}}$ is the typical number of COs in AGN disks, $t_{\mathrm{AGN}}$ is the average lifetime of AGN disks and $f_{\mathrm{FRB}}$ is the fraction of CO that can produce FRB. Using the fitting results of AGN number density $d n_{\mathrm{AGN}}/d M$ from \cite{Bartos2017}, the event rate in AGN disk is given by \citep{Tagawa2020,Perna2021}

\begin{equation}
\mathcal{R} \sim  5 ~\mathrm{Gpc}^{-3} \mathrm{yr}^{-1}\left(\frac{f_{\mathrm {FRB}}}{0.1}\right)\left(\frac{t_{\mathrm{AGN}}}{100 ~\mathrm{Myr}}\right)^{-1} \left(\frac{\eta_{\mathrm{n}, \mathrm{CO}}}{0.005~ M_{\odot}^{-1}}\right),
\end{equation}
where $\eta_{\mathrm{n}, \mathrm{CO}}$ is the number of COs per unit stellar mass. Considering different initial mass function (IMF, both the Salpeter IMF and the top-heavy IMF \citep{Lu2013}), the value of $\eta_{\mathrm{n}, \mathrm{NS}}$ is $\sim 0.005-0.03 M_{\odot}^{-1}$ for NSs \citep{Perna2021} and $\eta_{\mathrm{n}, \mathrm{BH}}$ is $\sim 0.002-0.02 M_{\odot}^{-1}$ for BHs \citep{Tagawa2020}. For magnetar models, taking the magnetar fraction as $f_{\mathrm{FRB}}\sim f_{\mathrm{m}} \sim 0.1$ \citep{Muno2008}, the magnetar origin FRB event rate is $\mathcal{R}_{\mathrm{m}}\sim 5-30$ \gpcyr. For accreting CO models, the $f_{\mathrm{FRB}}$ depends on the efficient accretion timescale. Taking $f_{\mathrm{FRB}}\sim t_{\mathrm{acc}}/t_{\mathrm{ref}}\sim 0.05$ (see Table \ref{tab:timescale2}), the accretion origin FRB event rate is $\mathcal{R}_{\mathrm{acc}}=\mathcal{R}_{\mathrm{acc,NS}}+\mathcal{R}_{\mathrm{acc,BH}}\sim 3.5-35$ \gpcyr. 

The FRB volumetric rate is found very high by assuming that the FRB 200428-like events are the same as the cosmological FRB population, e.g., $\sim 10^7-10^8$ \gpcyr \citep{Lu2020}, $>4 \times 10^6$ \gpcyr \citep{Bochenek2020}. The contribution of FRBs from AGN disks to the total FRB population is $\sim 10^{-5}-10^{-7}$. However, the FRBs with low energy can not be detected for cosmological origins. For accreting CO models, bright repeating FRBs are expected (see Section \ref{sec:burst}). The bright FRB rate is $\mathcal{R}_{\mathrm{FRB}}(L>10^{42} \mathrm{erg}~ \mathrm{s}^{-1})\sim 3.5 \times 10^4$ \gpcyr\citep{LR20}. The repeating FRBs fraction is about 10\%, estimated from the first CHIME FRB Catalog \citep{CHIME/FRBCollaboration2021}. Thus, the accreting-powered FRBs in AGN disks account for about $0.1\%-1\%$ of total bright repeating FRBs.

\subsection{Stochastic DM and RM variations, depolarization}

Many repeating FRBs show stochastic variations on DMs and RMs, e.g., FRB 20180916B \citep{Chawla2020,Pastor-Marazuela2021,Pleunis2021,Mckinven2023}, FRB 20190520B \citep{Niu2022}, FRB 20201124A \citep{Hilmarsson2021b,Kumar2022,Xu2022} and FRB 20220912A \citep{Zhang2023}. Inhomogeneous and turbulent media are introduced to explain stochastic DM and RM variations, e.g., clouds in SNRs \citep{Katz2021,Yang2023} or clumps in stellar wind or the decretion disk of massive stars \citep{Wang2022,Zhao2023}. The multipath propagation in turbulent magnetized plasma also causes frequency-dependent depolarization \citep{Bochenek2020,Yang2022}, which has been found in active repeating FRBs \citep{Feng2022,Lu2023}.

The AGN disk is also found to be inhomogeneous. Sonic-scale magneto-rotational and gravitational instabilities would commonly occur in AGN disks \citep[e.g.][]{Balbus1998, Gammie2001, Goodman2003, Lin2023}, both can excite inhomogeneous turbulence with locally chaotic eddies, leading to stochastic variations on DMs and RMs of FRBs.

\subsection{RM reversals}
Recently, RM reversals have been reported for some FRBs, such as FRB 20201124A \citep{Xu2022}, FRB 20190520B \citep{Anna-Thomas2023} and FRB 20180301A \citep{Kumar2023}. Possible explanations of magnetic field reversals are that the FRB source is in a massive binary system \citep{Wang2022,Zhao2023} or a magnetized turbulent environment \citep{Anna-Thomas2023}.

The large-scale toroidal magnetic fields of a magnetized accretion disk reverse with the magnetorotational instability (MRI) dynamo cycles \citep{Salvesen2016}. For the FRB sources in the AGN disks, the magnetic field reversal can be explained naturally.

\section{Conclusions}\label{sec:con}
In this work, we investigate the observational properties of FRBs to occur in the disks of AGNs. Two mainstream types of radiation mechanism models are considered, such as close-in models from the magnetosphere of magnetars and far-away models from relativistic outflows from accreting COs. Progenitors’ ejecta or accretion disks' outflows interact with the disk material to form a cavity. The cavity makes the FRB easier to escape. The propagation of FRBs in the disk or cavity causes dispersion, free-free absorption, Faraday rotation and Faraday conversion. Our conclusions are summarized as follows.  

\begin{itemize}
\item If the feedback of the ejecta or outflows is weak and there is no extra ionization source, whether the FRB is absorbed or not depends only on the optical depth of the disk. For an SG disk, the disk becomes optically thin at $r>10^4-10^5 R_{\mathrm{g}}$. For a TQM disk, the disk becomes optical thin at the distance $\sim 10^{3}-10^{4}R_g$.
\item For the inner disk or the outer disk ionized by the radiation or shocks, FRBs can be observed only from sources located at a few scale heights. The largest DM and RM contributed by the disk is about $10^3-10^5$ \pccm and $10^4-10^5$ \radm, respectively.
\item If the magnetic field in the AGN disks is toroidal field dominated, FC occurs. For the intrinsic 100\% linearly polarized radiation, FC converts linear polarization into circular polarization when radiation passes through the disk.
\item For close-in models of young magnetar born in SN explosions or the merger of two compact stars, the AGN disk environments do not affect the radiation properties. For accreting-powered models, the burst luminosity depends on the accretion rate. The event rate of magnetar-origin FRBs in AGN disks is $\sim 5-30$ \gpcyr. For accreting CO models, the FRB rate in AGN disks is $\sim 3.5-35$ \gpcyr. The hyper-Eddington accretion of COs in AGN disks makes FRB brighter, accounting for about $0.1\%-1\%$ of total bright repeating FRBs. 
\item Shock is generated during the interaction between the ejecta or outflows and the AGN disk materials. The shock is quenched by dense disk materials in the radial direction, but can break out in the vertical direction. The DM and RM during shock breakout show a non-power law evolution pattern over time, which is completely different from other environments (such as supernova remnants).
\item The gas in AGN disk is inhomogeneous and turbulent, resulting in stochastic DM and RM variations. The multipath propagation in turbulent magnetized plasma  also causes frequency-dependent depolarization \citep{Bochenek2020,Yang2022,Lu2023}.
\item The large-scale toroidal magnetic fields of a magnetized accretion disk reverse with the magnetorotational instability (MRI) dynamo cycles \citep{Salvesen2016}. For the FRB sources in the AGN disks, the magnetic field reversal can be explained naturally.

\end{itemize}

\section*{Acknowledgements}
We thank an anonymous referee for helpful comments, and Jian-Min Wang, Luis Ho, Jonathan Katz, and Shu-Qing Zhong for helpful discussions. This work was supported by the National SKA Program of China (grant Nos. 2020SKA0120300 and 2022SKA0130100), the National Natural Science Foundation of China (grant Nos. 12273009, and 12393812), and the China Manned Spaced Project (CMS-CSST-2021-A12).

%%%%%%%%%%%%%%%%%%%%%%%%%%%%%%%%%%%%%%%%%%%%%%%%%%
\section*{Data Availability}
There is no data generated from this theoretical work.



\bibliographystyle{mnras}
\bibliography{ms} % if your bibtex file is called example.bib

\begin{thebibliography}{}
\makeatletter
\relax
\def\mn@urlcharsother{\let\do\@makeother \do\$\do\&\do\#\do\^\do\_\do\%\do\~}
\def\mn@doi{\begingroup\mn@urlcharsother \@ifnextchar [ {\mn@doi@}
  {\mn@doi@[]}}
\def\mn@doi@[#1]#2{\def\@tempa{#1}\ifx\@tempa\@empty \href
  {http://dx.doi.org/#2} {doi:#2}\else \href {http://dx.doi.org/#2} {#1}\fi
  \endgroup}
\def\mn@eprint#1#2{\mn@eprint@#1:#2::\@nil}
\def\mn@eprint@arXiv#1{\href {http://arxiv.org/abs/#1} {{\tt arXiv:#1}}}
\def\mn@eprint@dblp#1{\href {http://dblp.uni-trier.de/rec/bibtex/#1.xml}
  {dblp:#1}}
\def\mn@eprint@#1:#2:#3:#4\@nil{\def\@tempa {#1}\def\@tempb {#2}\def\@tempc
  {#3}\ifx \@tempc \@empty \let \@tempc \@tempb \let \@tempb \@tempa \fi \ifx
  \@tempb \@empty \def\@tempb {arXiv}\fi \@ifundefined
  {mn@eprint@\@tempb}{\@tempb:\@tempc}{\expandafter \expandafter \csname
  mn@eprint@\@tempb\endcsname \expandafter{\@tempc}}}

\bibitem[\protect\citeauthoryear{{Anna-Thomas} et~al.,}{{Anna-Thomas}
  et~al.}{2023}]{Anna-Thomas2023}
{Anna-Thomas} R.,  et~al., 2023, \mn@doi [Science] {10.1126/science.abo6526},
  \href {https://ui.adsabs.harvard.edu/abs/2023Sci...380..599A} {380, 599}

\bibitem[\protect\citeauthoryear{{Balbus} \& {Hawley}}{{Balbus} \&
  {Hawley}}{1998}]{Balbus1998}
{Balbus} S.~A.,  {Hawley} J.~F.,  1998, \mn@doi [Reviews of Modern Physics]
  {10.1103/RevModPhys.70.1}, \href
  {https://ui.adsabs.harvard.edu/abs/1998RvMP...70....1B} {70, 1}

\bibitem[\protect\citeauthoryear{{Bartos}, {Kocsis}, {Haiman}  \&
  {M{\'a}rka}}{{Bartos} et~al.}{2017}]{Bartos2017}
{Bartos} I.,  {Kocsis} B.,  {Haiman} Z.,   {M{\'a}rka} S.,  2017, \mn@doi
  [\apj] {10.3847/1538-4357/835/2/165}, \href
  {https://ui.adsabs.harvard.edu/abs/2017ApJ...835..165B} {835, 165}

\bibitem[\protect\citeauthoryear{{Bauswein}, {Goriely}  \& {Janka}}{{Bauswein}
  et~al.}{2013}]{Bauswein2013}
{Bauswein} A.,  {Goriely} S.,   {Janka} H.~T.,  2013, \mn@doi [\apj]
  {10.1088/0004-637X/773/1/78}, \href
  {https://ui.adsabs.harvard.edu/abs/2013ApJ...773...78B} {773, 78}

\bibitem[\protect\citeauthoryear{{Bellovary}, {Mac Low}, {McKernan}  \&
  {Ford}}{{Bellovary} et~al.}{2016}]{Bellovary2016}
{Bellovary} J.~M.,  {Mac Low} M.-M.,  {McKernan} B.,   {Ford} K.~E.~S.,  2016,
  \mn@doi [\apjl] {10.3847/2041-8205/819/2/L17}, \href
  {https://ui.adsabs.harvard.edu/abs/2016ApJ...819L..17B} {819, L17}

\bibitem[\protect\citeauthoryear{{Beloborodov}}{{Beloborodov}}{2017}]{Beloborodov2017}
{Beloborodov} A.~M.,  2017, \mn@doi [\apjl] {10.3847/2041-8213/aa78f3}, \href
  {https://ui.adsabs.harvard.edu/abs/2017ApJ...843L..26B} {843, L26}

\bibitem[\protect\citeauthoryear{{Blandford} \& {Begelman}}{{Blandford} \&
  {Begelman}}{1999}]{Blandford1999}
{Blandford} R.~D.,  {Begelman} M.~C.,  1999, \mn@doi [\mnras]
  {10.1046/j.1365-8711.1999.02358.x}, \href
  {https://ui.adsabs.harvard.edu/abs/1999MNRAS.303L...1B} {303, L1}

\bibitem[\protect\citeauthoryear{{Blandford} \& {Znajek}}{{Blandford} \&
  {Znajek}}{1977}]{Blandford1977}
{Blandford} R.~D.,  {Znajek} R.~L.,  1977, \mn@doi [\mnras]
  {10.1093/mnras/179.3.433}, \href
  {https://ui.adsabs.harvard.edu/abs/1977MNRAS.179..433B} {179, 433}

\bibitem[\protect\citeauthoryear{{Bochenek}, {Ravi}, {Belov}, {Hallinan},
  {Kocz}, {Kulkarni}  \& {McKenna}}{{Bochenek} et~al.}{2020}]{Bochenek2020}
{Bochenek} C.~D.,  {Ravi} V.,  {Belov} K.~V.,  {Hallinan} G.,  {Kocz} J.,
  {Kulkarni} S.~R.,   {McKenna} D.~L.,  2020, \mn@doi [\nat]
  {10.1038/s41586-020-2872-x}, \href
  {https://ui.adsabs.harvard.edu/abs/2020Natur.587...59B} {587, 59}

\bibitem[\protect\citeauthoryear{{CHIME/FRB Collaboration} et~al.,}{{CHIME/FRB
  Collaboration} et~al.}{2018}]{CHIME2018}
{CHIME/FRB Collaboration} et~al., 2018, \mn@doi [\apj]
  {10.3847/1538-4357/aad188}, \href
  {https://ui.adsabs.harvard.edu/abs/2018ApJ...863...48C} {863, 48}

\bibitem[\protect\citeauthoryear{{CHIME/FRB Collaboration} et~al.,}{{CHIME/FRB
  Collaboration} et~al.}{2020}]{CHIME2020}
{CHIME/FRB Collaboration} et~al., 2020, \mn@doi [\nat]
  {10.1038/s41586-020-2863-y}, \href
  {https://ui.adsabs.harvard.edu/abs/2020Natur.587...54C} {587, 54}

\bibitem[\protect\citeauthoryear{{CHIME/FRB Collaboration} et~al.,}{{CHIME/FRB
  Collaboration} et~al.}{2021}]{CHIME/FRBCollaboration2021}
{CHIME/FRB Collaboration} et~al., 2021, \mn@doi [\apjs]
  {10.3847/1538-4365/ac33ab}, \href
  {https://ui.adsabs.harvard.edu/abs/2021ApJS..257...59C} {257, 59}

\bibitem[\protect\citeauthoryear{{Castor}, {McCray}  \& {Weaver}}{{Castor}
  et~al.}{1975}]{Castor1975}
{Castor} J.,  {McCray} R.,   {Weaver} R.,  1975, \mn@doi [\apjl]
  {10.1086/181908}, \href
  {https://ui.adsabs.harvard.edu/abs/1975ApJ...200L.107C} {200, L107}

\bibitem[\protect\citeauthoryear{{Chashkina}, {Lipunova}, {Abolmasov}  \&
  {Poutanen}}{{Chashkina} et~al.}{2019}]{Chashkina2019}
{Chashkina} A.,  {Lipunova} G.,  {Abolmasov} P.,   {Poutanen} J.,  2019,
  \mn@doi [\aap] {10.1051/0004-6361/201834414}, \href
  {https://ui.adsabs.harvard.edu/abs/2019A&A...626A..18C} {626, A18}

\bibitem[\protect\citeauthoryear{{Chatterjee} et~al.,}{{Chatterjee}
  et~al.}{2017}]{Chatterjee2017}
{Chatterjee} S.,  et~al., 2017, \mn@doi [\nat] {10.1038/nature20797}, \href
  {https://ui.adsabs.harvard.edu/abs/2017Natur.541...58C} {541, 58}

\bibitem[\protect\citeauthoryear{{Chawla} et~al.,}{{Chawla}
  et~al.}{2020}]{Chawla2020}
{Chawla} P.,  et~al., 2020, \mn@doi [\apjl] {10.3847/2041-8213/ab96bf}, \href
  {https://ui.adsabs.harvard.edu/abs/2020ApJ...896L..41C} {896, L41}

\bibitem[\protect\citeauthoryear{{Chen} \& {Dai}}{{Chen} \&
  {Dai}}{2024}]{Chen2024}
{Chen} K.,  {Dai} Z.-G.,  2024, \mn@doi [\apj] {10.3847/1538-4357/ad0dfd},
  \href {https://ui.adsabs.harvard.edu/abs/2024ApJ...961..206C} {961, 206}

\bibitem[\protect\citeauthoryear{{Chen} \& {Lin}}{{Chen} \&
  {Lin}}{2023}]{Lin2023}
{Chen} Y.-X.,  {Lin} D. N.~C.,  2023, \mn@doi [\mnras] {10.1093/mnras/stad992},
  \href {https://ui.adsabs.harvard.edu/abs/2023MNRAS.522..319C} {522, 319}

\bibitem[\protect\citeauthoryear{{Chen}, {Ren}  \& {Dai}}{{Chen}
  et~al.}{2023}]{Chen2023a}
{Chen} K.,  {Ren} J.,   {Dai} Z.-G.,  2023, \mn@doi [\apj]
  {10.3847/1538-4357/acc45f}, \href
  {https://ui.adsabs.harvard.edu/abs/2023ApJ...948..136C} {948, 136}

\bibitem[\protect\citeauthoryear{{Chevalier}}{{Chevalier}}{1982}]{Chevalier1982}
{Chevalier} R.~A.,  1982, \mn@doi [\apj] {10.1086/160126}, \href
  {https://ui.adsabs.harvard.edu/abs/1982ApJ...258..790C} {258, 790}

\bibitem[\protect\citeauthoryear{{Chime/Frb Collaboration} et~al.,}{{Chime/Frb
  Collaboration} et~al.}{2020}]{Chime/FrbCollaboration2020}
{Chime/Frb Collaboration} et~al., 2020, \mn@doi [\nat]
  {10.1038/s41586-020-2398-2}, \href
  {https://ui.adsabs.harvard.edu/abs/2020Natur.582..351C} {582, 351}

\bibitem[\protect\citeauthoryear{{Cruces} et~al.,}{{Cruces}
  et~al.}{2021}]{Cruces2021}
{Cruces} M.,  et~al., 2021, \mn@doi [\mnras] {10.1093/mnras/staa3223}, \href
  {https://ui.adsabs.harvard.edu/abs/2021MNRAS.500..448C} {500, 448}

\bibitem[\protect\citeauthoryear{{Dai} \& {Lu}}{{Dai} \& {Lu}}{1998}]{Dai1998}
{Dai} Z.~G.,  {Lu} T.,  1998, \mn@doi [\prl] {10.1103/PhysRevLett.81.4301},
  \href {https://ui.adsabs.harvard.edu/abs/1998PhRvL..81.4301D} {81, 4301}

\bibitem[\protect\citeauthoryear{{Dai}, {Wang}, {Wu}  \& {Zhang}}{{Dai}
  et~al.}{2006}]{Dai2006}
{Dai} Z.~G.,  {Wang} X.~Y.,  {Wu} X.~F.,   {Zhang} B.,  2006, \mn@doi [Science]
  {10.1126/science.1123606}, \href
  {https://ui.adsabs.harvard.edu/abs/2006Sci...311.1127D} {311, 1127}

\bibitem[\protect\citeauthoryear{{Dai}, {Wang}, {Wu}  \& {Huang}}{{Dai}
  et~al.}{2016}]{Dai2016}
{Dai} Z.~G.,  {Wang} J.~S.,  {Wu} X.~F.,   {Huang} Y.~F.,  2016, \mn@doi [\apj]
  {10.3847/0004-637X/829/1/27}, \href
  {https://ui.adsabs.harvard.edu/abs/2016ApJ...829...27D} {829, 27}

\bibitem[\protect\citeauthoryear{{Deng}, {Zhong}  \& {Dai}}{{Deng}
  et~al.}{2021}]{Deng2021}
{Deng} C.-M.,  {Zhong} S.-Q.,   {Dai} Z.-G.,  2021, \mn@doi [\apj]
  {10.3847/1538-4357/ac30db}, \href
  {https://ui.adsabs.harvard.edu/abs/2021ApJ...922...98D} {922, 98}

\bibitem[\protect\citeauthoryear{{Dessart}, {Burrows}, {Livne}  \&
  {Ott}}{{Dessart} et~al.}{2007}]{Dessart2007}
{Dessart} L.,  {Burrows} A.,  {Livne} E.,   {Ott} C.~D.,  2007, \mn@doi [\apj]
  {10.1086/521701}, \href
  {https://ui.adsabs.harvard.edu/abs/2007ApJ...669..585D} {669, 585}

\bibitem[\protect\citeauthoryear{{Dyson} \& {Williams}}{{Dyson} \&
  {Williams}}{1980}]{Dyson1980}
{Dyson} J.~E.,  {Williams} D.~A.,  1980, {Physics of the interstellar medium}

\bibitem[\protect\citeauthoryear{{Edgar}}{{Edgar}}{2004}]{Edgar2004}
{Edgar} R.,  2004, \mn@doi [\nar] {10.1016/j.newar.2004.06.001}, \href
  {https://ui.adsabs.harvard.edu/abs/2004NewAR..48..843E} {48, 843}

\bibitem[\protect\citeauthoryear{{Feng} et~al.,}{{Feng}
  et~al.}{2022}]{Feng2022}
{Feng} Y.,  et~al., 2022, \mn@doi [Science] {10.1126/science.abl7759}, \href
  {https://ui.adsabs.harvard.edu/abs/2022Sci...375.1266F} {375, 1266}

\bibitem[\protect\citeauthoryear{{Gajjar} et~al.,}{{Gajjar}
  et~al.}{2018}]{Gajjar2018}
{Gajjar} V.,  et~al., 2018, \mn@doi [\apj] {10.3847/1538-4357/aad005}, \href
  {https://ui.adsabs.harvard.edu/abs/2018ApJ...863....2G} {863, 2}

\bibitem[\protect\citeauthoryear{{Gallant}, {Hoshino}, {Langdon}, {Arons}  \&
  {Max}}{{Gallant} et~al.}{1992}]{Gallant1992}
{Gallant} Y.~A.,  {Hoshino} M.,  {Langdon} A.~B.,  {Arons} J.,   {Max} C.~E.,
  1992, \mn@doi [\apj] {10.1086/171326}, \href
  {https://ui.adsabs.harvard.edu/abs/1992ApJ...391...73G} {391, 73}

\bibitem[\protect\citeauthoryear{{Gammie}}{{Gammie}}{2001}]{Gammie2001}
{Gammie} C.~F.,  2001, \mn@doi [\apj] {10.1086/320631}, \href
  {https://ui.adsabs.harvard.edu/abs/2001ApJ...553..174G} {553, 174}

\bibitem[\protect\citeauthoryear{{Geng} \& {Huang}}{{Geng} \&
  {Huang}}{2015}]{Geng2015}
{Geng} J.~J.,  {Huang} Y.~F.,  2015, \mn@doi [\apj]
  {10.1088/0004-637X/809/1/24}, \href
  {https://ui.adsabs.harvard.edu/abs/2015ApJ...809...24G} {809, 24}

\bibitem[\protect\citeauthoryear{{Ghosh} \& {Lamb}}{{Ghosh} \&
  {Lamb}}{1979}]{Ghosh1979}
{Ghosh} P.,  {Lamb} F.~K.,  1979, \mn@doi [\apj] {10.1086/157498}, \href
  {https://ui.adsabs.harvard.edu/abs/1979ApJ...234..296G} {234, 296}

\bibitem[\protect\citeauthoryear{{Giacomazzo} \& {Perna}}{{Giacomazzo} \&
  {Perna}}{2013}]{Giacomazzo2013}
{Giacomazzo} B.,  {Perna} R.,  2013, \mn@doi [\apjl]
  {10.1088/2041-8205/771/2/L26}, \href
  {https://ui.adsabs.harvard.edu/abs/2013ApJ...771L..26G} {771, L26}

\bibitem[\protect\citeauthoryear{{Goodman}}{{Goodman}}{2003}]{Goodman2003}
{Goodman} J.,  2003, \mn@doi [\mnras] {10.1046/j.1365-8711.2003.06241.x}, \href
  {https://ui.adsabs.harvard.edu/abs/2003MNRAS.339..937G} {339, 937}

\bibitem[\protect\citeauthoryear{{Grishin}, {Bobrick}, {Hirai}, {Mandel}  \&
  {Perets}}{{Grishin} et~al.}{2021}]{Grishin2021}
{Grishin} E.,  {Bobrick} A.,  {Hirai} R.,  {Mandel} I.,   {Perets} H.~B.,
  2021, \mn@doi [\mnras] {10.1093/mnras/stab1957}, \href
  {https://ui.adsabs.harvard.edu/abs/2021MNRAS.507..156G} {507, 156}

\bibitem[\protect\citeauthoryear{{Grishin}, {Gilbaum}  \& {Stone}}{{Grishin}
  et~al.}{2023}]{Grishin2023}
{Grishin} E.,  {Gilbaum} S.,   {Stone} N.~C.,  2023, \mn@doi [arXiv e-prints]
  {10.48550/arXiv.2307.07546}, \href
  {https://ui.adsabs.harvard.edu/abs/2023arXiv230707546G} {p. arXiv:2307.07546}

\bibitem[\protect\citeauthoryear{{Gruzinov} \& {Levin}}{{Gruzinov} \&
  {Levin}}{2019}]{Gruzinov2019}
{Gruzinov} A.,  {Levin} Y.,  2019, \mn@doi [\apj] {10.3847/1538-4357/ab0fa3},
  \href {https://ui.adsabs.harvard.edu/abs/2019ApJ...876...74G} {876, 74}

\bibitem[\protect\citeauthoryear{{Hilmarsson}, {Spitler}, {Main}  \&
  {Li}}{{Hilmarsson} et~al.}{2021a}]{Hilmarsson2021b}
{Hilmarsson} G.~H.,  {Spitler} L.~G.,  {Main} R.~A.,   {Li} D.~Z.,  2021a,
  \mn@doi [\mnras] {10.1093/mnras/stab2936}, \href
  {https://ui.adsabs.harvard.edu/abs/2021MNRAS.508.5354H} {508, 5354}

\bibitem[\protect\citeauthoryear{{Hilmarsson} et~al.,}{{Hilmarsson}
  et~al.}{2021b}]{Hilmarsson2021}
{Hilmarsson} G.~H.,  et~al., 2021b, \mn@doi [\apjl] {10.3847/2041-8213/abdec0},
  \href {https://ui.adsabs.harvard.edu/abs/2021ApJ...908L..10H} {908, L10}

\bibitem[\protect\citeauthoryear{{Iwamoto}, {Matsumoto}, {Amano}, {Matsukiyo}
  \& {Hoshino}}{{Iwamoto} et~al.}{2024}]{Iwamoto2024}
{Iwamoto} M.,  {Matsumoto} Y.,  {Amano} T.,  {Matsukiyo} S.,   {Hoshino} M.,
  2024, \mn@doi [\prl] {10.1103/PhysRevLett.132.035201}, \href
  {https://ui.adsabs.harvard.edu/abs/2024PhRvL.132c5201I} {132, 035201}

\bibitem[\protect\citeauthoryear{{Jermyn}, {Dittmann}, {Cantiello}  \&
  {Perna}}{{Jermyn} et~al.}{2021}]{Jermyn2021}
{Jermyn} A.~S.,  {Dittmann} A.~J.,  {Cantiello} M.,   {Perna} R.,  2021,
  \mn@doi [\apj] {10.3847/1538-4357/abfb67}, \href
  {https://ui.adsabs.harvard.edu/abs/2021ApJ...914..105J} {914, 105}

\bibitem[\protect\citeauthoryear{{Kanagawa}, {Muto}, {Tanaka}, {Tanigawa},
  {Takeuchi}, {Tsukagoshi}  \& {Momose}}{{Kanagawa}
  et~al.}{2015}]{Kanagawa2015}
{Kanagawa} K.~D.,  {Muto} T.,  {Tanaka} H.,  {Tanigawa} T.,  {Takeuchi} T.,
  {Tsukagoshi} T.,   {Momose} M.,  2015, \mn@doi [\apjl]
  {10.1088/2041-8205/806/1/L15}, \href
  {https://ui.adsabs.harvard.edu/abs/2015ApJ...806L..15K} {806, L15}

\bibitem[\protect\citeauthoryear{{Kanagawa}, {Muto}, {Tanaka}, {Tanigawa},
  {Takeuchi}, {Tsukagoshi}  \& {Momose}}{{Kanagawa}
  et~al.}{2016}]{Kanagawa2016}
{Kanagawa} K.~D.,  {Muto} T.,  {Tanaka} H.,  {Tanigawa} T.,  {Takeuchi} T.,
  {Tsukagoshi} T.,   {Momose} M.,  2016, \mn@doi [\pasj] {10.1093/pasj/psw037},
  \href {https://ui.adsabs.harvard.edu/abs/2016PASJ...68...43K} {68, 43}

\bibitem[\protect\citeauthoryear{{Kashiyama} \& {Murase}}{{Kashiyama} \&
  {Murase}}{2017}]{Kashiyama2017}
{Kashiyama} K.,  {Murase} K.,  2017, \mn@doi [\apjl]
  {10.3847/2041-8213/aa68e1}, \href
  {https://ui.adsabs.harvard.edu/abs/2017ApJ...839L...3K} {839, L3}

\bibitem[\protect\citeauthoryear{{Katz}}{{Katz}}{2016}]{Katz2016}
{Katz} J.~I.,  2016, \mn@doi [\apj] {10.3847/0004-637X/826/2/226}, \href
  {https://ui.adsabs.harvard.edu/abs/2016ApJ...826..226K} {826, 226}

\bibitem[\protect\citeauthoryear{{Katz}}{{Katz}}{2020}]{Katz2020}
{Katz} J.~I.,  2020, \mn@doi [\mnras] {10.1093/mnrasl/slaa038}, \href
  {https://ui.adsabs.harvard.edu/abs/2020MNRAS.494L..64K} {494, L64}

\bibitem[\protect\citeauthoryear{{Katz}}{{Katz}}{2021}]{Katz2021}
{Katz} J.~I.,  2021, \mn@doi [\mnras] {10.1093/mnrasl/slaa202}, \href
  {https://ui.adsabs.harvard.edu/abs/2021MNRAS.501L..76K} {501, L76}

\bibitem[\protect\citeauthoryear{{Katz}}{{Katz}}{2022}]{Katz2022}
{Katz} J.~I.,  2022, \mn@doi [\mnras] {10.1093/mnrasl/slac051}, \href
  {https://ui.adsabs.harvard.edu/abs/2022MNRAS.514L..27K} {514, L27}

\bibitem[\protect\citeauthoryear{{King}, {Pringle}  \& {Wickramasinghe}}{{King}
  et~al.}{2001}]{King2001}
{King} A.~R.,  {Pringle} J.~E.,   {Wickramasinghe} D.~T.,  2001, \mn@doi
  [\mnras] {10.1046/j.1365-8711.2001.04184.x}, \href
  {https://ui.adsabs.harvard.edu/abs/2001MNRAS.320L..45K} {320, L45}

\bibitem[\protect\citeauthoryear{{Kitaki}, {Mineshige}, {Ohsuga}  \&
  {Kawashima}}{{Kitaki} et~al.}{2021}]{Kitaki2021}
{Kitaki} T.,  {Mineshige} S.,  {Ohsuga} K.,   {Kawashima} T.,  2021, \mn@doi
  [\pasj] {10.1093/pasj/psab011}, \href
  {https://ui.adsabs.harvard.edu/abs/2021PASJ...73..450K} {73, 450}

\bibitem[\protect\citeauthoryear{{Kocsis}, {Yunes}  \& {Loeb}}{{Kocsis}
  et~al.}{2011}]{Kocsis2011}
{Kocsis} B.,  {Yunes} N.,   {Loeb} A.,  2011, \mn@doi [\prd]
  {10.1103/PhysRevD.84.024032}, \href
  {https://ui.adsabs.harvard.edu/abs/2011PhRvD..84b4032K} {84, 024032}

\bibitem[\protect\citeauthoryear{{Kumar} \& {Bo{\v{s}}njak}}{{Kumar} \&
  {Bo{\v{s}}njak}}{2020}]{Kumar2020}
{Kumar} P.,  {Bo{\v{s}}njak} {\v{Z}}.,  2020, \mn@doi [\mnras]
  {10.1093/mnras/staa774}, \href
  {https://ui.adsabs.harvard.edu/abs/2020MNRAS.494.2385K} {494, 2385}

\bibitem[\protect\citeauthoryear{{Kumar}, {Shannon}, {Lower}, {Bhandari},
  {Deller}, {Flynn}  \& {Keane}}{{Kumar} et~al.}{2022}]{Kumar2022}
{Kumar} P.,  {Shannon} R.~M.,  {Lower} M.~E.,  {Bhandari} S.,  {Deller} A.~T.,
  {Flynn} C.,   {Keane} E.~F.,  2022, \mn@doi [\mnras] {10.1093/mnras/stac683},
  \href {https://ui.adsabs.harvard.edu/abs/2022MNRAS.512.3400K} {512, 3400}

\bibitem[\protect\citeauthoryear{{Kumar} et~al.,}{{Kumar}
  et~al.}{2023}]{Kumar2023}
{Kumar} P.,  et~al., 2023, \mn@doi [\mnras] {10.1093/mnras/stad2969}, \href
  {https://ui.adsabs.harvard.edu/abs/2023MNRAS.526.3652K} {526, 3652}

\bibitem[\protect\citeauthoryear{{Li} et~al.,}{{Li} et~al.}{2021a}]{Li2021}
{Li} D.,  et~al., 2021a, \mn@doi [\nat] {10.1038/s41586-021-03878-5}, \href
  {https://ui.adsabs.harvard.edu/abs/2021Natur.598..267L} {598, 267}

\bibitem[\protect\citeauthoryear{{Li}, {Yang}, {Wang}, {Xu}, {Shao}, {Liu}  \&
  {Dai}}{{Li} et~al.}{2021b}]{LiQC2021}
{Li} Q.-C.,  {Yang} Y.-P.,  {Wang} F.~Y.,  {Xu} K.,  {Shao} Y.,  {Liu} Z.-N.,
  {Dai} Z.-G.,  2021b, \mn@doi [\apjl] {10.3847/2041-8213/ac1922}, \href
  {https://ui.adsabs.harvard.edu/abs/2021ApJ...918L...5L} {918, L5}

\bibitem[\protect\citeauthoryear{{Li}, {Chen}, {Lin}  \& {Wang}}{{Li}
  et~al.}{2022}]{Li2022}
{Li} Y.-P.,  {Chen} Y.-X.,  {Lin} D. N.~C.,   {Wang} Z.,  2022, \mn@doi [\apjl]
  {10.3847/2041-8213/ac5b61}, \href
  {https://ui.adsabs.harvard.edu/abs/2022ApJ...928L...1L} {928, L1}

\bibitem[\protect\citeauthoryear{{Li}, {Bilous}, {Ransom}, {Main}  \&
  {Yang}}{{Li} et~al.}{2023a}]{LDZ2023}
{Li} D.,  {Bilous} A.,  {Ransom} S.,  {Main} R.,   {Yang} Y.-P.,  2023a,
  \mn@doi [\nat] {10.1038/s41586-023-05983-z}, \href
  {https://ui.adsabs.harvard.edu/abs/2023Natur.618..484L} {618, 484}

\bibitem[\protect\citeauthoryear{{Li}, {Liu}, {Fan}, {Hu}, {Yang}, {Geng}  \&
  {Wu}}{{Li} et~al.}{2023b}]{Li2023}
{Li} F.-L.,  {Liu} Y.,  {Fan} X.,  {Hu} M.-K.,  {Yang} X.,  {Geng} J.-J.,
  {Wu} X.-F.,  2023b, \mn@doi [\apj] {10.3847/1538-4357/acd2d1}, \href
  {https://ui.adsabs.harvard.edu/abs/2023ApJ...950..161L} {950, 161}

\bibitem[\protect\citeauthoryear{{Lorimer}, {Bailes}, {McLaughlin}, {Narkevic}
  \& {Crawford}}{{Lorimer} et~al.}{2007}]{Lorimer2007}
{Lorimer} D.~R.,  {Bailes} M.,  {McLaughlin} M.~A.,  {Narkevic} D.~J.,
  {Crawford} F.,  2007, \mn@doi [Science] {10.1126/science.1147532}, \href
  {https://ui.adsabs.harvard.edu/abs/2007Sci...318..777L} {318, 777}

\bibitem[\protect\citeauthoryear{{Lu}, {Do}, {Ghez}, {Morris}, {Yelda}  \&
  {Matthews}}{{Lu} et~al.}{2013}]{Lu2013}
{Lu} J.~R.,  {Do} T.,  {Ghez} A.~M.,  {Morris} M.~R.,  {Yelda} S.,   {Matthews}
  K.,  2013, \mn@doi [\apj] {10.1088/0004-637X/764/2/155}, \href
  {https://ui.adsabs.harvard.edu/abs/2013ApJ...764..155L} {764, 155}

\bibitem[\protect\citeauthoryear{{Lu}, {Kumar}  \& {Zhang}}{{Lu}
  et~al.}{2020}]{Lu2020}
{Lu} W.,  {Kumar} P.,   {Zhang} B.,  2020, \mn@doi [\mnras]
  {10.1093/mnras/staa2450}, \href
  {https://ui.adsabs.harvard.edu/abs/2020MNRAS.498.1397L} {498, 1397}

\bibitem[\protect\citeauthoryear{{Lu}, {Zhao}, {Wang}  \& {Dai}}{{Lu}
  et~al.}{2023}]{Lu2023}
{Lu} W.-J.,  {Zhao} Z.-Y.,  {Wang} F.~Y.,   {Dai} Z.~G.,  2023, \mn@doi [\apjl]
  {10.3847/2041-8213/acf8cb}, \href
  {https://ui.adsabs.harvard.edu/abs/2023ApJ...956L...9L} {956, L9}

\bibitem[\protect\citeauthoryear{{Luo}, {Men}, {Lee}, {Wang}, {Lorimer}  \&
  {Zhang}}{{Luo} et~al.}{2020a}]{LR20}
{Luo} R.,  {Men} Y.,  {Lee} K.,  {Wang} W.,  {Lorimer} D.~R.,   {Zhang} B.,
  2020a, \mn@doi [\mnras] {10.1093/mnras/staa704}, \href
  {https://ui.adsabs.harvard.edu/abs/2020MNRAS.494..665L} {494, 665}

\bibitem[\protect\citeauthoryear{{Luo} et~al.,}{{Luo} et~al.}{2020b}]{Luo2020}
{Luo} R.,  et~al., 2020b, \mn@doi [\nat] {10.1038/s41586-020-2827-2}, \href
  {https://ui.adsabs.harvard.edu/abs/2020Natur.586..693L} {586, 693}

\bibitem[\protect\citeauthoryear{{Luo}, {Wu}, {Zhang}, {Wang}, {Ho}  \&
  {Yuan}}{{Luo} et~al.}{2023}]{Luo2023}
{Luo} Y.,  {Wu} X.-J.,  {Zhang} S.-R.,  {Wang} J.-M.,  {Ho} L.~C.,   {Yuan}
  Y.-F.,  2023, \mn@doi [\mnras] {10.1093/mnras/stad2188}, \href
  {https://ui.adsabs.harvard.edu/abs/2023MNRAS.524.6015L} {524, 6015}

\bibitem[\protect\citeauthoryear{{Lyubarsky}}{{Lyubarsky}}{2014}]{Lyubarsky2014}
{Lyubarsky} Y.,  2014, \mn@doi [\mnras] {10.1093/mnrasl/slu046}, \href
  {https://ui.adsabs.harvard.edu/abs/2014MNRAS.442L...9L} {442, L9}

\bibitem[\protect\citeauthoryear{{Margalit} \& {Metzger}}{{Margalit} \&
  {Metzger}}{2018}]{Margalit2018}
{Margalit} B.,  {Metzger} B.~D.,  2018, \mn@doi [\apjl]
  {10.3847/2041-8213/aaedad}, \href
  {https://ui.adsabs.harvard.edu/abs/2018ApJ...868L...4M} {868, L4}

\bibitem[\protect\citeauthoryear{{Matzner} \& {McKee}}{{Matzner} \&
  {McKee}}{1999}]{Matzner1999}
{Matzner} C.~D.,  {McKee} C.~F.,  1999, \mn@doi [\apj] {10.1086/306571}, \href
  {https://ui.adsabs.harvard.edu/abs/1999ApJ...510..379M} {510, 379}

\bibitem[\protect\citeauthoryear{{McKernan}, {Ford}  \&
  {O'Shaughnessy}}{{McKernan} et~al.}{2020}]{Mckernan2020}
{McKernan} B.,  {Ford} K.~E.~S.,   {O'Shaughnessy} R.,  2020, \mn@doi [\mnras]
  {10.1093/mnras/staa2681}, \href
  {https://ui.adsabs.harvard.edu/abs/2020MNRAS.498.4088M} {498, 4088}

\bibitem[\protect\citeauthoryear{{Mckinven} et~al.,}{{Mckinven}
  et~al.}{2023}]{Mckinven2023}
{Mckinven} R.,  et~al., 2023, \mn@doi [\apj] {10.3847/1538-4357/acc65f}, \href
  {https://ui.adsabs.harvard.edu/abs/2023ApJ...950...12M} {950, 12}

\bibitem[\protect\citeauthoryear{{Melrose}}{{Melrose}}{2010}]{Melrose2010}
{Melrose} D.~B.,  2010, \mn@doi [\apj] {10.1088/0004-637X/725/2/1600}, \href
  {https://ui.adsabs.harvard.edu/abs/2010ApJ...725.1600M} {725, 1600}

\bibitem[\protect\citeauthoryear{{Melrose} \& {McPhedran}}{{Melrose} \&
  {McPhedran}}{1991}]{Melrose1991}
{Melrose} D.~B.,  {McPhedran} R.~C.,  1991, {Electromagnetic Processes in
  Dispersive Media}

\bibitem[\protect\citeauthoryear{{Melrose} \& {Robinson}}{{Melrose} \&
  {Robinson}}{1994}]{Melrose1994}
{Melrose} D.~B.,  {Robinson} P.~A.,  1994, \mn@doi [\pasa]
  {10.1017/S1323358000019597}, \href
  {https://ui.adsabs.harvard.edu/abs/1994PASA...11...16M} {11, 16}

\bibitem[\protect\citeauthoryear{{Metzger}, {Berger}  \& {Margalit}}{{Metzger}
  et~al.}{2017}]{Metzger2017}
{Metzger} B.~D.,  {Berger} E.,   {Margalit} B.,  2017, \mn@doi [\apj]
  {10.3847/1538-4357/aa633d}, \href
  {https://ui.adsabs.harvard.edu/abs/2017ApJ...841...14M} {841, 14}

\bibitem[\protect\citeauthoryear{{Metzger}, {Margalit}  \& {Sironi}}{{Metzger}
  et~al.}{2019}]{Metzger2019}
{Metzger} B.~D.,  {Margalit} B.,   {Sironi} L.,  2019, \mn@doi [\mnras]
  {10.1093/mnras/stz700}, \href
  {https://ui.adsabs.harvard.edu/abs/2019MNRAS.485.4091M} {485, 4091}

\bibitem[\protect\citeauthoryear{{Michilli} et~al.,}{{Michilli}
  et~al.}{2018}]{Michilli2018}
{Michilli} D.,  et~al., 2018, \mn@doi [\nat] {10.1038/nature25149}, \href
  {https://ui.adsabs.harvard.edu/abs/2018Natur.553..182M} {553, 182}

\bibitem[\protect\citeauthoryear{{Moranchel-Basurto}, {S{\'a}nchez-Salcedo},
  {Chametla}  \& {Vel{\'a}zquez}}{{Moranchel-Basurto}
  et~al.}{2021}]{Moranchel-Basurto2021}
{Moranchel-Basurto} A.,  {S{\'a}nchez-Salcedo} F.~J.,  {Chametla} R.~O.,
  {Vel{\'a}zquez} P.~F.,  2021, \mn@doi [\apj] {10.3847/1538-4357/abca88},
  \href {https://ui.adsabs.harvard.edu/abs/2021ApJ...906...15M} {906, 15}

\bibitem[\protect\citeauthoryear{{Muno}, {Gaensler}, {Nechita}, {Miller}  \&
  {Slane}}{{Muno} et~al.}{2008}]{Muno2008}
{Muno} M.~P.,  {Gaensler} B.~M.,  {Nechita} A.,  {Miller} J.~M.,   {Slane}
  P.~O.,  2008, \mn@doi [\apj] {10.1086/527316}, \href
  {https://ui.adsabs.harvard.edu/abs/2008ApJ...680..639M} {680, 639}

\bibitem[\protect\citeauthoryear{{Murase}, {Kashiyama}  \&
  {M{\'e}sz{\'a}ros}}{{Murase} et~al.}{2016}]{Murase2016}
{Murase} K.,  {Kashiyama} K.,   {M{\'e}sz{\'a}ros} P.,  2016, \mn@doi [\mnras]
  {10.1093/mnras/stw1328}, \href
  {https://ui.adsabs.harvard.edu/abs/2016MNRAS.461.1498M} {461, 1498}

\bibitem[\protect\citeauthoryear{{Niu} et~al.,}{{Niu} et~al.}{2022}]{Niu2022}
{Niu} C.~H.,  et~al., 2022, \mn@doi [\nat] {10.1038/s41586-022-04755-5}, \href
  {https://ui.adsabs.harvard.edu/abs/2022Natur.606..873N} {606, 873}

\bibitem[\protect\citeauthoryear{{Nomoto} \& {Kondo}}{{Nomoto} \&
  {Kondo}}{1991}]{Nomoto1991}
{Nomoto} K.,  {Kondo} Y.,  1991, \mn@doi [\apjl] {10.1086/185922}, \href
  {https://ui.adsabs.harvard.edu/abs/1991ApJ...367L..19N} {367, L19}

\bibitem[\protect\citeauthoryear{{Ostriker} \& {McKee}}{{Ostriker} \&
  {McKee}}{1988}]{Ostriker1988}
{Ostriker} J.~P.,  {McKee} C.~F.,  1988, \mn@doi [Reviews of Modern Physics]
  {10.1103/RevModPhys.60.1}, \href
  {https://ui.adsabs.harvard.edu/abs/1988RvMP...60....1O} {60, 1}

\bibitem[\protect\citeauthoryear{{Pan} \& {Yang}}{{Pan} \&
  {Yang}}{2021a}]{Pan2021}
{Pan} Z.,  {Yang} H.,  2021a, \mn@doi [\prd] {10.1103/PhysRevD.103.103018},
  \href {https://ui.adsabs.harvard.edu/abs/2021PhRvD.103j3018P} {103, 103018}

\bibitem[\protect\citeauthoryear{{Pan} \& {Yang}}{{Pan} \&
  {Yang}}{2021b}]{Pan2021b}
{Pan} Z.,  {Yang} H.,  2021b, \mn@doi [\apj] {10.3847/1538-4357/ac249c}, \href
  {https://ui.adsabs.harvard.edu/abs/2021ApJ...923..173P} {923, 173}

\bibitem[\protect\citeauthoryear{{Parfrey}, {Spitkovsky}  \&
  {Beloborodov}}{{Parfrey} et~al.}{2016}]{Parfrey2016}
{Parfrey} K.,  {Spitkovsky} A.,   {Beloborodov} A.~M.,  2016, \mn@doi [\apj]
  {10.3847/0004-637X/822/1/33}, \href
  {https://ui.adsabs.harvard.edu/abs/2016ApJ...822...33P} {822, 33}

\bibitem[\protect\citeauthoryear{{Pastor-Marazuela} et~al.,}{{Pastor-Marazuela}
  et~al.}{2021}]{Pastor-Marazuela2021}
{Pastor-Marazuela} I.,  et~al., 2021, \mn@doi [\nat]
  {10.1038/s41586-021-03724-8}, \href
  {https://ui.adsabs.harvard.edu/abs/2021Natur.596..505P} {596, 505}

\bibitem[\protect\citeauthoryear{{Perna}, {Tagawa}, {Haiman}  \&
  {Bartos}}{{Perna} et~al.}{2021}]{Perna2021}
{Perna} R.,  {Tagawa} H.,  {Haiman} Z.,   {Bartos} I.,  2021, \mn@doi [\apj]
  {10.3847/1538-4357/abfdb4}, \href
  {https://ui.adsabs.harvard.edu/abs/2021ApJ...915...10P} {915, 10}

\bibitem[\protect\citeauthoryear{{Petroff}, {Hessels}  \& {Lorimer}}{{Petroff}
  et~al.}{2022}]{Petroff2022}
{Petroff} E.,  {Hessels} J.~W.~T.,   {Lorimer} D.~R.,  2022, \mn@doi [\aapr]
  {10.1007/s00159-022-00139-w}, \href
  {https://ui.adsabs.harvard.edu/abs/2022A&ARv..30....2P} {30, 2}

\bibitem[\protect\citeauthoryear{{Piro} \& {Gaensler}}{{Piro} \&
  {Gaensler}}{2018}]{Piro2018}
{Piro} A.~L.,  {Gaensler} B.~M.,  2018, \mn@doi [\apj]
  {10.3847/1538-4357/aac9bc}, \href
  {https://ui.adsabs.harvard.edu/abs/2018ApJ...861..150P} {861, 150}

\bibitem[\protect\citeauthoryear{{Pleunis} et~al.,}{{Pleunis}
  et~al.}{2021}]{Pleunis2021}
{Pleunis} Z.,  et~al., 2021, \mn@doi [\apjl] {10.3847/2041-8213/abec72}, \href
  {https://ui.adsabs.harvard.edu/abs/2021ApJ...911L...3P} {911, L3}

\bibitem[\protect\citeauthoryear{{Plotnikov} \& {Sironi}}{{Plotnikov} \&
  {Sironi}}{2019}]{Plotnikov2019}
{Plotnikov} I.,  {Sironi} L.,  2019, \mn@doi [\mnras] {10.1093/mnras/stz640},
  \href {https://ui.adsabs.harvard.edu/abs/2019MNRAS.485.3816P} {485, 3816}

\bibitem[\protect\citeauthoryear{{Price} \& {Rosswog}}{{Price} \&
  {Rosswog}}{2006}]{Price2006}
{Price} D.~J.,  {Rosswog} S.,  2006, \mn@doi [Science]
  {10.1126/science.1125201}, \href
  {https://ui.adsabs.harvard.edu/abs/2006Sci...312..719P} {312, 719}

\bibitem[\protect\citeauthoryear{{Qu} \& {Zhang}}{{Qu} \&
  {Zhang}}{2023}]{Qu2023}
{Qu} Y.,  {Zhang} B.,  2023, \mn@doi [\mnras] {10.1093/mnras/stad1072}, \href
  {https://ui.adsabs.harvard.edu/abs/2023MNRAS.522.2448Q} {522, 2448}

\bibitem[\protect\citeauthoryear{{Radice}, {Perego}, {Hotokezaka}, {Fromm},
  {Bernuzzi}  \& {Roberts}}{{Radice} et~al.}{2018}]{Radice2018}
{Radice} D.,  {Perego} A.,  {Hotokezaka} K.,  {Fromm} S.~A.,  {Bernuzzi} S.,
  {Roberts} L.~F.,  2018, \mn@doi [\apj] {10.3847/1538-4357/aaf054}, \href
  {https://ui.adsabs.harvard.edu/abs/2018ApJ...869..130R} {869, 130}

\bibitem[\protect\citeauthoryear{{Rajwade} et~al.,}{{Rajwade}
  et~al.}{2020}]{Rajwade2020}
{Rajwade} K.~M.,  et~al., 2020, \mn@doi [\mnras] {10.1093/mnras/staa1237},
  \href {https://ui.adsabs.harvard.edu/abs/2020MNRAS.495.3551R} {495, 3551}

\bibitem[\protect\citeauthoryear{{Raynaud}, {Guilet}, {Janka}  \&
  {Gastine}}{{Raynaud} et~al.}{2020}]{Raynaud2020}
{Raynaud} R.,  {Guilet} J.,  {Janka} H.-T.,   {Gastine} T.,  2020, \mn@doi
  [Science Advances] {10.1126/sciadv.aay2732}, \href
  {https://ui.adsabs.harvard.edu/abs/2020SciA....6.2732R} {6, eaay2732}

\bibitem[\protect\citeauthoryear{{Ren}, {Chen}, {Wang}  \& {Dai}}{{Ren}
  et~al.}{2022}]{Ren2022}
{Ren} J.,  {Chen} K.,  {Wang} Y.,   {Dai} Z.-G.,  2022, \mn@doi [\apjl]
  {10.3847/2041-8213/aca025}, \href
  {https://ui.adsabs.harvard.edu/abs/2022ApJ...940L..44R} {940, L44}

\bibitem[\protect\citeauthoryear{{Rosswog}, {Ramirez-Ruiz}  \&
  {Davies}}{{Rosswog} et~al.}{2003}]{Rosswog2003}
{Rosswog} S.,  {Ramirez-Ruiz} E.,   {Davies} M.~B.,  2003, \mn@doi [\mnras]
  {10.1046/j.1365-2966.2003.07032.x}, \href
  {https://ui.adsabs.harvard.edu/abs/2003MNRAS.345.1077R} {345, 1077}

\bibitem[\protect\citeauthoryear{{Rybicki} \& {Lightman}}{{Rybicki} \&
  {Lightman}}{1986}]{Rybicki1986}
{Rybicki} G.~B.,  {Lightman} A.~P.,  1986, {Radiative Processes in
  Astrophysics}

\bibitem[\protect\citeauthoryear{Sakurai}{Sakurai}{1960}]{Sakurai1960}
Sakurai A.,  1960, \mn@doi [Communications on Pure and Applied Mathematics]
  {https://doi.org/10.1002/cpa.3160130303}, 13, 353

\bibitem[\protect\citeauthoryear{{Salvesen}, {Simon}, {Armitage}  \&
  {Begelman}}{{Salvesen} et~al.}{2016}]{Salvesen2016}
{Salvesen} G.,  {Simon} J.~B.,  {Armitage} P.~J.,   {Begelman} M.~C.,  2016,
  \mn@doi [\mnras] {10.1093/mnras/stw029}, \href
  {https://ui.adsabs.harvard.edu/abs/2016MNRAS.457..857S} {457, 857}

\bibitem[\protect\citeauthoryear{{Sari} \& {Piran}}{{Sari} \&
  {Piran}}{1995}]{Sari1995}
{Sari} R.,  {Piran} T.,  1995, \mn@doi [\apjl] {10.1086/309835}, \href
  {https://ui.adsabs.harvard.edu/abs/1995ApJ...455L.143S} {455, L143}

\bibitem[\protect\citeauthoryear{{Sazonov}}{{Sazonov}}{1969}]{Sazonov1969}
{Sazonov} V.~N.,  1969, \sovast, \href
  {https://ui.adsabs.harvard.edu/abs/1969SvA....13..396S} {13, 396}

\bibitem[\protect\citeauthoryear{{Schwab}, {Quataert}  \& {Bildsten}}{{Schwab}
  et~al.}{2015}]{Schwab2015}
{Schwab} J.,  {Quataert} E.,   {Bildsten} L.,  2015, \mn@doi [\mnras]
  {10.1093/mnras/stv1804}, \href
  {https://ui.adsabs.harvard.edu/abs/2015MNRAS.453.1910S} {453, 1910}

\bibitem[\protect\citeauthoryear{{Schwab}, {Quataert}  \& {Kasen}}{{Schwab}
  et~al.}{2016}]{Schwab2016}
{Schwab} J.,  {Quataert} E.,   {Kasen} D.,  2016, \mn@doi [\mnras]
  {10.1093/mnras/stw2249}, \href
  {https://ui.adsabs.harvard.edu/abs/2016MNRAS.463.3461S} {463, 3461}

\bibitem[\protect\citeauthoryear{{Sedov}}{{Sedov}}{1959}]{Sedov1959}
{Sedov} L.~I.,  1959, {Similarity and Dimensional Methods in Mechanics}

\bibitem[\protect\citeauthoryear{{Sirko} \& {Goodman}}{{Sirko} \&
  {Goodman}}{2003}]{SG03}
{Sirko} E.,  {Goodman} J.,  2003, \mn@doi [\mnras]
  {10.1046/j.1365-8711.2003.06431.x}, \href
  {https://ui.adsabs.harvard.edu/abs/2003MNRAS.341..501S} {341, 501}

\bibitem[\protect\citeauthoryear{{Sridhar} \& {Metzger}}{{Sridhar} \&
  {Metzger}}{2022}]{Sridhar2022}
{Sridhar} N.,  {Metzger} B.~D.,  2022, \mn@doi [\apj]
  {10.3847/1538-4357/ac8a4a}, \href
  {https://ui.adsabs.harvard.edu/abs/2022ApJ...937....5S} {937, 5}

\bibitem[\protect\citeauthoryear{{Sridhar}, {Metzger}, {Beniamini}, {Margalit},
  {Renzo}, {Sironi}  \& {Kovlakas}}{{Sridhar} et~al.}{2021}]{Sridhar2021}
{Sridhar} N.,  {Metzger} B.~D.,  {Beniamini} P.,  {Margalit} B.,  {Renzo} M.,
  {Sironi} L.,   {Kovlakas} K.,  2021, \mn@doi [\apj]
  {10.3847/1538-4357/ac0140}, \href
  {https://ui.adsabs.harvard.edu/abs/2021ApJ...917...13S} {917, 13}

\bibitem[\protect\citeauthoryear{{Stone}, {Metzger}  \& {Haiman}}{{Stone}
  et~al.}{2017}]{Stone2017}
{Stone} N.~C.,  {Metzger} B.~D.,   {Haiman} Z.,  2017, \mn@doi [\mnras]
  {10.1093/mnras/stw2260}, \href
  {https://ui.adsabs.harvard.edu/abs/2017MNRAS.464..946S} {464, 946}

\bibitem[\protect\citeauthoryear{{Tagawa}, {Haiman}  \& {Kocsis}}{{Tagawa}
  et~al.}{2020}]{Tagawa2020}
{Tagawa} H.,  {Haiman} Z.,   {Kocsis} B.,  2020, \mn@doi [\apj]
  {10.3847/1538-4357/ab9b8c}, \href
  {https://ui.adsabs.harvard.edu/abs/2020ApJ...898...25T} {898, 25}

\bibitem[\protect\citeauthoryear{{Tagawa}, {Kimura}, {Haiman}, {Perna},
  {Tanaka}  \& {Bartos}}{{Tagawa} et~al.}{2022}]{Tagawa2022}
{Tagawa} H.,  {Kimura} S.~S.,  {Haiman} Z.,  {Perna} R.,  {Tanaka} H.,
  {Bartos} I.,  2022, \mn@doi [\apj] {10.3847/1538-4357/ac45f8}, \href
  {https://ui.adsabs.harvard.edu/abs/2022ApJ...927...41T} {927, 41}

\bibitem[\protect\citeauthoryear{{Tagawa}, {Kimura}, {Haiman}, {Perna}  \&
  {Bartos}}{{Tagawa} et~al.}{2023}]{Tagawa23}
{Tagawa} H.,  {Kimura} S.~S.,  {Haiman} Z.,  {Perna} R.,   {Bartos} I.,  2023,
  \mn@doi [\apjl] {10.3847/2041-8213/acc103}, \href
  {https://ui.adsabs.harvard.edu/abs/2023ApJ...946L...3T} {946, L3}

\bibitem[\protect\citeauthoryear{{Takahashi}, {Mineshige}  \&
  {Ohsuga}}{{Takahashi} et~al.}{2018}]{Takahashi2018}
{Takahashi} H.~R.,  {Mineshige} S.,   {Ohsuga} K.,  2018, \mn@doi [\apj]
  {10.3847/1538-4357/aaa082}, \href
  {https://ui.adsabs.harvard.edu/abs/2018ApJ...853...45T} {853, 45}

\bibitem[\protect\citeauthoryear{{Tanigawa} \& {Tanaka}}{{Tanigawa} \&
  {Tanaka}}{2016}]{Tanigawa2016}
{Tanigawa} T.,  {Tanaka} H.,  2016, \mn@doi [\apj]
  {10.3847/0004-637X/823/1/48}, \href
  {https://ui.adsabs.harvard.edu/abs/2016ApJ...823...48T} {823, 48}

\bibitem[\protect\citeauthoryear{{Tanigawa}, {Ohtsuki}  \&
  {Machida}}{{Tanigawa} et~al.}{2012}]{Tanigawa2012}
{Tanigawa} T.,  {Ohtsuki} K.,   {Machida} M.~N.,  2012, \mn@doi [\apj]
  {10.1088/0004-637X/747/1/47}, \href
  {https://ui.adsabs.harvard.edu/abs/2012ApJ...747...47T} {747, 47}

\bibitem[\protect\citeauthoryear{{Tauris}, {Sanyal}, {Yoon}  \&
  {Langer}}{{Tauris} et~al.}{2013}]{Tauris2013}
{Tauris} T.~M.,  {Sanyal} D.,  {Yoon} S.~C.,   {Langer} N.,  2013, \mn@doi
  [\aap] {10.1051/0004-6361/201321662}, \href
  {https://ui.adsabs.harvard.edu/abs/2013A&A...558A..39T} {558, A39}

\bibitem[\protect\citeauthoryear{{Taylor}}{{Taylor}}{1946}]{Taylor1946}
{Taylor} G.~I.,  1946, \mn@doi [Proceedings of the Royal Society of London
  Series A] {10.1098/rspa.1946.0044}, \href
  {https://ui.adsabs.harvard.edu/abs/1946RSPSA.186..273T} {186, 273}

\bibitem[\protect\citeauthoryear{{Tchekhovskoy}, {Narayan}  \&
  {McKinney}}{{Tchekhovskoy} et~al.}{2011}]{Tchekhovskoy2011}
{Tchekhovskoy} A.,  {Narayan} R.,   {McKinney} J.~C.,  2011, \mn@doi [\mnras]
  {10.1111/j.1745-3933.2011.01147.x}, \href
  {https://ui.adsabs.harvard.edu/abs/2011MNRAS.418L..79T} {418, L79}

\bibitem[\protect\citeauthoryear{{Thompson}, {Quataert}  \&
  {Murray}}{{Thompson} et~al.}{2005}]{Thompson2005}
{Thompson} T.~A.,  {Quataert} E.,   {Murray} N.,  2005, \mn@doi [\apj]
  {10.1086/431923}, \href
  {https://ui.adsabs.harvard.edu/abs/2005ApJ...630..167T} {630, 167}

\bibitem[\protect\citeauthoryear{{Toomre}}{{Toomre}}{1964}]{Toomre1964}
{Toomre} A.,  1964, \mn@doi [\apj] {10.1086/147861}, \href
  {https://ui.adsabs.harvard.edu/abs/1964ApJ...139.1217T} {139, 1217}

\bibitem[\protect\citeauthoryear{{Wang} \& {Yu}}{{Wang} \&
  {Yu}}{2017}]{Wang2017}
{Wang} F.~Y.,  {Yu} H.,  2017, \mn@doi [\jcap] {10.1088/1475-7516/2017/03/023},
  \href {https://ui.adsabs.harvard.edu/abs/2017JCAP...03..023W} {2017, 023}

\bibitem[\protect\citeauthoryear{{Wang}, {Wang}, {Yang}, {Yu}, {Zuo}  \&
  {Dai}}{{Wang} et~al.}{2020}]{Wang2020}
{Wang} F.~Y.,  {Wang} Y.~Y.,  {Yang} Y.-P.,  {Yu} Y.~W.,  {Zuo} Z.~Y.,   {Dai}
  Z.~G.,  2020, \mn@doi [\apj] {10.3847/1538-4357/ab74d0}, \href
  {https://ui.adsabs.harvard.edu/abs/2020ApJ...891...72W} {891, 72}

\bibitem[\protect\citeauthoryear{{Wang}, {Liu}, {Ho}  \& {Du}}{{Wang}
  et~al.}{2021a}]{Wang2021a}
{Wang} J.-M.,  {Liu} J.-R.,  {Ho} L.~C.,   {Du} P.,  2021a, \mn@doi [\apjl]
  {10.3847/2041-8213/abee81}, \href
  {https://ui.adsabs.harvard.edu/abs/2021ApJ...911L..14W} {911, L14}

\bibitem[\protect\citeauthoryear{{Wang}, {Liu}, {Ho}, {Li}  \& {Du}}{{Wang}
  et~al.}{2021b}]{Wang2021b}
{Wang} J.-M.,  {Liu} J.-R.,  {Ho} L.~C.,  {Li} Y.-R.,   {Du} P.,  2021b,
  \mn@doi [\apjl] {10.3847/2041-8213/ac0b46}, \href
  {https://ui.adsabs.harvard.edu/abs/2021ApJ...916L..17W} {916, L17}

\bibitem[\protect\citeauthoryear{{Wang}, {Zhang}, {Dai}  \& {Cheng}}{{Wang}
  et~al.}{2022}]{Wang2022}
{Wang} F.~Y.,  {Zhang} G.~Q.,  {Dai} Z.~G.,   {Cheng} K.~S.,  2022, \mn@doi
  [Nature Communications] {10.1038/s41467-022-31923-y}, \href
  {https://ui.adsabs.harvard.edu/abs/2022NatCo..13.4382W} {13, 4382}

\bibitem[\protect\citeauthoryear{{Ward}}{{Ward}}{1997}]{Ward1997}
{Ward} W.~R.,  1997, \mn@doi [\icarus] {10.1006/icar.1996.5647}, \href
  {https://ui.adsabs.harvard.edu/abs/1997Icar..126..261W} {126, 261}

\bibitem[\protect\citeauthoryear{{Weaver}, {McCray}, {Castor}, {Shapiro}  \&
  {Moore}}{{Weaver} et~al.}{1977}]{Weaver1977}
{Weaver} R.,  {McCray} R.,  {Castor} J.,  {Shapiro} P.,   {Moore} R.,  1977,
  \mn@doi [\apj] {10.1086/155692}, \href
  {https://ui.adsabs.harvard.edu/abs/1977ApJ...218..377W} {218, 377}

\bibitem[\protect\citeauthoryear{{Wu}, {Zhang}, {Wang}  \& {Dai}}{{Wu}
  et~al.}{2020}]{Wu2020}
{Wu} Q.,  {Zhang} G.~Q.,  {Wang} F.~Y.,   {Dai} Z.~G.,  2020, \mn@doi [\apjl]
  {10.3847/2041-8213/abaef1}, \href
  {https://ui.adsabs.harvard.edu/abs/2020ApJ...900L..26W} {900, L26}

\bibitem[\protect\citeauthoryear{{Xia}, {Yang}, {Li}, {Wang}, {Liu}  \&
  {Dai}}{{Xia} et~al.}{2023}]{Xia2023}
{Xia} Z.-Y.,  {Yang} Y.-P.,  {Li} Q.-C.,  {Wang} F.-Y.,  {Liu} B.-Y.,   {Dai}
  Z.-G.,  2023, \mn@doi [\apj] {10.3847/1538-4357/acf5eb}, \href
  {https://ui.adsabs.harvard.edu/abs/2023ApJ...957....1X} {957, 1}

\bibitem[\protect\citeauthoryear{{Xiao}, {Wang}  \& {Dai}}{{Xiao}
  et~al.}{2021}]{Xiao2021}
{Xiao} D.,  {Wang} F.,   {Dai} Z.,  2021, \mn@doi [Science China Physics,
  Mechanics, and Astronomy] {10.1007/s11433-020-1661-7}, \href
  {https://ui.adsabs.harvard.edu/abs/2021SCPMA..6449501X} {64, 249501}

\bibitem[\protect\citeauthoryear{{Xu} et~al.,}{{Xu} et~al.}{2022}]{Xu2022}
{Xu} H.,  et~al., 2022, \mn@doi [\nat] {10.1038/s41586-022-05071-8}, \href
  {https://ui.adsabs.harvard.edu/abs/2022Natur.609..685X} {609, 685}

\bibitem[\protect\citeauthoryear{{Yang} \& {Dai}}{{Yang} \&
  {Dai}}{2019}]{YYH2019}
{Yang} Y.-H.,  {Dai} Z.-G.,  2019, \mn@doi [\apj] {10.3847/1538-4357/ab48dd},
  \href {https://ui.adsabs.harvard.edu/abs/2019ApJ...885..149Y} {885, 149}

\bibitem[\protect\citeauthoryear{{Yang} \& {Zhang}}{{Yang} \&
  {Zhang}}{2017}]{Yang2017}
{Yang} Y.-P.,  {Zhang} B.,  2017, \mn@doi [\apj] {10.3847/1538-4357/aa8721},
  \href {https://ui.adsabs.harvard.edu/abs/2017ApJ...847...22Y} {847, 22}

\bibitem[\protect\citeauthoryear{{Yang} \& {Zhang}}{{Yang} \&
  {Zhang}}{2018}]{Yang2018}
{Yang} Y.-P.,  {Zhang} B.,  2018, \mn@doi [\apj] {10.3847/1538-4357/aae685},
  \href {https://ui.adsabs.harvard.edu/abs/2018ApJ...868...31Y} {868, 31}

\bibitem[\protect\citeauthoryear{{Yang}, {Yuan}, {Ohsuga}  \& {Bu}}{{Yang}
  et~al.}{2014}]{Yang2014}
{Yang} X.-H.,  {Yuan} F.,  {Ohsuga} K.,   {Bu} D.-F.,  2014, \mn@doi [\apj]
  {10.1088/0004-637X/780/1/79}, \href
  {https://ui.adsabs.harvard.edu/abs/2014ApJ...780...79Y} {780, 79}

\bibitem[\protect\citeauthoryear{{Yang}, {Bartos}, {Haiman}, {Kocsis},
  {M{\'a}rka}, {Stone}  \& {M{\'a}rka}}{{Yang} et~al.}{2019}]{Yang2019}
{Yang} Y.,  {Bartos} I.,  {Haiman} Z.,  {Kocsis} B.,  {M{\'a}rka} Z.,  {Stone}
  N.~C.,   {M{\'a}rka} S.,  2019, \mn@doi [\apj] {10.3847/1538-4357/ab16e3},
  \href {https://ui.adsabs.harvard.edu/abs/2019ApJ...876..122Y} {876, 122}

\bibitem[\protect\citeauthoryear{{Yang}, {Lu}, {Feng}, {Zhang}  \& {Li}}{{Yang}
  et~al.}{2022}]{Yang2022}
{Yang} Y.-P.,  {Lu} W.,  {Feng} Y.,  {Zhang} B.,   {Li} D.,  2022, \mn@doi
  [\apjl] {10.3847/2041-8213/ac5f46}, \href
  {https://ui.adsabs.harvard.edu/abs/2022ApJ...928L..16Y} {928, L16}

\bibitem[\protect\citeauthoryear{{Yang}, {Xu}  \& {Zhang}}{{Yang}
  et~al.}{2023}]{Yang2023}
{Yang} Y.-P.,  {Xu} S.,   {Zhang} B.,  2023, \mn@doi [\mnras]
  {10.1093/mnras/stad168}, \href
  {https://ui.adsabs.harvard.edu/abs/2023MNRAS.520.2039Y} {520, 2039}

\bibitem[\protect\citeauthoryear{{Yoon}, {Podsiadlowski}  \& {Rosswog}}{{Yoon}
  et~al.}{2007}]{Yoon2007}
{Yoon} S.~C.,  {Podsiadlowski} P.,   {Rosswog} S.,  2007, \mn@doi [\mnras]
  {10.1111/j.1365-2966.2007.12161.x}, \href
  {https://ui.adsabs.harvard.edu/abs/2007MNRAS.380..933Y} {380, 933}

\bibitem[\protect\citeauthoryear{{Yuan}, {Murase}, {Guetta}, {Pe'er}, {Bartos}
  \& {M{\'e}sz{\'a}ros}}{{Yuan} et~al.}{2022}]{Yuan2022}
{Yuan} C.,  {Murase} K.,  {Guetta} D.,  {Pe'er} A.,  {Bartos} I.,
  {M{\'e}sz{\'a}ros} P.,  2022, \mn@doi [\apj] {10.3847/1538-4357/ac6ddf},
  \href {https://ui.adsabs.harvard.edu/abs/2022ApJ...932...80Y} {932, 80}

\bibitem[\protect\citeauthoryear{{Zenati}, {Perets}  \& {Toonen}}{{Zenati}
  et~al.}{2019}]{Zenati2019}
{Zenati} Y.,  {Perets} H.~B.,   {Toonen} S.,  2019, \mn@doi [\mnras]
  {10.1093/mnras/stz316}, \href
  {https://ui.adsabs.harvard.edu/abs/2019MNRAS.486.1805Z} {486, 1805}

\bibitem[\protect\citeauthoryear{{Zhang}}{{Zhang}}{2020}]{ZB2020}
{Zhang} B.,  2020, \mn@doi [\nat] {10.1038/s41586-020-2828-1}, \href
  {https://ui.adsabs.harvard.edu/abs/2020Natur.587...45Z} {587, 45}

\bibitem[\protect\citeauthoryear{{Zhang}}{{Zhang}}{2023}]{ZB2023}
{Zhang} B.,  2023, \mn@doi [Reviews of Modern Physics]
  {10.1103/RevModPhys.95.035005}, \href
  {https://ui.adsabs.harvard.edu/abs/2023RvMP...95c5005Z} {95, 035005}

\bibitem[\protect\citeauthoryear{{Zhang} et~al.,}{{Zhang}
  et~al.}{2023}]{Zhang2023}
{Zhang} Y.-K.,  et~al., 2023, \mn@doi [\apj] {10.3847/1538-4357/aced0b}, \href
  {https://ui.adsabs.harvard.edu/abs/2023ApJ...955..142Z} {955, 142}

\bibitem[\protect\citeauthoryear{{Zhao} \& {Wang}}{{Zhao} \&
  {Wang}}{2021}]{Zhao2021b}
{Zhao} Z.~Y.,  {Wang} F.~Y.,  2021, \mn@doi [\apjl] {10.3847/2041-8213/ac3f2f},
  \href {https://ui.adsabs.harvard.edu/abs/2021ApJ...923L..17Z} {923, L17}

\bibitem[\protect\citeauthoryear{{Zhao}, {Zhang}, {Wang}, {Tu}  \&
  {Wang}}{{Zhao} et~al.}{2021}]{Zhao2021a}
{Zhao} Z.~Y.,  {Zhang} G.~Q.,  {Wang} Y.~Y.,  {Tu} Z.-L.,   {Wang} F.~Y.,
  2021, \mn@doi [\apj] {10.3847/1538-4357/abd321}, \href
  {https://ui.adsabs.harvard.edu/abs/2021ApJ...907..111Z} {907, 111}

\bibitem[\protect\citeauthoryear{{Zhao}, {Zhang}, {Wang}  \& {Dai}}{{Zhao}
  et~al.}{2023}]{Zhao2023}
{Zhao} Z.~Y.,  {Zhang} G.~Q.,  {Wang} F.~Y.,   {Dai} Z.~G.,  2023, \mn@doi
  [\apj] {10.3847/1538-4357/aca66b}, \href
  {https://ui.adsabs.harvard.edu/abs/2023ApJ...942..102Z} {942, 102}

\bibitem[\protect\citeauthoryear{{Zhong} \& {Dai}}{{Zhong} \&
  {Dai}}{2020}]{Zhong2020}
{Zhong} S.-Q.,  {Dai} Z.-G.,  2020, \mn@doi [\apj] {10.3847/1538-4357/ab7bdf},
  \href {https://ui.adsabs.harvard.edu/abs/2020ApJ...893....9Z} {893, 9}

\bibitem[\protect\citeauthoryear{{Zhou}, {Zhu}  \& {Wang}}{{Zhou}
  et~al.}{2023}]{Zhou2023}
{Zhou} Z.-H.,  {Zhu} J.-P.,   {Wang} K.,  2023, \mn@doi [\apj]
  {10.3847/1538-4357/acd380}, \href
  {https://ui.adsabs.harvard.edu/abs/2023ApJ...951...74Z} {951, 74}

\bibitem[\protect\citeauthoryear{{Zhu}, {Zhang}, {Yu}  \& {Gao}}{{Zhu}
  et~al.}{2021a}]{Zhu2021b}
{Zhu} J.-P.,  {Zhang} B.,  {Yu} Y.-W.,   {Gao} H.,  2021a, \mn@doi [\apjl]
  {10.3847/2041-8213/abd412}, \href
  {https://ui.adsabs.harvard.edu/abs/2021ApJ...906L..11Z} {906, L11}

\bibitem[\protect\citeauthoryear{{Zhu}, {Yang}, {Zhang}, {Liu}, {Yu}  \&
  {Gao}}{{Zhu} et~al.}{2021b}]{Zhu2021a}
{Zhu} J.-P.,  {Yang} Y.-P.,  {Zhang} B.,  {Liu} L.-D.,  {Yu} Y.-W.,   {Gao} H.,
   2021b, \mn@doi [\apjl] {10.3847/2041-8213/abff5a}, \href
  {https://ui.adsabs.harvard.edu/abs/2021ApJ...914L..19Z} {914, L19}

\makeatother
\end{thebibliography}

\clearpage
%%%%%%%%%%%%%%%%% APPENDICES %%%%%%%%%%%%%%%%%%%%%
\appendix
\section{The AGN disk structure}\label{sec:disk_model}
\subsection{SG disk}
First, we discuss the disk structure for the $\alpha$-viscosity prescription. Viscosity is $\nu=\alpha c_{\mathrm{s}} H \beta^b$, where $\alpha$ is the viscosity parameter and $\beta=p_{\mathrm{gas}}/p_{\mathrm{tot}}$ is the ratio of gas pressure to total pressure. If the viscosity is proportional to the total pressure, we have $b=0$ ($\alpha$ disk). If the viscosity is proportional to the gas pressure, we have $b=1$ ($\beta$ disk). Following \cite{SG03} (hereafter \citetalias{SG03} model), the structure of the $\alpha$ viscosity prescription disk is determined by equations

\begin{subequations}\label{eq: SG model}
\begin{align}
& \sigma T_{\mathrm {eff}}^4=\frac{3}{8 \pi} \dot{M}^{\prime} \Omega^2 \label{eq:Teff} \\
& T^4=\left(\frac{3}{8} \tau_{\mathrm{d}}+\frac{1}{2}+\frac{1}{4 \tau_{\mathrm{d}}}\right) T_{\mathrm {eff }}^4 \\
& \tau_{\mathrm{d}}=\frac{\kappa \Sigma}{2} \\
& \beta^b c_{\mathrm{s}}^2 \Sigma=\frac{\dot{M}^{\prime} \Omega}{3 \pi \alpha} \\
& p_{\mathrm {rad}}=\frac{\tau_{\mathrm{d}} \sigma}{2 c} T_{\mathrm {eff}}^4 \\
& p_{\mathrm {gas}}=\frac{\rho_0 k_{\mathrm{B}} T}{m} \\
& \beta=\frac{p_{\mathrm {gas}}}{p_{\mathrm {gas }}+p_{\mathrm{rad}}}\\
&\Sigma=2 \rho_0 H\\
&H=\frac{c_{\mathrm{s}}}{\Omega}\\
&c_{\mathrm{s}}^2=\frac{p_{\mathrm{gas}}+p_{\mathrm{rad}}}{\rho_0}\\
&\kappa=\kappa(\rho_0, T), 
\end{align}
\end{subequations}
where $\Omega=\sqrt{G M / r^3}$ is the Keplerian angular velocity. $\dot{M}^{\prime}=\dot{M}\left(1-\sqrt{r_{\min } / r}\right)$ with $\dot{M}$ being the accretion rate. In this work, the inner radius of the disk$r_{\mathrm {min }}=6r_g$ and the mean molecular mass $m=0.62 m_{\mathrm{H}}$ is taken. We adopt the approximation function of the opacity $\kappa(\rho_0, T)$ given by \cite{Yang2019} (also see Fig. 1 in \citealt{Thompson2005}). For given parameters $M, \dot{M}, \alpha$, the structure for the inner disk (disk temperature $T$, effective blackbody temperature $T_{\mathrm {eff}}$, optical depth $\tau_{\mathrm{d}}$, disk surface density $\Sigma$, mid-plane disk density $\rho_0$, scale height $H$, gas pressure $p_{\mathrm {gas}}$, radiation pressure $p_{\mathrm {rad}}$, sound speed $c_{\mathrm{s}}$, gas-total pressure ratio $\beta$ and opacity $\kappa$) at different radius can be obtained by solving Equations (\ref{eq: SG model}).

However, the outer disk can be self-gravity unstable if the Toomre parameter \citep{Toomre1964}
\begin{equation}
    Q=\frac{c_s \Omega}{\pi \Sigma} \simeq \frac{\Omega^2}{2 \pi \rho_0}
\end{equation}
is smaller than unity. To maintain stability, the outer parts must be heated sufficiently by some feedback mechanism (e.g., star formation). Thus, the assumption that energy comes only from accretion becomes invalid for the outer parts (Equation (\ref{eq:Teff})). To solve the structure of outer parts, we replace Equation (\ref{eq:Teff}) with
\begin{equation}\label{eq:rho_out}
    \rho_0=\frac{\Omega^2}{2 \pi G Q_{\min }},
\end{equation}
where a minimum value is fixed at $Q_{\min } \approx 1$. The structures of SG disks with $\alpha=0.1, \beta=0, \dot{M}=0.5 \dot{M}_{\mathrm {Edd}}$ are shown in Figures \ref{fig:disk1} (for $M= 10^8 M_{\odot}$) and \ref{fig:disk2} (for $M=4 \times 10^6 M_{\odot}$).

\subsection{TQM disk}
\begin{figure*}
    \centering
    \includegraphics[width = 1\textwidth]{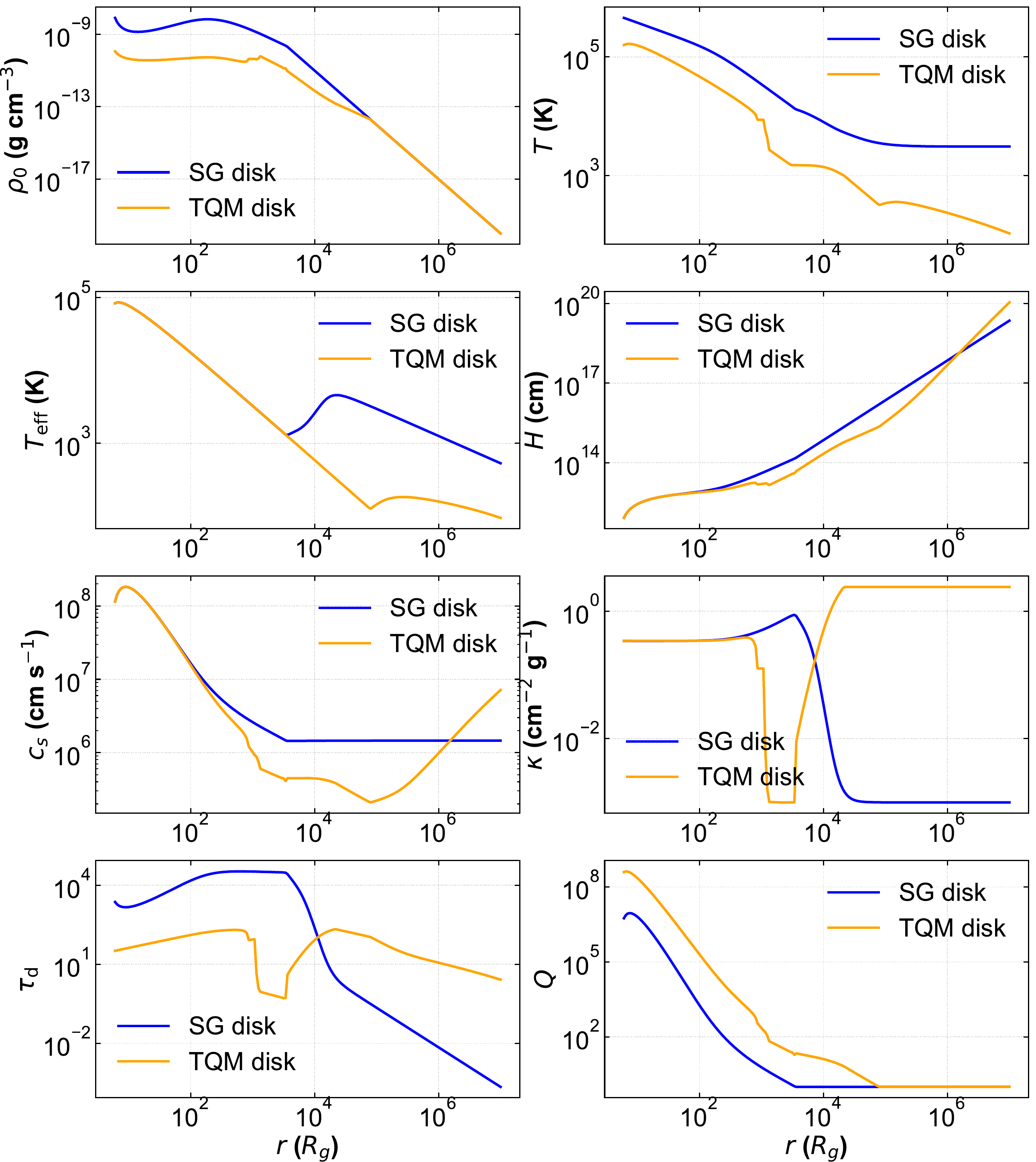}
    \caption{AGN disk structures for the SG disk (blue lines) and the TQM disk (magenta lines). The mass of the SMBH is $M=10^8 M_{\odot}$. The parameters of the SG disk are $\alpha=0.1, \beta=0, \dot{M}=0.5 \dot{M}_{\mathrm {Edd }}$. The parameters of the TQM disk are $ \dot{M}=0.1 \dot{M}_{\mathrm {Edd }}, X=0.1$, $\epsilon_{\star}=10^{-3}$.}
    \label{fig:disk1}
\end{figure*}

\begin{figure*}
    \centering
    \includegraphics[width = 1\textwidth]{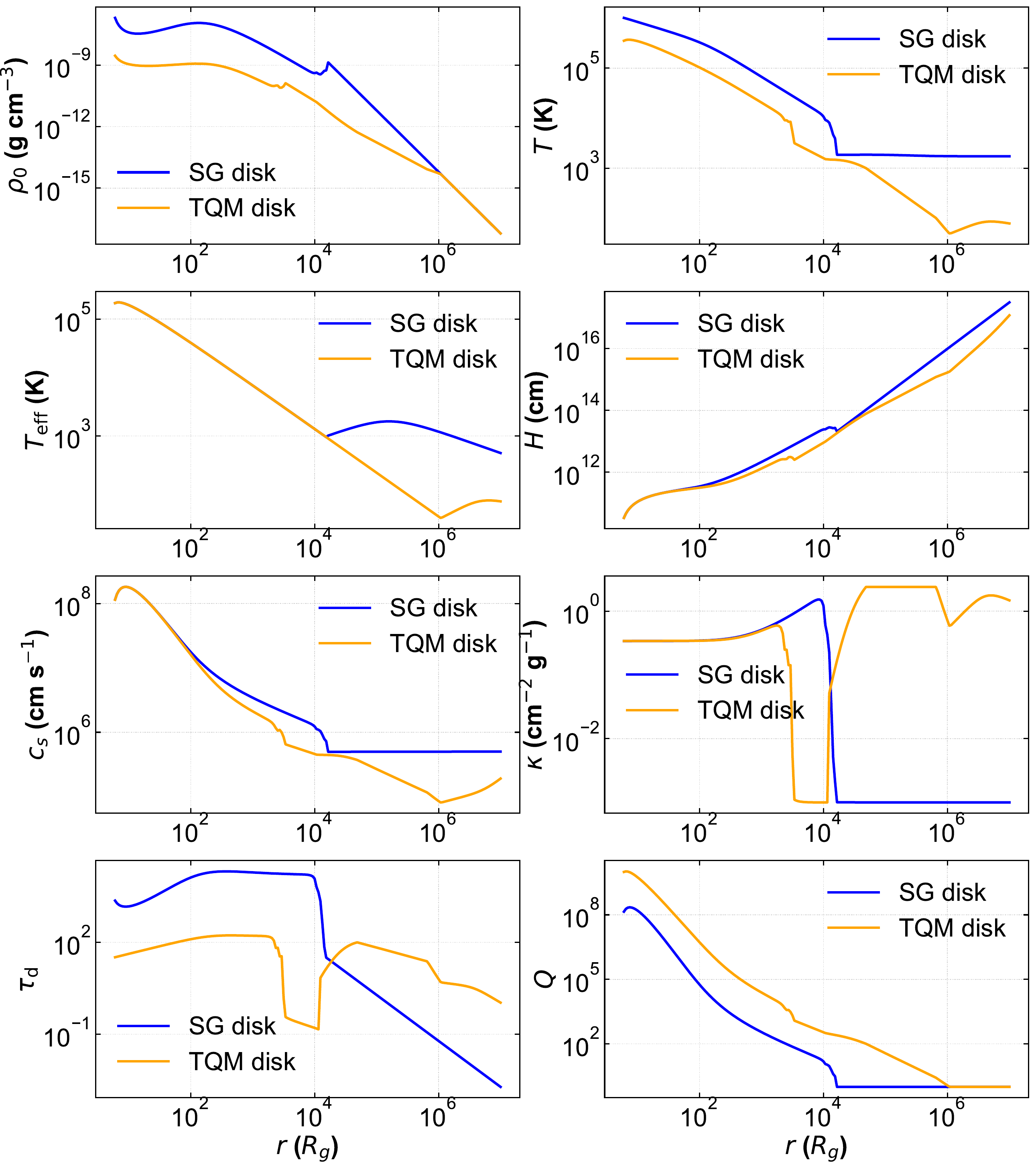}
    \caption{Same as Figure \ref{fig:disk1}, but $M=4\times 10^6 M_{\odot}$.}
    \label{fig:disk2}
\end{figure*}

Different from the $\alpha$-viscosity prescription, the disk angular momentum transfer is assumed to be caused by global torques in the \citetalias{Thompson2005} disk model \citep{Thompson2005}. In outer parts, the disk is assumed to be heated by the star formation process $\sigma T_{\mathrm{eff},\star}^4=\frac{1}{2} \epsilon_{\star}\dot{\Sigma}_{\star} c^2$, where $\dot{\Sigma}_{\star}$ is the star formation rate per unit area and $\epsilon_{\star}$ is the star formation radiation efficiency. The turbulent pressure related to the star formation is also introduced $p_{\mathrm{tb}}=\epsilon_{\star} \dot{\Sigma}_{\star}c$. The disk structures of the TQM model are governed by the following equations \citep{Thompson2005,Pan2021}

\begin{subequations}
\begin{align}
&\sigma T_{\mathrm{eff}}^4=\frac{3}{8 \pi} \dot{M}^{\prime} \Omega^2+\frac{1}{2} \epsilon_{\star} \dot{\Sigma}_{\star}c^2, \\
&T^4=\left(\frac{3}{8} \tau_{\mathrm{d}}+\frac{1}{2}+\frac{1}{4 \tau_{\mathrm{d}}}\right) T_{\mathrm{eff}}^4, \\
&\tau_{\mathrm{d}}=\frac{\kappa \Sigma}{2},\\
&\dot{M}(r)=X c_s(2 \pi r \Sigma) \\
&c_s=H \Omega=\sqrt{p_{\mathrm{tot}} / \rho} \\
&p_{\mathrm{gas}}=\frac{\rho_0 k_{\mathrm{B}} T}{m} \\
&p_{\mathrm{rad}}=\frac{\tau_{\mathrm{d}}}{2c} \sigma T_{\mathrm{eff}}^4 \\
&p_{\mathrm{tb}}=\epsilon_{\star} \dot{\Sigma}_{\star}c \\
&\Sigma=2 \rho_0 H \\
&\dot{M}(r)=\dot{M}^{\prime}+\int_{r_{\mathrm{min}}}^r 2 \pi r \dot{\Sigma}_{\star} d r \\
&\kappa=\kappa\left(\rho_0, T\right),
\end{align}
\end{subequations}
where $X$ is the gas inflow Mach number. For the inner disk, the star formation term vanishes ($\dot{\Sigma}_{\star}=0$). For the outer disk, the disk density is given by Equation (\ref{eq:rho_out}). The structures of TQM disks with $\dot{M}=0.5 \dot{M}_{\mathrm {Edd }}, X=0.1$ and $\epsilon_{\star}=10^{-3}$ are shown in Figures \ref{fig:disk1} (for $M= \times 10^8 M_{\odot}$) and \ref{fig:disk2} (for $M=4 \times 10^6 M_{\odot}$).

%%%%%%%%%%%%%%%%%%%%%%%%%%%%%%%%%%%%%%%%%%%%%%%%%%

% Don't change these lines
\bsp	% typesetting comment
\label{lastpage}
\end{document}